\documentclass[10pt, preprint2]{aastex}
\usepackage{float}
\usepackage{hyperref}
\def\iso#1#2{\mbox{${}^{#2}{\rm #1}$}}
\def\he#1{\iso{He}{#1}}
\def\li#1{\iso{Li}{#1}}
\begin{document}
\title{
Parameters and pitfalls in dark energy models with time varying equation of state}
\begin{abstract}
Are geometrical summaries of the CMB and LSS sufficient for estimating 
cosmological parameters? And how does our
choice of a dark energy model impact the current constraints on standard
cosmological parameters?
We address these questions in the context of the widely used CPL 
parametrization
of a time varying equation of state $w$ in a cosmology allowing spatial
curvature. 
We study examples of different behavior allowed in a CPL 
parametrization in a phase diagram, and relate these to effects on the 
observables. 
We examine parameter constraints in such a cosmology by combining 
WMAP5, SDSS, SNe, HST data sets by comparing the power spectra. We 
carefully 
quantify the differences of these constraints to those obtained by using 
geometrical summaries for the same data sets. 
We find that
(a) using summary parameters instead of the full data sets give parameter
constraints that are similar, but with discernible differences, (b) due 
to degeneracies, the constraints on the standard parameters broaden 
significantly for the same data sets. In particular, we find that in the 
context of CPL dark energy, (i) a Harrison-Zeldovich spectrum
cannot be ruled out at $2\sigma$ levels 
with our current data sets. and
(ii) the SNe IA, HST, and WMAP 5 data are not sufficient to constrain 
spatial curvature; we additionally require the SDSS DR4 data to achieve this. 
\end{abstract} 
\author{Rahul~Biswas\altaffilmark{1}~,
~Benjamin~D.~Wandelt\altaffilmark{1,2,3}}
\affil{$^1$ Department of Physics, 
University of Illinois at Urbana-Champaign, 
1110 W. Green Street, Urbana, \\IL 61801, USA}
\affil{$^2$ 
Department of Astronomy, 
University of Illinois at Urbana-Champaign, 
1002 W.Green Street, 
Urbana,\\ IL 61801, USA}
\affil{$^3$
California Institute of Technology, 
MC 130-33, 
Pasadena,\\CA 91125, USA
}
\maketitle
\section{Introduction}
A number of observations~\citep{1999ApJ...517..565P,1998AJ....116.1009R,1998ApJ...509...74G,2003ApJ...598..102K,2003ApJ...594....1T,2004ApJ...607..665R,2006A&A...447...31A,2007ApJ...666..694W,2009arXiv0901.4804H}have established that the expansion of 
the universe is accelerating. 
The cause is usually attributed to a currently 
dominant component called dark energy. 
Current data is 
consistent with a standard cosmological model called $\Lambda$CDM, with 
dark energy in the form of a cosmological constant $\Lambda$. However, 
dark energy might, 
in fact, be a dynamical component. 
Indeed, there exist numerous models of
cosmology which produce the observed acceleration, 
either by postulating the
the existence of one or more otherwise unobserved fields, 
or as the effects of a departure of 
gravity from General Relativity at large scales that cannot be ruled out
by current data.
Therefore an important objective of
current and future observational efforts is to study the acceleration of 
the universe in different ways and detect departures in the behavior from 
that expected in a standard $\Lambda$CDM  model. 
To this end it is usual to parametrize dark energy as a `fluid' with its 
equation of state (EoS), and the speed of sound in the fluid specified 
independently. 
Generally such a  description  encompasses a large variety of physical 
models if the equation  of state is assumed to depend on the 
density, and the speed of sound depends on both the background density 
and the wavelength of perturbation. 
The idea is that constraining these phenomenological parameters of a fluid 
will narrow down the class of physical models causing the acceleration, and
in particular, study the differences with $\Lambda$CDM.

A specific time dependent parametrization of the EoS, 
and a constant speed of sound describes a subclass of these phenomenological
models. A simple example is the 
CPL parametrization of the equation of state
~\citep{2001IJMPD..10..213C,2003PhRvL..90i1301L}
\begin{equation}
w(a) \equiv w_0 + w_1 (1 - a) 
\label{eq:CPLwa}
\end{equation}
of a non-interacting dark energy, which has been adopted by the Dark Energy
Task force ~\citep{2006astro.ph..9591A} in determining the relative 
importance of future experiments studying dark energy. Conveniently, it includes the case of a constant EoS (wCDM) with $(w_0=w,w_1 =0)$, and the 
$\Lambda CDM$ model ($w_0=-1,w_1=0$).

Parameter constraints on dark energy parameters with different time varying
equation of state, including the CPL parametrization have been 
investigated in the context of similar recent data sets 
~\citep{2007ApJ...664..633W,2008PhRvD..77l3525W,2008JCAP...07..012L,2007ApJ...666..716D}
. These analysis use certain summary parameters 
(the shift parameters $R$ and $l$ for the CMB, and $D$ parameter for the Baryon
 Acoustic Oscillations, see Sec.~\ref{Data} and references therein), 
 which are intended to capture most of the information in data sets. 
In contrast, we use a likelihood analysis that compares the theoretical 
power spectra to spectra inferred from data, as is the standard practice for $\Lambda CDM$ models. 

In this paper, we investigate the constraints on the parameters of a 
non-flat cosmology with a CPL dark energy from current data sets.
These include 
the WMAP five year data set ~\citep{Dunkley:2008ie} for the 
anisotropies in the Cosmic Microwave Background (CMB), 
SDSS DR4 data set for  Luminous Red Galaxies (LRG)
~\citep{2004PhRvD..69j3501T,2006PhRvD..74l3507T}, 
the Union data set for supernovae ~\citep{2008ApJ...686..749K}. 
We point out the differences in 
dark energy parameter constraints computed from the two approaches. 
Evidently, dark energy with  a time varying equation of state 
behaves very differently from $\Lambda$CDM cosmology, where the fractional 
density of dark energy is tied to acceleration at a particular redshift. 
Even in the simple and comparatively benign CPL model this gives rise to counter-intuitive effects on general parameter 
constraints, which we study here. While we discuss the behavior of 
generic time varying EoS, we restrict our calculations to the specific case
 of the CPL parametrization. This is an appropriate example, not only 
because of the current consensus of using the CPL as a standard 
~\citep{2006astro.ph..9591A}, but also because the it is a fairly benign 
evolution. Model independent constraints one might hope to study
will include much stiffer variation of the EoS, where the dynamical effects
we describe (see Sec.~\ref{motivation}) could be more pronounced possibly 
leading to larger differences with parameter constraints computed from 
summary parameters.  

In Sec.~\ref{motivation}, we discuss the behavior of different relevant 
regions of the CPL parameter space, their observable signatures and their
possible consequences for parameter estimation. In Sec.~\ref{methods} 
we describe the details of our method to investigate parameter constraints
derived using summary parameters and power spectra. The results of our
investigations are presented in Sec.~\ref{ResultsSection}. We discuss
our conclusions and possible implications in Sec.~\ref{Disc}.

\section{Characteristics of Dark Energy with Time Varying Equation of State }
\label{motivation}

The density of a dark energy component with a 
variable equation of state $w(a)$ evolves with scale factor as 
$\rho_{DE}~\sim~a^{-3(1+w_{\rm{eff}}(a))},$ while the pressure $P(a)~\sim~(1+w(a))\rho_{\rm{DE}}(a),$ where
\begin{eqnarray}
w_{\rm{eff}}(a)~=~-1 +\frac{\int_1^a da (1+w(a))/a}{ln(a)}~\nonumber\\
=~(w_0 + w_1) + (1 - a)w_1/ln(a)\nonumber
\end{eqnarray}
with the last expression being valid for a CPL EoS. 
Dark energy of current density 
(in units of critical density today) $\Omega_{DE}$ contributes an 
amount
$\ddot{a}=a H_0^2\Omega_{DE}(1+ 3 w(a))a^{-3(w_{\rm{eff}}(a) +1)}$ towards the 
acceleration of the universe at a redshift $z~=~1/a -1,$ while its density 
in units of matter density grows as $\sim a^{3 w_{
\rm{eff}}}.$
Thus, the presence of dark energy at a particular redshift modifies the 
background expansion. This (along with the presence of curvature) 
affects the CMB and 
the matter power spectrum (a) 
geometrically by altering the angular position of 
the peaks and  (b) dynamically by altering the magnitude of the spectrum. 
Dark 
energy is believed to be smooth 
(ie. its density perturbations do not grow at scales smaller than the 
Hubble scale), leading to less clustering of density perturbations 
at a particular redshift
if a 
significant fraction of the background density is made up of dark energy. 
The gravitational potentials also evolve differently 
because the background evolves differently from a matter dominated 
universe. This leads to weaker sourcing of the growth of 
perturbations, thereby suppressing the number of galaxies formed at a 
particular redshift reducing the matter power spectrum inferred from
galaxy surveys~\citep{2001PhRvD..64l3520D}. The change in gravitational potentials also affects the CMB
power spectrum through the Integrated Sachs-Wolfe (ISW) effect. On a plot 
of the matter power spectrum or CMB power spectrum 
the geometrical effects
show up as horizontal differences (change in scales), while the dynamical 
effects show up on the vertical differences (magnitudes). 

For a dark energy characterized by a constant EoS, $w_{
\rm{eff}}(a)~=~w(a)~=w$. 
To explain the acceleration observed today from SN IA data, $w < -1/3$ and
is close to $-1$ if all data sets are considered. This also implies that 
the 
dark energy density (in units of matter density) grows as $\approx a^{3},$
so that it is negligible at earlier times. Hence, its dynamical effect on 
the matter power spectrum is small, and its effect on the CMB power 
spectrum is limited to a late time ISW effect, which only affects the 
low multipoles of the spectrum. Since the cosmic variance 
at the low multipoles is high, the observable imprints of a redshift 
independent equation of state are limited to the geometrical effects.
\begin{figure}[!tp]
\begin{center}
\includegraphics[width=0.45\textwidth]{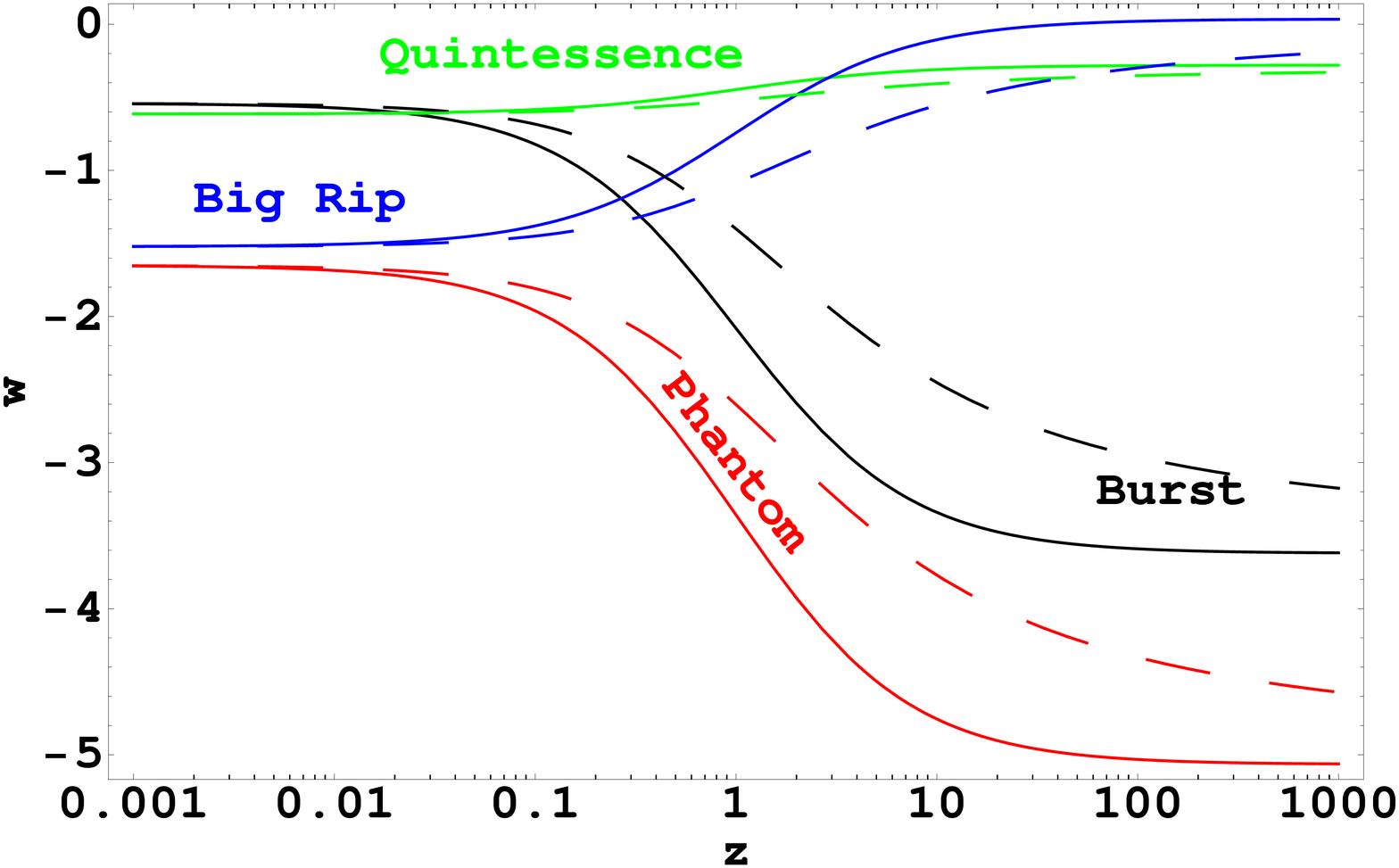}
\hfill
\includegraphics[width=0.45\textwidth]{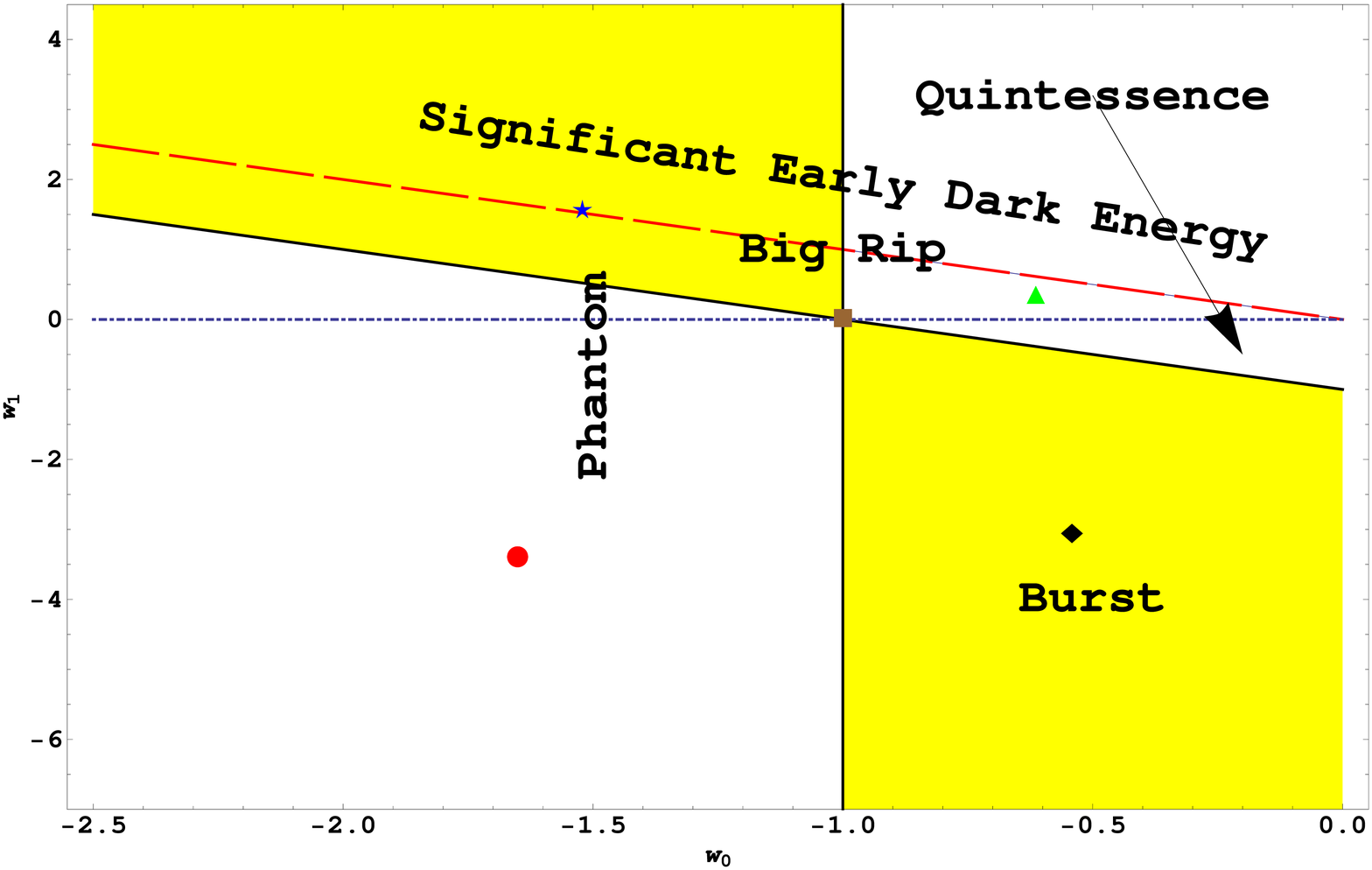}
\end{center}
\caption{~The upper panel 
shows the evolution of w (solid) and 
$w_{ \rm{eff}}$ 
(dashed) for the set  ($w_0,w_1$) = (-1.65116, -3.41463)  
(\rm{Red}), (-0.541648, -3.07916) (\rm{Black}) 
(0.613754, 0.333922) (\rm{Green}),(-1.52168, 1.5576) (\rm{Blue}). 
The lower panel shows the Phase Diagram 
of the CPL model. The regions above the horizontal solid black line 
are parameters that will eventually go over to the Phantom phase causing a Big Rip, 
while the regions to the left of the solid black vertical line are in the 
phantom phase now. The region above the red diagonal line have significant
dark energy at early times. The shaded regions `cross the phantom divide',
while the regions in the South Eastern quadrant lead to a short burst of
acceleration. The dark energy parameters plotted in the upper panel are 
also marked.
\label{PhaseDiagram}
} 
\end{figure}
For a redshift dependent equation of state, $w(a)$ and therefore 
$w_{
\rm{eff}}(a)$ 
can be quite different from $w(0)$ for smaller values of $a$. Hence, it is
possible to simultaneously have the observed acceleration due to very 
negative values of $w(a=1),$ while the ratio of densities of dark energy to
matter $\sim a^{3 w_{
\rm{eff}}(a)}$ remains significant at 
early times, if $w_{
\rm{eff}}(a)\approx 0.$ Thus dark energy 
with redshift dependent EoS can cause acceleration today, and also be 
non-negligible at early times, thereby suppressing the growth of 
structures,
and affecting the CMB by an Early ISW Effect. It is possible
for models with a time dependent EoS to have regions in 
parameter space which are relatively indistinguishable 
in terms of the geometric effects, but distinguishable in terms 
of dynamical effects.  
This kind of geometrical degeneracy is 
likely to be more pronounced for parametrizations which allow a strong 
variation of $w(a)$, as that would facilitate a quicker transition from 
an equation of  state around $-1$ to $0$. Here, we will work out the 
consequences in the specific example of the widely used CPL 
parametrization in Eq.~\ref{eq:CPLwa}, which is a fairly gentle variation.

\begin{figure}[!ht]
\begin{center}
\includegraphics[width=0.45\textwidth]{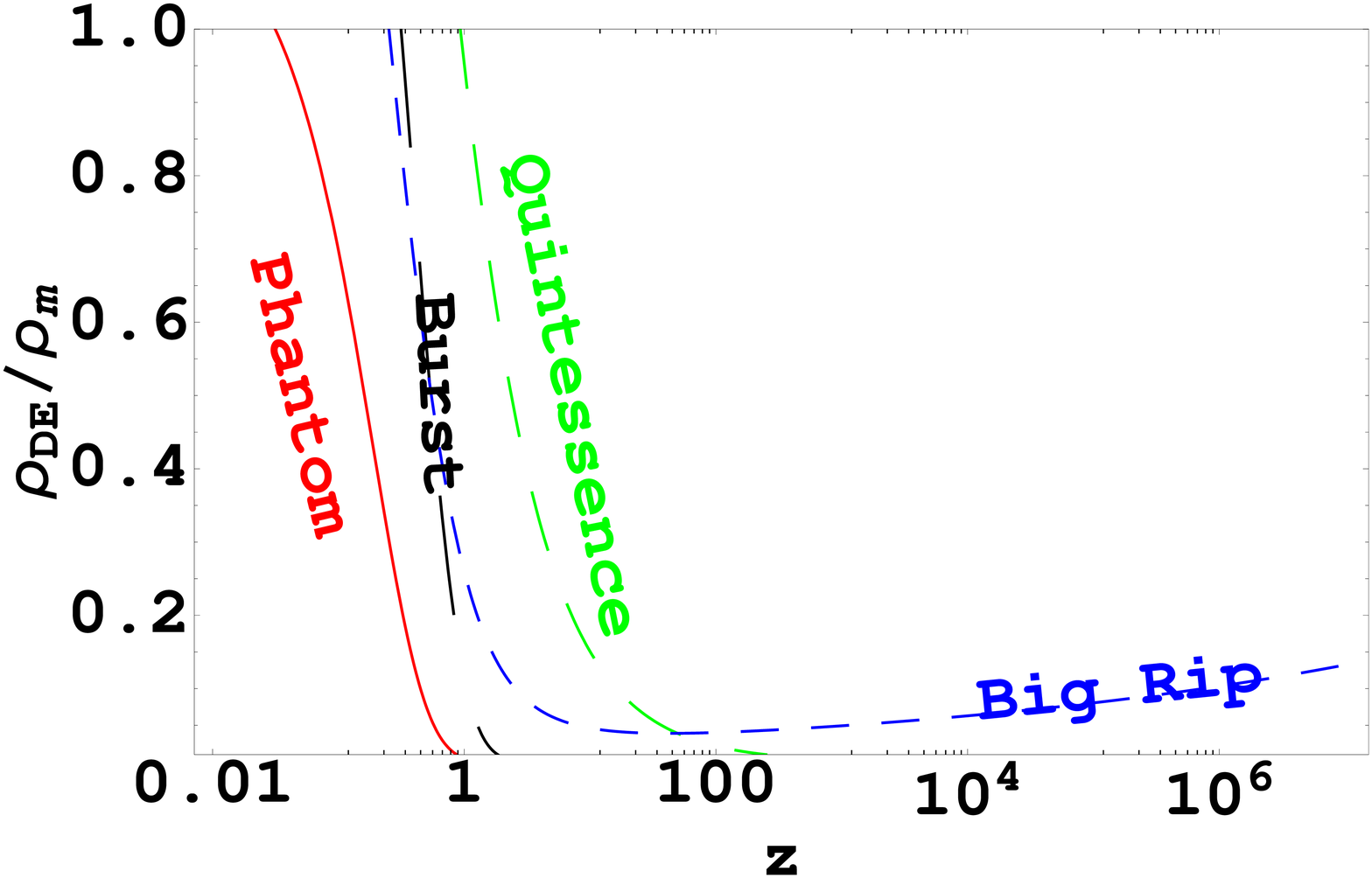}
\hfill
\includegraphics[width=0.45\textwidth]{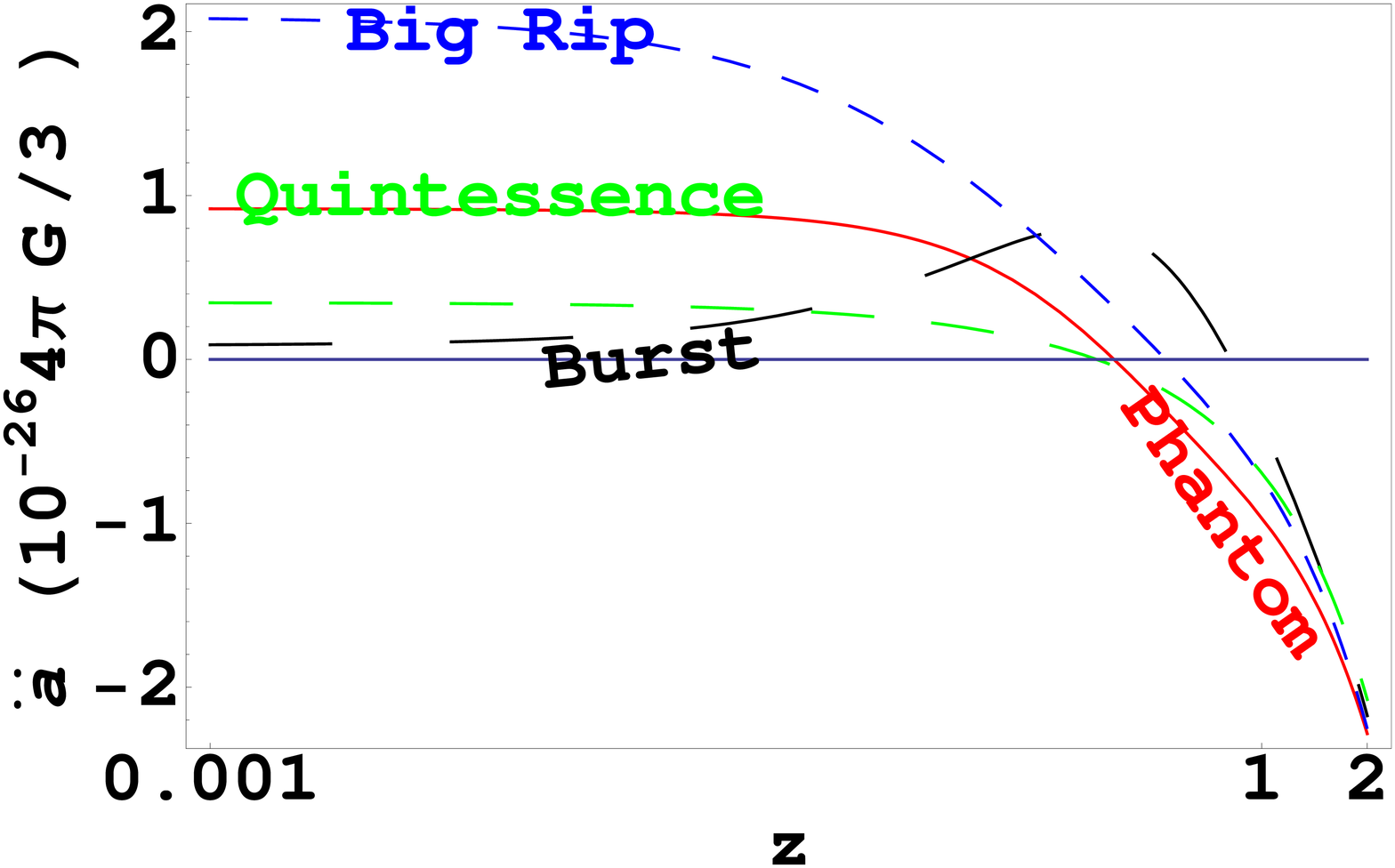}
\hfill
\includegraphics[width=0.45\textwidth]{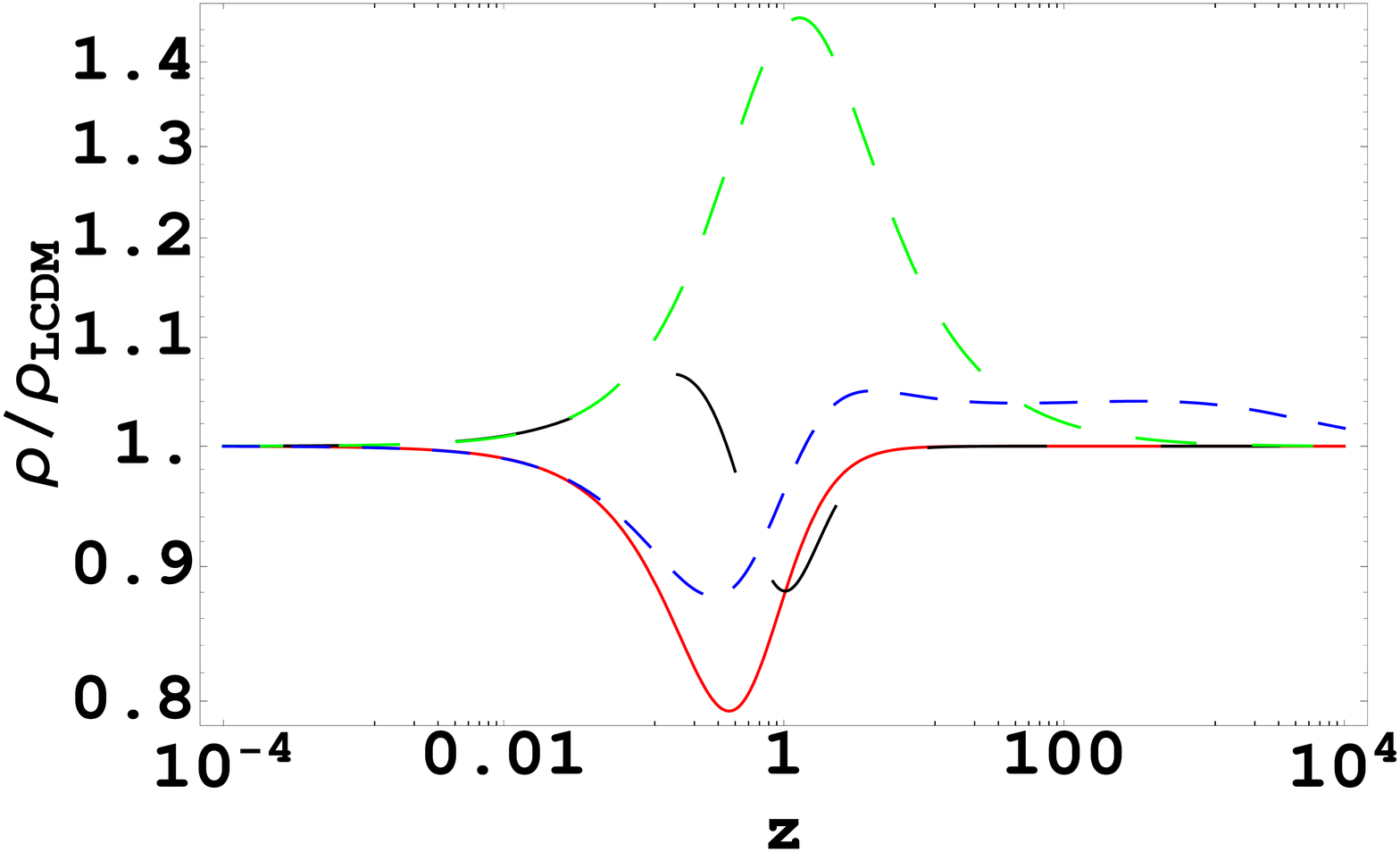}
\caption{~The upper panel
shows the ratio of the densities of dark energy
to the density of dark matter in models selected from different parts of 
the phase diagram in Fig.~\ref{PhaseDiagram}. 
The middle panel shows that even though dark energy dominates at 
large redshifts for the (Big Rip, Phantom) kind of dark energy models
(blue line), it does not accelerate the universe at early times. 
The lower panel shows the ratio of the total density of components
(matter, dark energy, relativistic components) in units of the total 
density assuming $(w_0,w_1)$=$(-1,0)$ of the models selected 
in Fig.~\ref{PhaseDiagram}
\label{EffectonBackground}
}
\end{center}
\end{figure}
The CPL parametrization is an ad-hoc parametrization which allows the 
dark energy EoS $w(a)$ (and $w_{
\rm{eff}}(a)$) 
to asymptote between two constant values $w_0 +w_1$ and $w_0$ at early 
and late times. In order to understand the behavior of this parametrization
and the physical models it may represent, we can study a phase
diagram on the $w_0,w_1$ plane. In the upper panel of 
Fig.~\ref{PhaseDiagram}, 
we show the 
evolution of the CPL EoS $w$ (solid lines) and $w_{\rm{eff}}$ (dashed). 
It is 
useful to study this parametrization as it captures general features of
dynamical dark energies, and CPL models that correspond to $(w_0,w_1)$ 
different from $(-1,0)$ are definitely dynamical models distinct from
a cosmological constant.
From the asymptotic behavior of the upper panel of Fig.~\ref{PhaseDiagram},
we can study different phases of the CPL EoS, where the evolution of
dark energy can be asymptotically similar to different models of 
dark energy. 
This is studied in the lower panel of Fig.~\ref{PhaseDiagram}.
Firstly we can separate the regions where the dark energy is in the
Phantom phase with equation of state ($w(a) <-1$) which occurs for
scalar field theories with tachyonic instabilities, or with non canonical 
kinetic energy terms inspired by higher derivative theories
~\citep{2002PhLB..545...23C,2003PhRvD..68b3509C}. 
The two black lines (solid) bound the regions where the dark 
energy behaves like a phantom model ($w(a) <-1$): the parameter region 
left of the vertical black line is currently in the phantom phase, while
the parameter region below the diagonal black line was in a phantom phase 
at early times. 
The region above the horizontal dot-dashed blue line will eventually 
become a phantom model, resulting in a `Big Rip'. The region below 
the horizontal dot-dashed line will eventually become `normal' with
$w >-1$. The region between the dot-dashed blue line and the diagonal 
black line (labeled quintessence), to the right of the vertical black 
line is the only 
region where $w(a) >-1$ at all times. Thus this is
the only phase which may be similar to models of a single, non-interacting 
stable canonical scalar field.
The North Western and South Eastern quadrants (shaded yellow)
marked out by the solid black lines exhibit the crossing behavior from 
the `phantom phase' to the `normal phase' where $w$ crosses $-1$; this 
leads to instabilities in the evolution of perturbations at the crossing 
(see Sec.~\ref{computationPS} and references therein for details).
The South Eastern quadrant shows a rapid change of $w(a)$ at 
recent times, but decays rapidly with redshift.
The dashed red line marks the boundary at which $w(a)=0$ at early 
times; the regions above this line have significant 
dark energy density at early times like recombination. In order to 
study the distinct effects, we mark five points on this phase diagram. 
The brown point represents a cosmological constant with 
$\Omega_{\rm{m}}=0.3, \Omega_{\rm{DE}}=0.7, H_0=72 \rm{km/s/Mpc}.$
The other points were chosen from our chains representing the posterior
of current CMB , HST and SNe data using summary parameters 
(see Sec.~\ref{Data} for details). 
The red filled circle is a `Phantom', the black diamond is `Burst DE',
the green triangle is similar to a `Quintessence' till now, while the 
blue star will have a `Big Rip' and is chosen to have significant early
dark energy. We will stick with this color code in discussing effects, 
and use these names to label the plots when possible.

The effect of dark energy parameters of these distinctive types can be 
seen in Fig.~\ref{EffectonBackground}, where we show the impact 
on the background 
for the parameter points marked out in Fig.~\ref{PhaseDiagram}. 
The upper panel shows the ratio of dark energy density to 
dark matter density as a function of redshift. We notice that 
the dark energy density decays with redshift, except for the model
`Big Rip', which had significant early dark energy (cf lower panel 
Fig.~\ref{PhaseDiagram}).
The middle panel shows the acceleration of the
universe  for dark energy with the same parameters. 
Clearly, the `Phantom' and `Big Rip' models with very low values
of $w_0$ exhibit a super-accelerated phase. 
It is interesting to note that even for the `Big Rip' 
where dark energy density is significant at early times, the 
universe is decelerating before a redshift of $2$ because its EoS 
(Blue solid curve in the upper panel of Fig.~\ref{PhaseDiagram})
keeps increasing.
Hence for these models with significant early dark energy, 
the universe can 
be `dark energy dominated' without accelerating, with
the effect of dark energy on background expansion being similar to matter.
The lower 
panel shows the evolution of the total density of matter, dark energy and
relativistic components.
$\rho_{tot}= \rho_{m} + \rho_{DE} +\rho_{r}$ with redshift, in units 
of the corresponding quantity assuming that dark energy was a cosmological 
constant.

\begin{figure}[!htp]
\begin{center}
\includegraphics[width=0.45\textwidth]{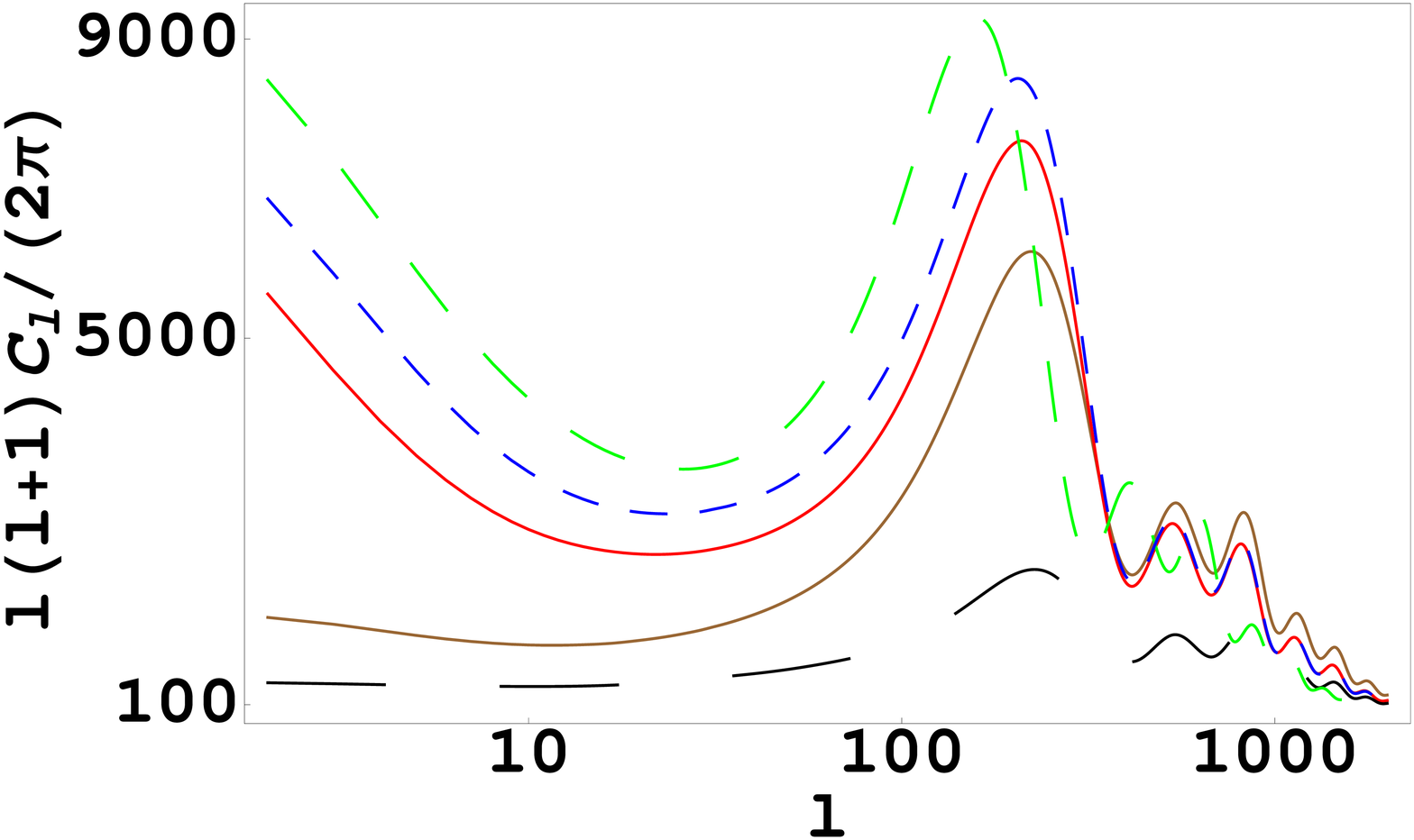}
\hfill
\includegraphics[width=0.45\textwidth]{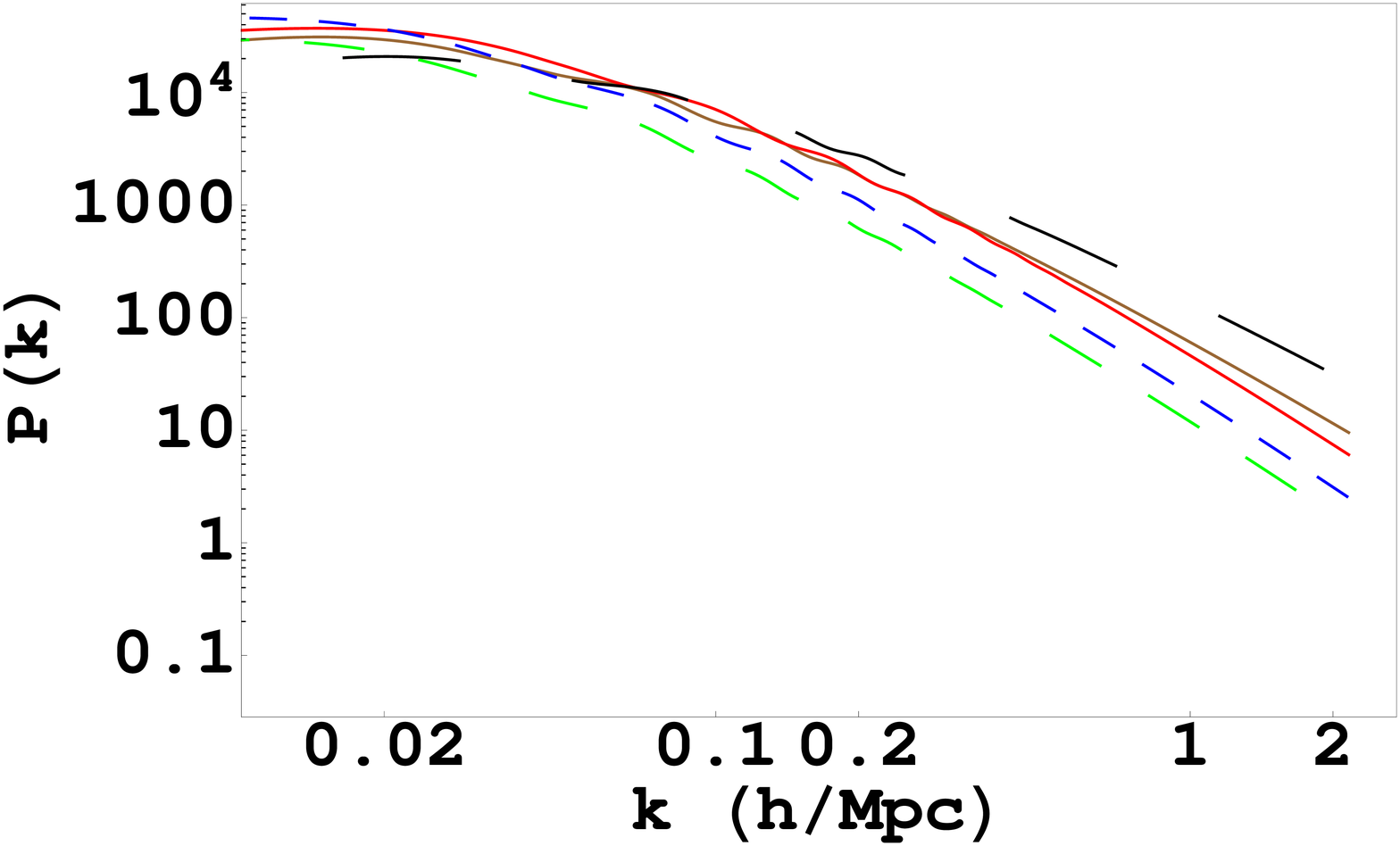}
\hfill
\includegraphics[width=0.45\textwidth]{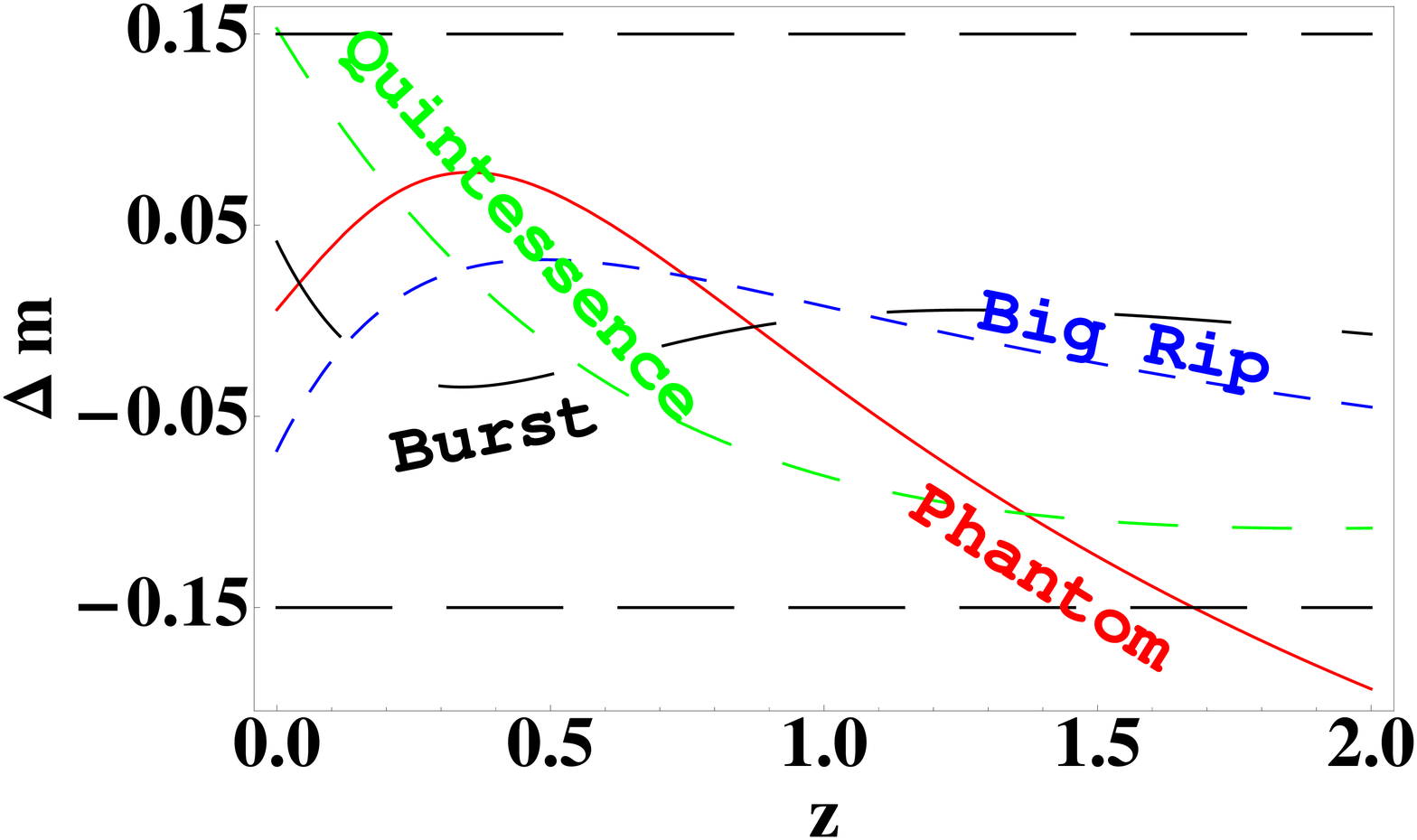}
\caption{~Effect on observables for models with different values of dark 
energy parameters. The parameters are the points marked in the 
phase diagram of Fig.~\ref{PhaseDiagram}, chosen from different regions 
marked on the phase diagram.: 
The upper panel shows the CMB power spectrum for the parameters 
shown. 
The middle panel shows the matter power spectrum as a 
function of wave-number in units of $h^{-1} Mpc$. 
The lower panel shows the change $\Delta m$ in the apparent magnitude 
with 
redshift $z.$ However, we use a likelihood which is analytically 
marginalized over 
the absolute magnitude of the supernovae; hence we only show $\Delta m$ 
up to a constant for each cosmology. The horizontal dashed black lines
are at $\pm 0.15$ which is the usual intrinsic dispersion for supernovae.
\label{EffectonObservables}
}
\end{center}
\end{figure}

The effect of these parameters on observables is studied in 
Fig.~\ref{EffectonObservables} for the models chosen in 
Fig.~\ref{PhaseDiagram}. Since we are interested in comparing the
effect on the magnitudes of the spectra due to dark energy, 
we set the amplitude of the primordial power spectrum to be the 
same for each model. This does not change the value 
of the posterior (using Likelihood B, see Sec.~\ref{Data}) probability of 
the model plotted, as the posterior is independent of the amplitude.
The upper panel shows the CMB power spectrum,
the middle panel shows the matter power spectrum, and the lower panel
shows the apparent magnitude of the SNe IA, which is a purely geometrical effect. The difference in positions of
the angular positions of peaks in both the CMB and the oscillations in 
matter power spectrum, as also the differences in the redshift-magnitude 
plot is due to geometrical effects. The difference in the magnitudes of 
the power spectrum is due to dynamical effects that are not captured in
summary parameters (see Sec.~\ref{Data}). It is interesting to note that 
the models for which the CMB power spectrum is enhanced have suppressed 
matter power spectrum.

The comparison of the difference of the CMB and Matter Power Spectra
inferred from the data with their theoretically computed counterparts with 
the scale of expected deviations leads to constraints on cosmological 
parameters.
It is possible that a subset of this information pertaining to 
specific features in the spectra is almost as useful in constraining
parameters. For example, 
many features of CMB power spectra depend on the 
cosmological parameters through very specific functions of the these 
parameters ~\citep{2001ApJ...549..669H,2002ARA&A..40..171H,2001ApJ...559..501D,2002MNRAS.330..965D}.
This motivates the idea of using CMB shift parameters to 
summarize the information content of the CMB anisotropies, introduced in 
~\citet{1997MNRAS.291L..33B} in the context of forecasting for 
standard cosmologies. Further work on this subject 
~\citep{2007A&A...471...65E,2007PhRvD..76j3533W} suggests that 
one can summarize the information the in the spectra 
efficiently in terms of the summary parameters $R, l_a$, which relate to 
the position of the first peak of the CMB spectrum, and the spacing of the
peaks due to acoustic oscillations.
Recently such summary parameters have become popular in 
studying the constraints on dark energy parameters. 
If the effect of dark energy is mostly geometrical, it is
tempting to speed up the calculation by summarizing the CMB/LSS data in 
terms of a few geometrical summary parameters that describe these effects, 
instead of going through the 
time consuming process of theoretically computing the angular power 
spectrum. 

This approach of using summary parameters has been studied in 
~\citet{2007PhRvD..76j3533W,2007A&A...471...65E,2002MNRAS.330..965D}, and 
the procedure nicely outlined in ~\citet{2008arXiv0803.0547K}. 
One 
finds samples of the posterior distribution of the data set concerned for 
cosmological models with a specific form ($\Lambda CDM$) of dark energy. One
then uses these samples to estimate  the set of summary parameters over
the posterior distribution as well as the covariance over these summary 
parameters. 
~\citet{2007PhRvD..76j3533W} also show that the summary 
parameters are comparatively weakly correlated to the other cosmological 
parameters. 
One then 
approximates the likelihood of the CMB data set, by a Gaussian distribution
 over these summary parameters
It can be seen that there are three crucial assumptions in this 
procedure: 
\begin{enumerate}
\item The Likelihood of the CMB spectra are well described by CMB 
summary parameters. 
\item The mean and covariance of the summary parameters for a particular
data set do not change appreciably when the model space is enlarged, and do
not develop correlations.
\item The summary parameters are relatively weakly correlated with the 
other cosmological parameters.
\end{enumerate}
We examine these by studying the differences in the parameter 
constraints in 
comparison of the power spectra and the use of the summary parameters.

\medskip
\section{Method}
\label{methods}
We study parameter constraints 
using a 
suitably modified version of the Markov Chain Monte Carlo (MCMC) engine
CosmoMC
~\citep{2002PhRvD..66j3511L}
to explore the Bayesian posteriors for the cosmological parameters. We
assume an isotropic and homogeneous universe with dynamics dictated by 
standard general relativity with the densities of the background components
to be determined, take the non-baryonic dark matter to be entirely cold
 and neglect effects of neutrino mass, assuming 
$N_{\rm{eff}}=3.04$ species of massless neutrinos. 
The current background density
of photons is also assumed to be fixed, by neglecting any error-bars 
associated with the measurement of the CMB temperature. We then
explore possible values for the background densities of other components 
(baryons, cold dark matter, curvature) with broad, flat 
priors over $\omega_{\rm{b}}, \omega_{\rm{c}}, \Omega_{\rm{k}},$ 
where $\Omega_{\rm{i}}$ is the density
of the component  in units of the current critical density, and 
$\omega_{\rm{i}} = h^2 \Omega_{\rm{i}},$ with the Hubble constant 
$H_0 = 100 h$ Km/s/Mpc.
The primordial perturbations are assumed to be adiabatic and Gaussian 
distributed with 
a power spectrum $P_{\rm{j}}(k)= A_{\rm{j}} (k/k_{\rm{p}})^{(n_{\rm{j}}(k) -1)}.$ We neglect the 
effect of tensor perturbations setting the ratio $r$ of its amplitude 
$A_{\rm{t}}$ to that of the scalar perturbations $A_{\rm{s}}$ to be zero, and $n_{\rm{t}}=0$. The
scalar spectral index $n_{\rm{s}}$ is assumed to be scale independent (running of 
spectral index) $n_{\rm{r}}=0$ and we choose a pivot scale 
$k_{\rm{p}} = 0.05/Mpc$. 
We also ignore the effect of the the SZ amplitude.
We allow for a single re-ionization taking place at an optical depth of 
$\tau$, and $\theta$ parametrizing the angle subtended by the sound 
horizon at the surface of last scattering instead of the Hubble constant 
$H_0$.
Therefore, our model space has the fixed values  
$\omega_{{\nu}}= n_{\rm{t}} =n_{\rm{r}} =r= 0,$ and priors on the 
parameters 
$\{\Omega_{\rm{b}}h^2, \Omega_{\rm{c}}h^2, \theta, \tau, \Omega_{\rm{k}}, w_0, w_1,n_s, \rm{Log}(10^10 A_s) \}$ 
summarized in Table~\ref{PriorsTable}.
The Markov chains are assumed to have converged when the $R-1$ statistic
had been below 0.03 for a few tens of thousand chain steps; this results in
a final $R-1$ statistic of about $\sim 4 \times 10^{-3}- 1\times 10^{-2}$. 
\begin{table*}
\caption{Parameters used in the MCMC: Cosmological parameters used and the lower and upper limit on the flat priors on these parameters. For a $w$CDM 
model, $w_1$ is fixed to be 0.}
\begin{center}
\begin{tabular}{|c|c|c|c|c|c|c|c|c|}
\hline
$\Omega_{\rm{b}}\rm{h}^2$ & $\Omega_{\rm{c}}\rm{h}^2$
& $\theta$ & $\tau$ & $\Omega_{\rm{k}}$ & $w_0$ &$w_1$ & $n_s$ &
$log(10^{10}A_s)$ \\
\hline
(0.005,1) & (0.01,0.99) & (0.5,10) & (0.01,0.6) & (-0.2,0.2) & (-3,1.5) &
(-7.0,4.5) & (0.5,1.5) & (2.7,4)\\
\hline
\end{tabular}
\end{center}
\label{PriorsTable}
\end{table*}

\begin{table*}
\caption{Definitions of data sets I and II and Likelihoods A and B}
\begin{center}
\begin{tabular}{|c|c|c|c|}
\hline
Data Set I& Data Set II& Likelihood A & Likelihood B\\
\hline
WMAP5 + SNE + HST & WMAP5 + SNE + HST  & Summary Parameters & Power Spectra\\
 & + SDSS + BBN & $l_{\rm{A}}, R_{\rm{A}}, z_{\star}, D_{\rm{A}}$ & \\ 
\hline
\end{tabular}
\label{LikesTable}
\end{center}
\end{table*}

\subsection{Data and Likelihoods}
\label{Data}
We
summarize our usage of different data sets and likelihoods in 
Table~\ref{LikesTable} and explain it in detail below.\\ 

\noindent
{\it{CMB data:}} We use the WMAP 5 year data in two different ways:
\\
(A) By using the publicly available WMAP likelihood code (version 3)
~\citep{Dunkley:2008ie,2008arXiv0803.0732H,2008arXiv0803.0593N}, 
which compares the observation to a 
our theoretical computation of the  CMB power spectrum.
and\\ 
(B) By using a  Gaussian likelihood in the summary parameters 
$\{l_A, R, z_{\star}\}$ as recommended by~\citet{2008arXiv0803.0547K}
\begin{eqnarray}
l_A~=~(1+z_{\star})\frac{\pi D_A(z_\star)}{r_s(z_\star)}\nonumber\\
R~=~\sqrt{\Omega_m H_0^2}(1+z_\star ) D_A(z_\star) 
\end{eqnarray}
where $z_{\star}$ is the redshift of the surface of the last scattering 
computed from the fitting formula 
~\citep{1996ApJ...471..542H}
in terms of only densities of baryons ($\omega_b$) 
and matter ($\omega_m$)
with the mean of the distribution taken to be the maximum Likelihood values
in Table 10, and the inverse covariance matrix in Table 11 of
~\citet{2008arXiv0803.0547K}. There are slight differences in the literature
about how the parameters $l_a, R $ are best defined, and we adopt the 
definition of~\citet{2008arXiv0803.0547K} as we use their numerical values
for the likelihood.\\ 

\noindent
{\it{Galaxy Power Spectrum Data:}} We use the galaxy power spectrum data
from the 
Luminous Red Galaxy (LRG) sample of the 
Sloan Digital Sky Survey DR4 (SDSS)\\  
(A) By using a modified version of the  publicly available likelihood
code which compares the matter power spectrum inferred from the data with 
the theoretical computation after analytic marginalization over a scale 
independent linear bias~\citep{2006PhRvD..74l3507T}. 
We modified the code in order to recompute the 
geometric scaling at the redshift of the sample related to 
$$D_V~=~\left( (1+z)^2 D_A^2(z) c z/H(z) \right)^{1/3} $$
for the CPL model, and also to compute the growth function to
account for the scale independent change in the  matter power spectrum at 
the mean redshift $0.35$ of the LRG sample from the matter power spectrum 
at a redshift of $z=0$.\\ 
(B) By using only the geometric distance measure $r_s(z_d)/D_V(z),$ where 
$r_s$ is the sound horizon at the redshift of drag epoch $z_d$ where the
baryons were released from the photons. Following
~\citet{2008arXiv0803.0547K},
the redshift $z_d$ is computed 
through a fitting function 
~\citep{1998ApJ...498..497H}
which again depends only on $\omega_{\rm{b}}$ and 
$\omega_{\rm{m}}.$ As recommended, we use a
Gaussian Likelihood with a mean of $0.1094$ and a standard deviation of 
$0.0033$. 
\\

\noindent
{\it{Supernovae Data:}} We use the Union data set 
~\citep{2008ApJ...686..749K} 
which is a compilation of  307 supernovae IA discovered in different 
surveys. This combines high redshift supernovae by the ESSENCE, SNLS and 
HST Goods surveys, with low redshift ones ($z\approx 0.02 -0.1$)  
~\citet{2006A&A...447...31A,1999ApJ...517..565P,2004ApJ...607..665R}
using a weighting scheme to take into account the heterogeneous sources. The 
Likelihood is Gaussian distributed in the magnitude space and includes covariance 
contributions due to systematic effects. However, we ignore lensing of 
supernovae.\\ 
{\it{HST:}}
We  also incorporate the results of the Hubble Space Telescope Survey 
measurements of the Hubble constant ~\citep{2001ApJ...553...47F} as a 
Gaussian prior on the value of the  Hubble constant 
$H_0 = 72 \pm 8 km/s/Mpc.$ \\ 
{\it{Big Bang Nucleosynthesis:}}
The primordial abundance of light nuclei, determined at the time of 
Big Bang Nucleosynthesis (BBN) depends on the baryon to photon 
ratio, as well as the expansion rate at this time
~\citep{Amsler:2008zzb,2002PhRvD..65l3503C}. Since, in time varying 
equation of state dark energy models, dark energy can be non-negligible 
around the time of BBN ($z\sim 10^{7}$), we need to examine the effect of
BBN on both these parameters. 
Among the abundances of light elements, the abundance of primordial 
Helium is most sensitive to the expansion rate during BBN; 
the abundance of primordial Deuterium, while almost insensitive to 
the expansion rate, is  extremely sensitive to the baryon to photon ratio.
We ignore the abundance of \li7 even though it is extremely sensitive 
to the baryon photon ratio, because of the controversies regarding 
systematic uncertainties in determining the observed abundances of 
\li7, stemming from uncertainties in measurement of effective 
temperature of stars, or an unaccounted correlation between the 
metallicity and estimated abundance ~\citep{2008JCAP...11..012C}.
We constrain the dark energy density at the time of BBN by 
using the fit equations for the primordial abundance of Deuterium, and 
Helium from ~\citep{2008JCAP...06..016S} 
in terms of the baryon-to-photon ratio 
$\eta_{10},$ and the ratio $S$ of the Hubble parameter at times of BBN
($z\sim 10^7$)
with a Hubble parameter for a universe completely dominated by relativistic
 degrees of freedom (photons and massless neutrinos). The parameter 
$\eta_{10}$ depends on the mass fractions of the light nuclei, but the 
dependence is extremely weak. We adopt the values of 
$\eta_{10} =273.9 \omega_b$ for our Likelihood calculations. 
We ignore the theoretical errors in the fitting functions, and 
write a Gaussian likelihood using the error 
estimates of observed abundances for De and \he4. We also note that 
since the \he4 fraction monotonically increases with time, the lowest
values detected are a robust upper bound to the Helium fraction. We impose 
 a much weaker hard prior on the equation of state parameters, 
as will be described and further justified in the 
Sec.~\ref{ResultsSection}. 
\subsection{Theoretical Computation of Power Spectra}
\label{computationPS}
The Likelihoods (A) for the CMB and Matter Power Spectra data require 
theoretically computed values of these power spectra. 
We evaluate these by modifying the background expansion, and 
the perturbation equations for this Dark Energy model using CAMB 
~\citep{2000ApJ...538..473L}, 
which uses RECFAST ~\citep{Seager:1999bc,Seager:1999km,Wong:2007ym} 
to compute the recombination history. 
The CPL parametrization allows for the case where the dark energy equation 
of state is less than $-1$, and allows crossing of  $-1$ during the 
evolution. For a single non-interacting scalar field, $w \geq -1$. 
However, lower values of $w$ occur for scalar fields with non-canonical 
kinetic energy terms, or due to interaction between more than one field 
~\citep{2005PhRvD..71d7301H,2006PhRvD..74b3519H,2006MPLA...21..231Z,Li:2005fm,2008PhRvD..78h7303F}. 
Thus, we follow the 
standard practice and do not rule out  values of $w(a) < -1.$ 
It is known, that in the fluid model perturbation theory of dark 
energy, there is a runaway problem associated with the models which 
allow crossing of $-1$. 
We follow the prescription of 
~\citep{2004astro.ph.11102H,2005PhRvD..72d3527C}, 
which essentially assumes that this is a numerical artifact. 
While, a negative dark energy sound speed would lead to large clustering
of structure that  is unobserved in data, 
we find that current data is insufficient
to put any meaningful constraints in the open 
interval (0,1) (in natural units) 
on the speed of sound $c_s$ for dark energy. 
Therefore, we fix the value of the speed of sound to a reasonable choice
of $c_s^2=1$ as would be the case for quintessence~\citep{2003MNRAS.346..987W,2005PhRvD..72d3527C}. 

\medskip
\section{Results}
\label{ResultsSection}
\begin{figure*}[!hp]
\begin{centering}
\includegraphics[width=1.0\textwidth]{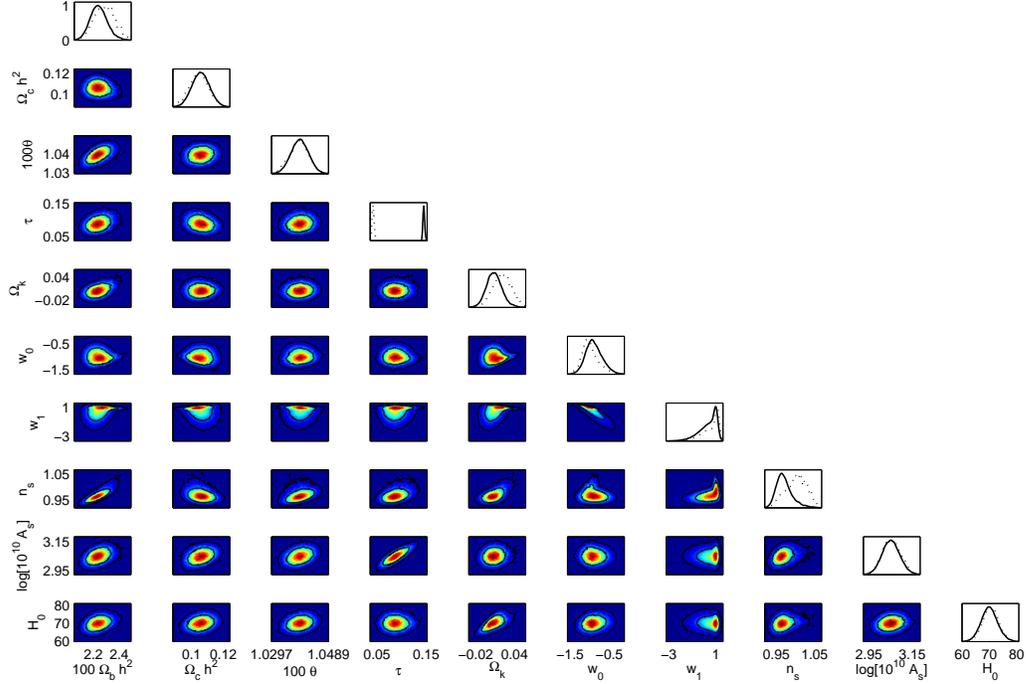}
\caption{
~Marginalized one dimensional and two dimensional joint posterior 
distributions of our maximal data set (WMAP5, SDSS, HST, SNe , 
and BBN) on the cosmological parameters. The solid lines show the
posteriors, while the dotted lines show the mean likelihoods.}
\label{FL_maximalconstraints}
\end{centering}
\end{figure*}

We study the posterior distributions on the cosmological parameters 
and the joint posterior distributions of pairs of cosmological parameters. 
Use of the maximal data set available today exploits the complementarity of
different probes to obtain the tightest constraints on the parameters. 
On the other hand, it is useful to examine meaningful
constraints from different subsets of the maximal data set. Constraints 
from subsets allow one to check for self-consistency of cosmological 
models since different data sets actually constrain different aspects of 
physics; and also the separate the constraints from assumptions inherent 
to different data sets. We use two sets of data. (I) WMAP 5 yr data, the 
Supernovae Union data set, the HST constraints on the Hubble constant, and 
(II) additionally BBN constraints, and the SDSS LRG power spectrum. The 
data combination (II) represents the maximal data set we use here. 
We present the 
constraints obtained for our maximal data set in Fig.~\ref{FL_maximalconstraints}.\\ 

\subsection{Features of the Posterior Distribution}
\begin{figure*}[!hp]
\begin{centering}
\includegraphics[width=0.4\textwidth]{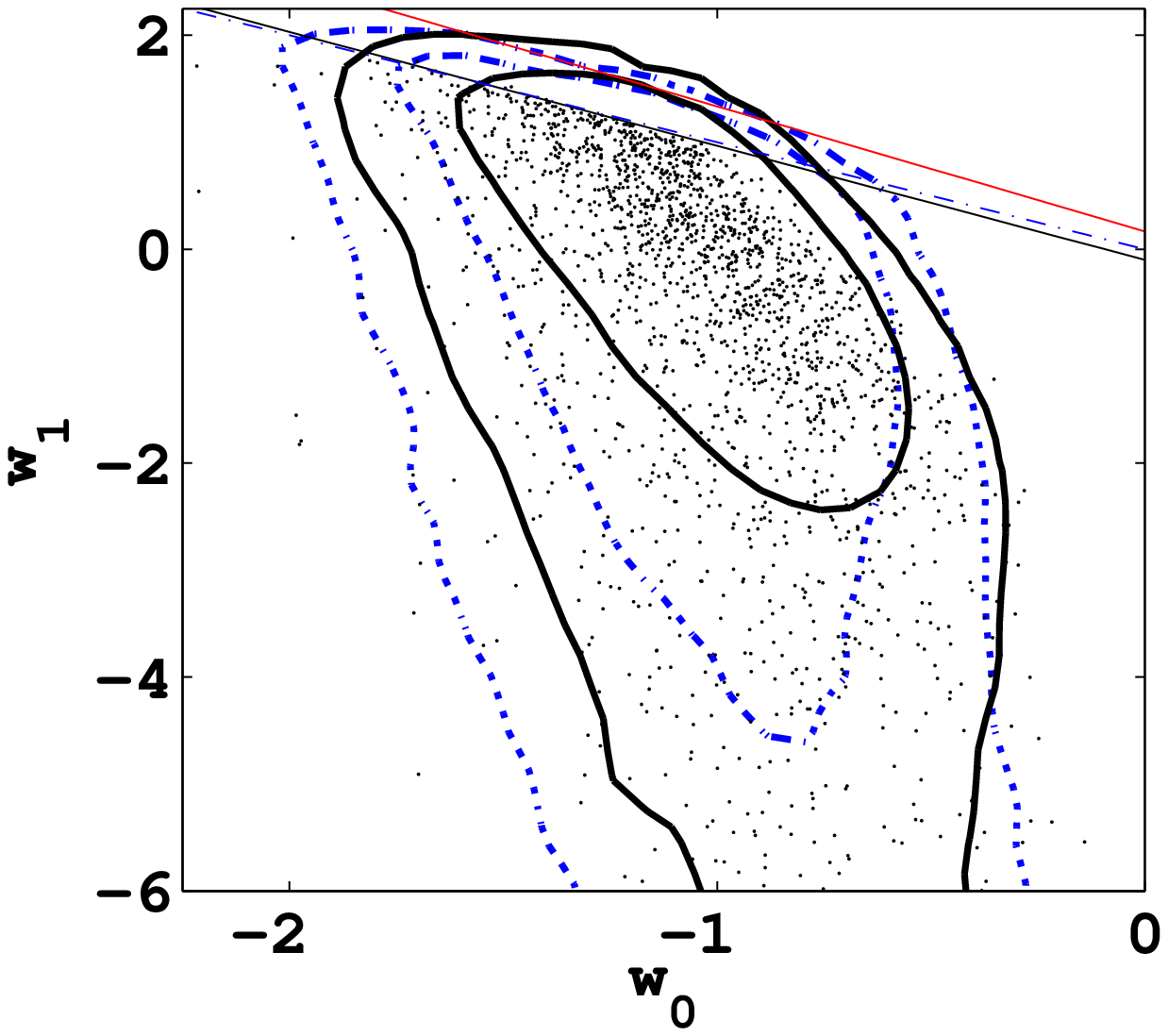}
\hfill
\includegraphics[width=0.4\textwidth]{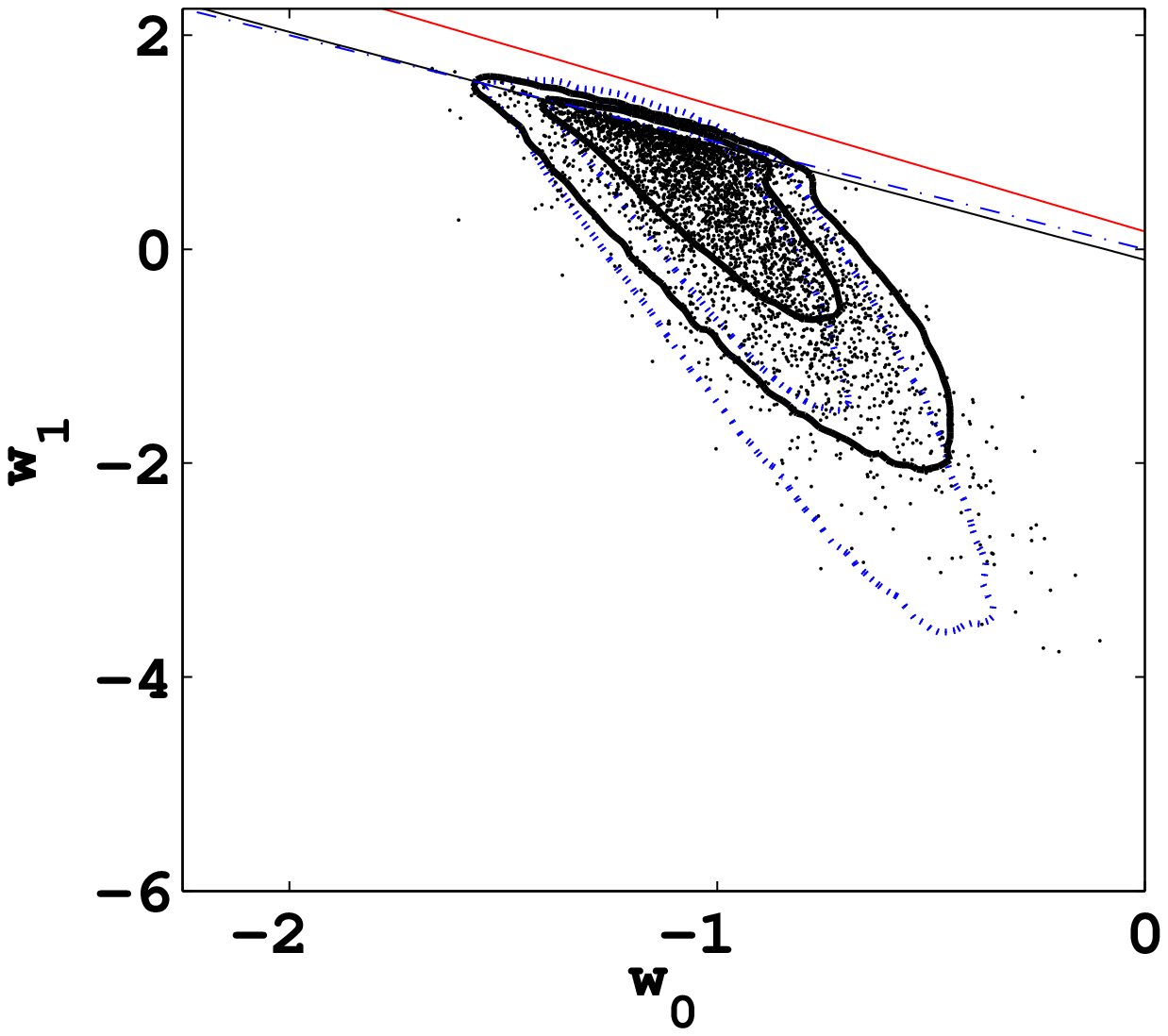}
\caption{~2D Joint Posterior distributions given (left) data set I 
(WMAP5, SNe, HST) and (right) data II (WMAP5, SDSS, SNe, BBN, HST). 
The black (solid) lines were computed using the 
Likelihoods A comparing spectra, while the blue (dash-dot) lines were 
computed using summary parameters B.} 
\label{w0w1Posteriors}
\end{centering}
\end{figure*}
\begin{figure}[!h]
\begin{centering}
\includegraphics[width=0.45\textwidth]{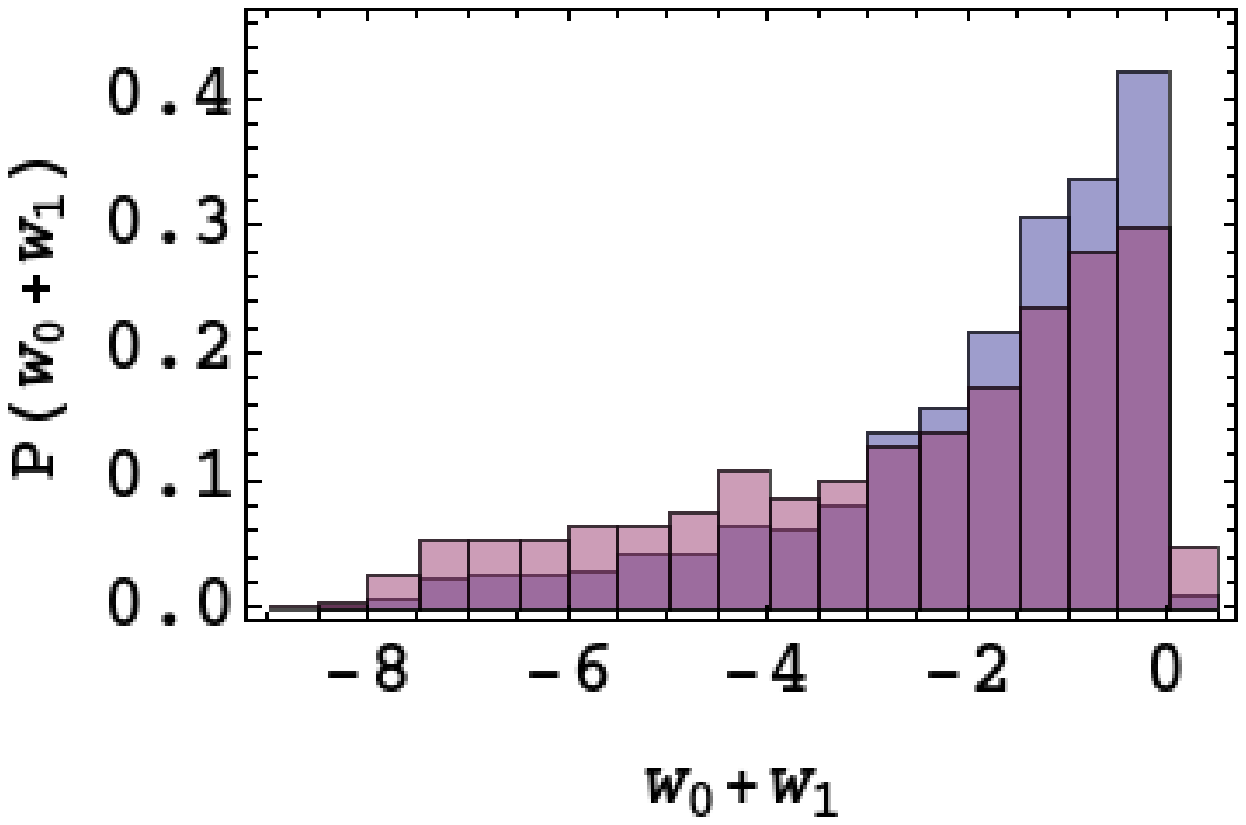}
\hfill
\includegraphics[width=0.45\textwidth]{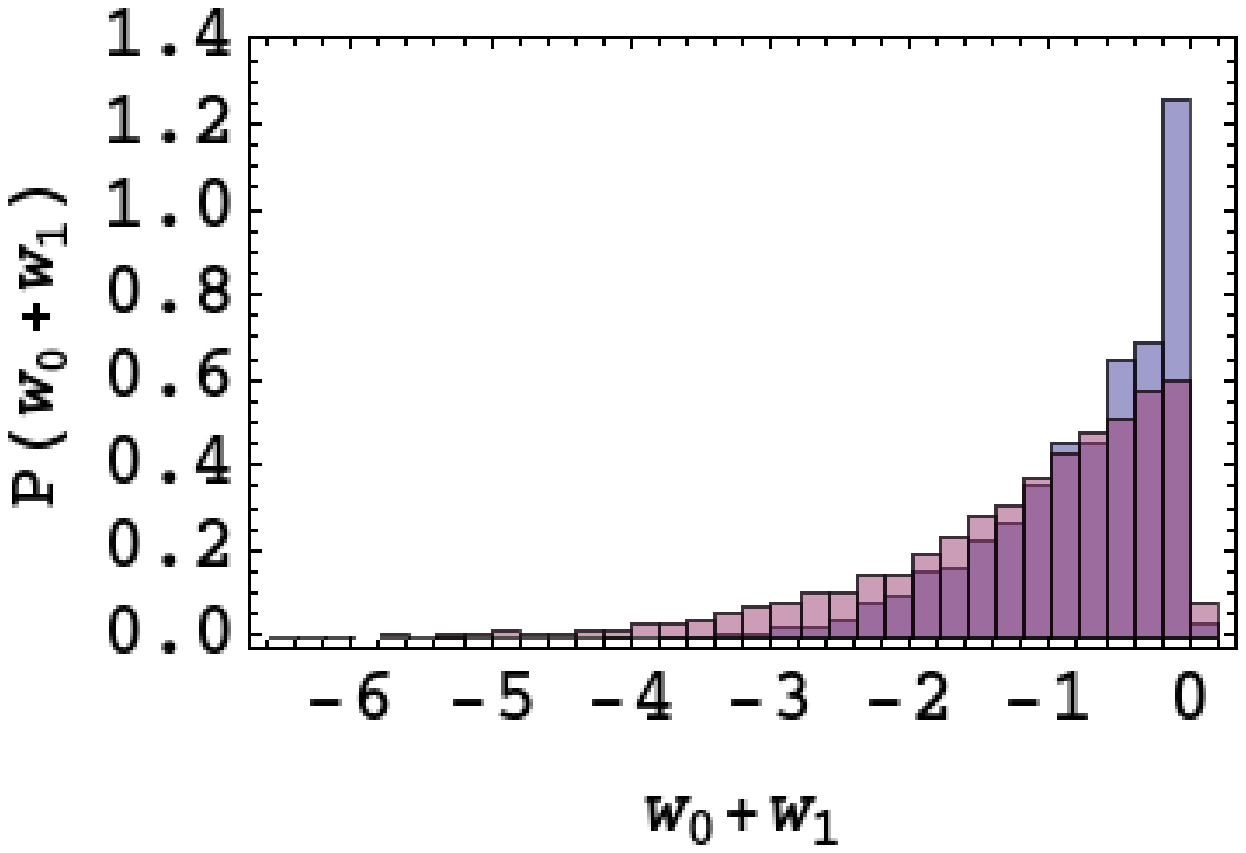}
\caption{~Binned one dimensional posterior on the dark energy EoS at very 
early times $w(0)\rightarrow w_0 +w_1$  marginalized over all other 
parameters showing the sudden drop in the 
posterior for values of $w(0) +w(1)=0$.  
The upper panel shows the posterior given the combination data set (I), 
while the lower panel shows the posterior given the combination of 
data sets (II). The blue histogram shows the distribution due to the likelihood (A)
while the magenta histogram is computed with the use of the likelihood (B).}
\label{EarlyEOS}
\end{centering}
\end{figure}

Focusing attention on the two dimensional joint posterior on $w_0,w_1,$
the left panel of Fig.~\ref{w0w1Posteriors} shows the posteriors of 
data set (I), while the right panel shows the posteriors for data set (II).
The black (solid) contours are computed by the use of likelihood A,
while the blue contours are computed using Likelihood B. The blue contours
may be compared to Fig.~1 of ~\citet{2008PhRvD..77l3525W}. 
We note that the posterior distribution is highly non-Gaussian. 
Firstly, the posterior contours are fairly 
banana-shaped rather than ellipsoidal without the SDSS data. With the 
addition of the SDSS data, the elongated banana 
shape gets pinched off.
However, the distribution is still elongated, and
even in the high posterior region the
posterior peak does not seem to be well-centered. Further, the plots show 
that the posterior falls abruptly around the blue (dot-dashed) line 
$w_1 = -w_0.$ The diagonal (solid) black line represents the $2 \sigma$ 
limit from BBN constraints; $w_0,w_1$ values lying above the black line
are unlikely due to the BBN constraints. The red diagonal line is a hard
prior we used to limit the exploration.
The use of both Likelihoods (A) and (B) result in a dramatically lower 
number of points beyond the blue (dot-dashed) line $w_0 +w_1=0$, 
though the change in B is less sharp. 
This edge in the likelihood, 
is entirely due to the CMB data and represents the edge of parameter 
space beyond which the `dark energy' dominates at the redshift of the 
surface of last scattering, rather than being related to the BBN data.
We can see this better in a histogram of asymptotically early equation 
of state in Fig.~\ref{EarlyEOS}. It shows the drop in the posterior sample 
for both Likelihoods B as well as A for data set I (Upper Panel) and data 
set II (Lower Panel), even though data set Idoes not include the 
BBN constraints. 
The lack of points above the (dot-dashed) blue line, further justifies our
use of the hard prior, since the posterior is well disconnected from the 
hard prior.
In our contour plot, we therefore could cut off the posterior contours 
that we 
would have drawn by smoothing the posterior densities with a 
Gaussian kernel. This sharp edge in the 2D joint posteriors is unlikely to go away with the addition of 
further data. 
\begin{figure}[!h]
\begin{centering}
\includegraphics[width=7.5cm]{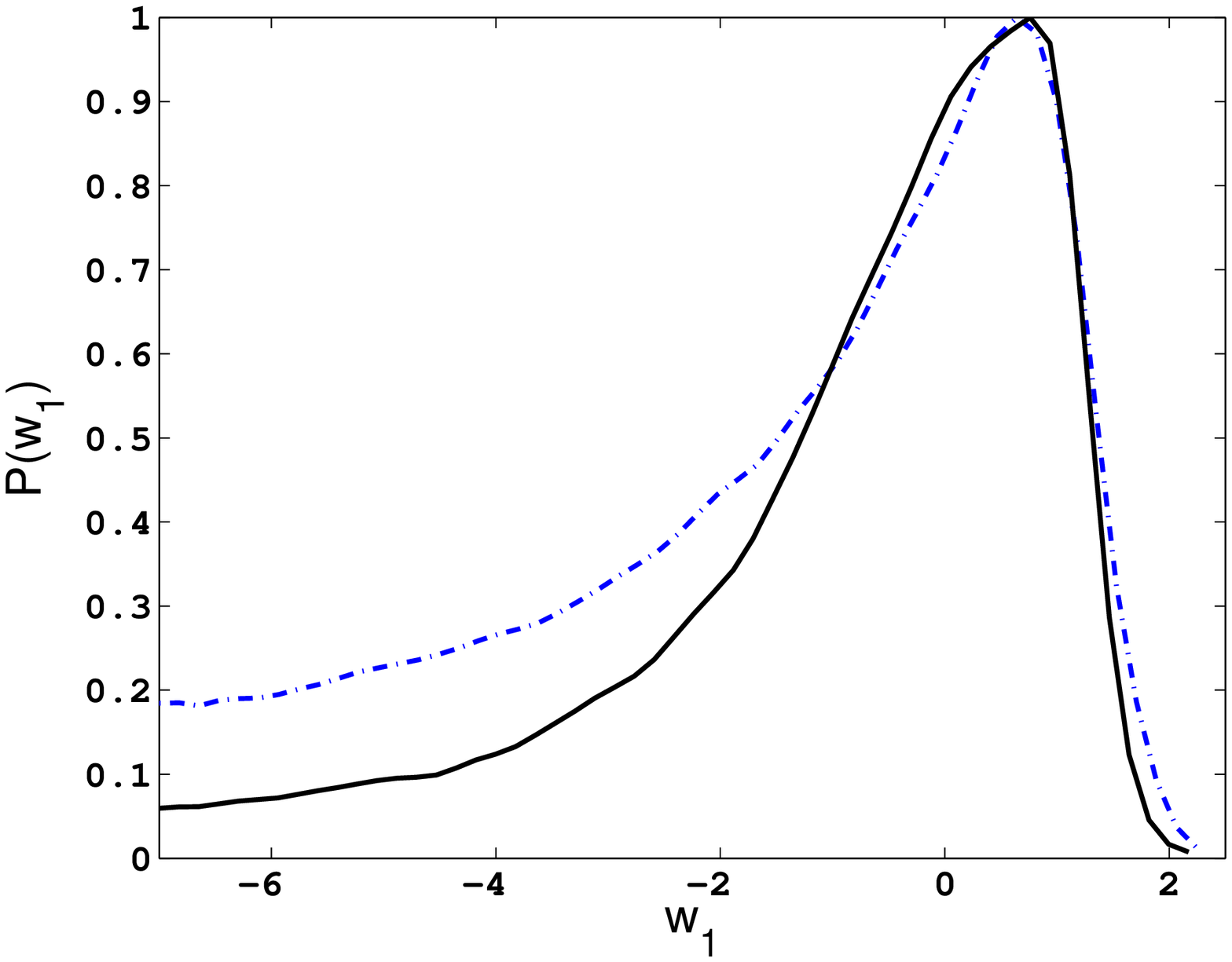}
\hfill
\includegraphics[width=7.5cm]{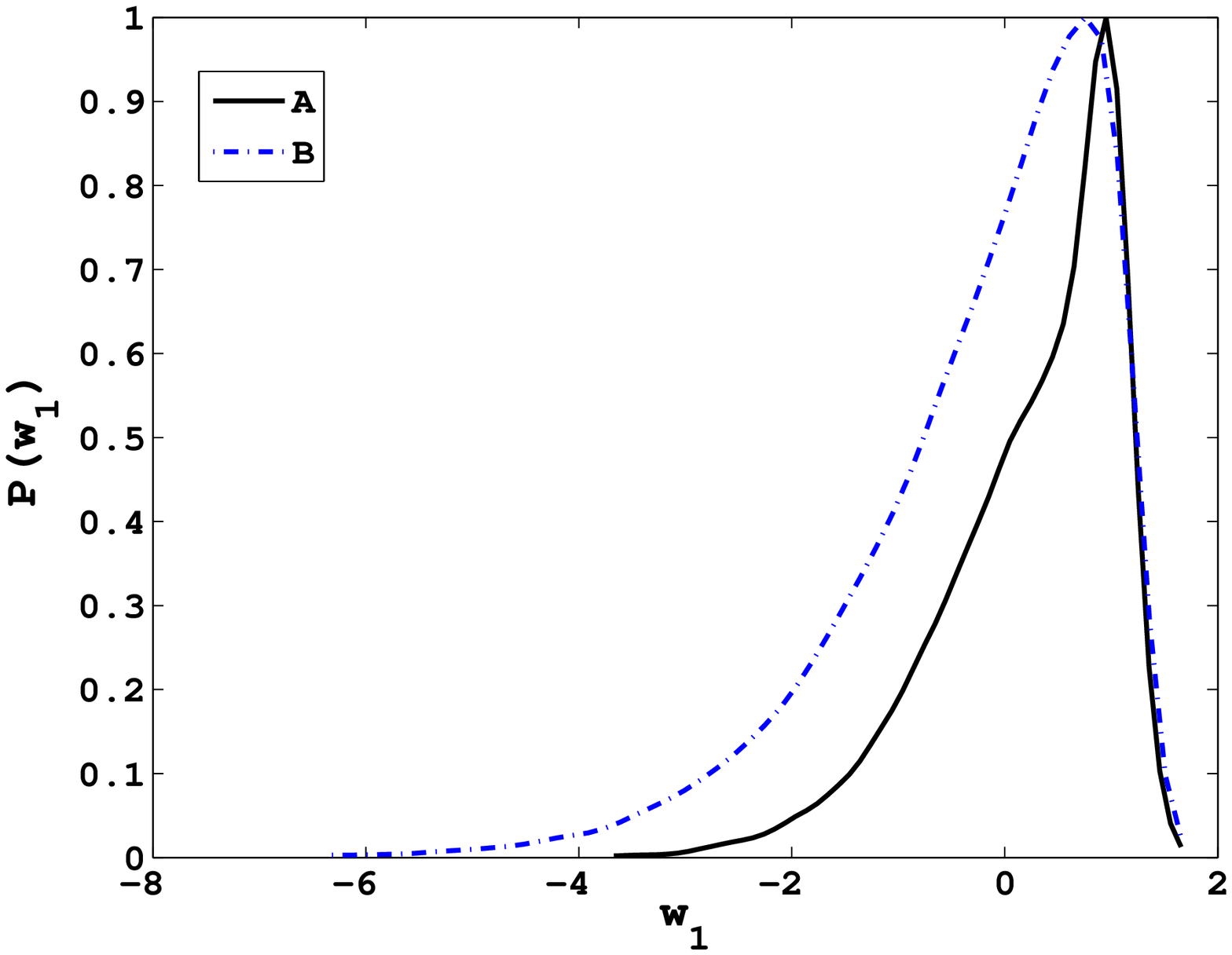}
\hfill
\caption{~Marginalized one dimensional posterior distributions on 
$w_1$: The upper panel  shows the posteriors for  data I (WMAP5, SNe, 
HST) and lower panel shows the posteriors for  
data II (WMAP5, SDSS, SNe, BBN , HST). 
The black (solid) lines were computed using the Likelihoods (A) 
comparing spectra, while the blue (dash-dot) lines were computed 
using summary parameters (B).} 
\label{w1Posteriors}
\end{centering}
\end{figure}

Next, we look at the constraints on the dark energy parameter $w_1$. 
In Fig.~\ref{w1Posteriors}, 
we show the one dimensional marginalized 
posterior distribution on $w_1$ computed from the data combinations 
I (left) and II  (right) described above, 
according to the likelihoods A (black solid) and B (dot-dashed blue). 
The posteriors computed using Likelihood B may be compared to the 
Fig.~2 of ~\citet{2008PhRvD..77l3525W}
where they use likelihoods using $z_{\star}$ as a parameter; 
as expected these posteriors match to extremely small differences that 
may be attributed to the slightly different data or numerical procedures. 
We note that these posteriors are asymmetric, a result of the fact that
significant early dark energy is essentially ruled out by the CMB data.
From our one dimensional distributions, 
we can see that while the approximate posterior (B) computed by using the 
summary have the same shape  as the posteriors (A) computed by 
comparison of power spectra, there are significant differences 
between likelihoods A and B in the extent of the tails: 
the posteriors using the power spectra are sharper and narrower than the 
posteriors using the summary parameters. For our maximal data set (II), this
translates to the tails (computed by using (B)) extending about 
twice as much as the tails using summary parameters (A).
From the one-dimensional distributions, we see that there is a sharp edge
in the distribution of $w_1$ at the higher tail, while low $w_1$ values are
are allowed. The edge at the tail is related to the edge in the joint 
posterior.
The use of summary parameters using the 
likelihood (B) has a similar effect, except the distributions are
broader, and the distinction more significant in the low $w_1$ tails.

\begin{figure}[!h]
\centering
\includegraphics[width=0.45\textwidth]{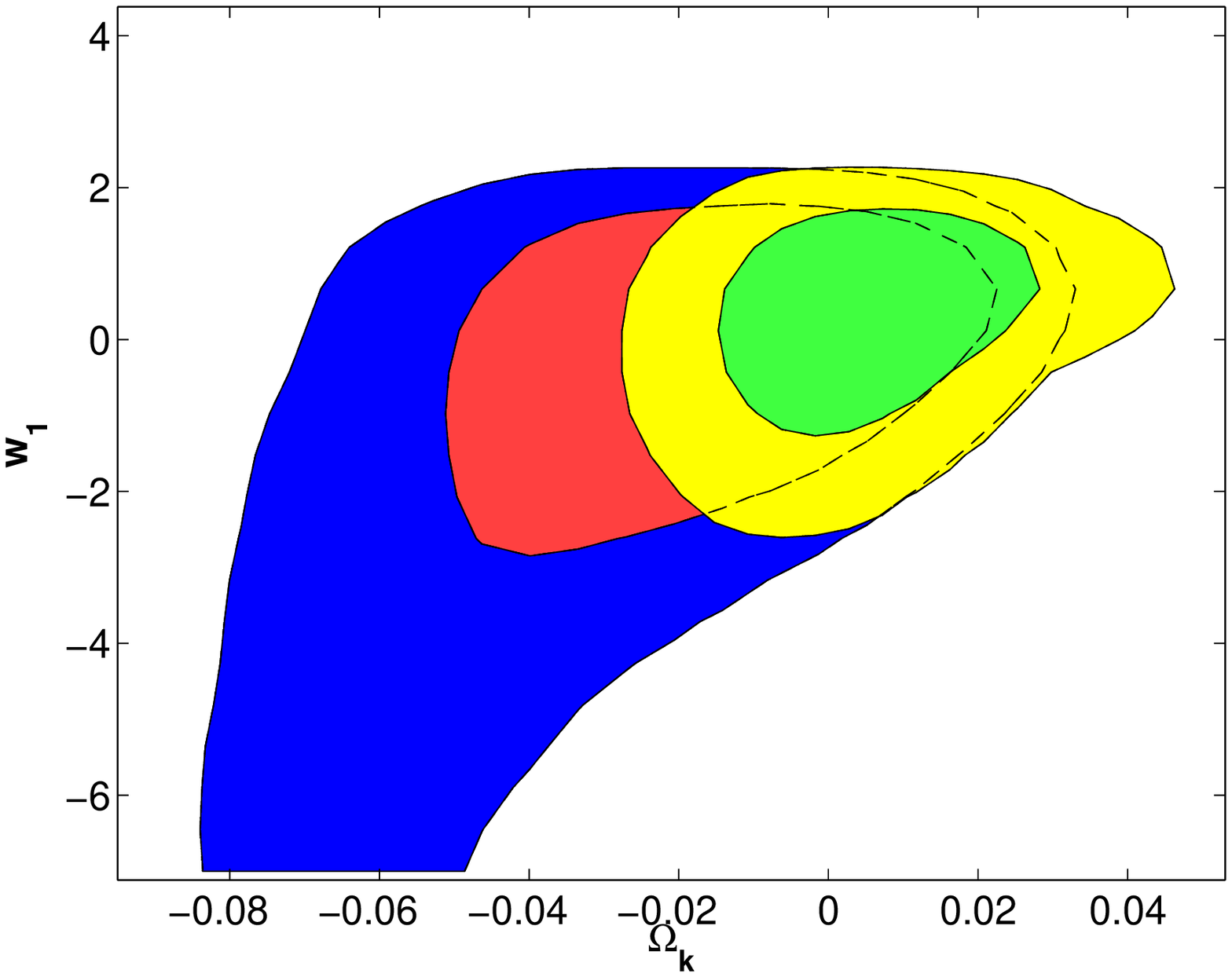}
\hfill
\includegraphics[width=0.45\textwidth]{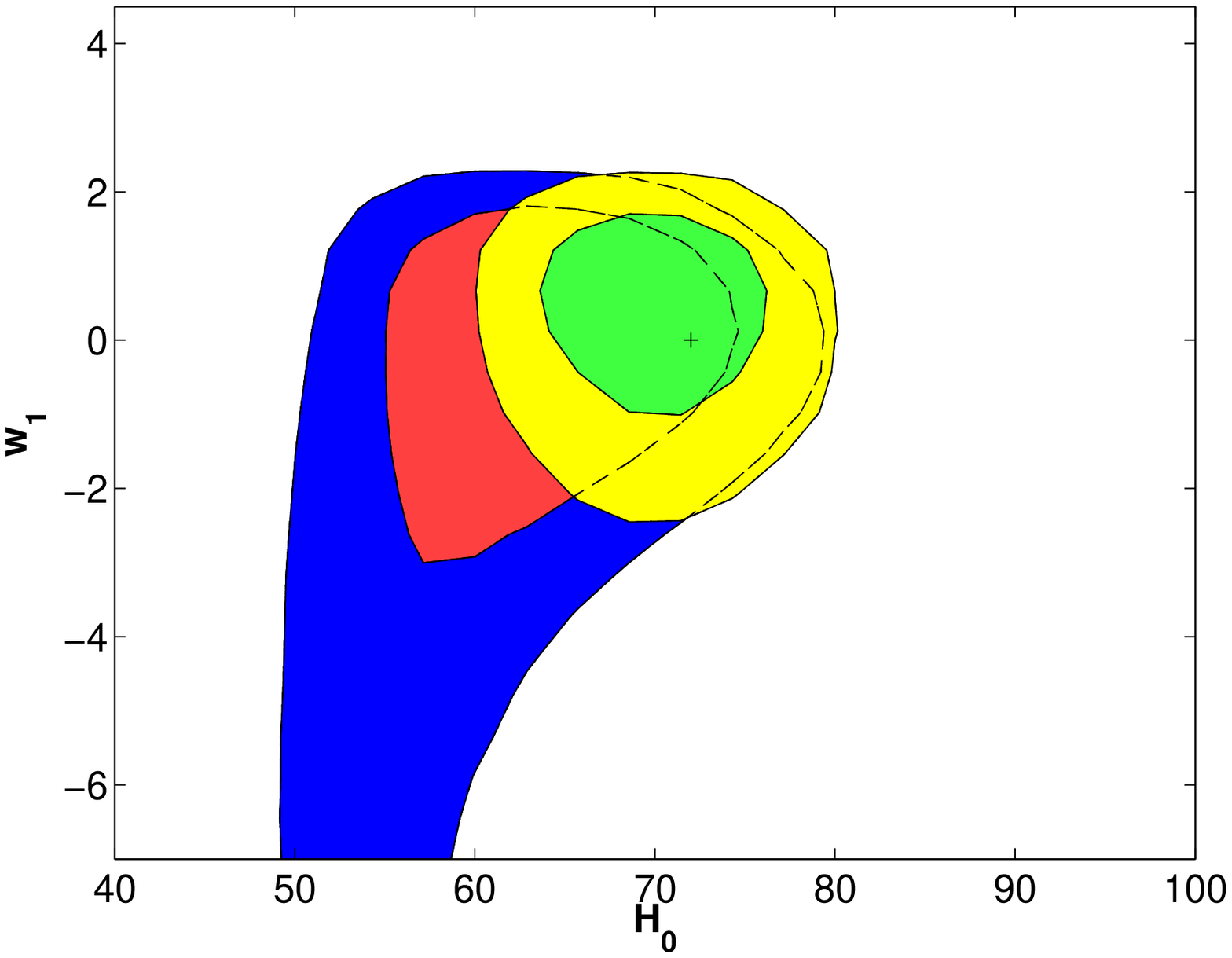}
\caption{
~Correlations of $w_1$ with background parameters: 2D Joint posteriors
of $w_1 $ values with $\Omega_k$ (upper panel) and $H_0$ 
(lower panel) showing the 68 (green and red) and 95 (yellow and blue) 
percent contours 
for data sets I (WMAP5, SNe, HST) and 
II (WMAP5, SNe, SDSS, HST, BBN) using the Likelihood A comparing the power spectra}
\label{correlations}
\end{figure}

From both the one dimensional and two dimensional posteriors, we note 
(I) allow fairly low values of $w_1$ at levels, which are ruled 
out by the SDSS data (II), or if the model is constrained to be flat. 
This is because (I) allows models with low values of $w_1, H_0,$ and 
$\Omega_k$, which are ruled out by the simultaneous use of the SDSS data. 
This can be 
seen by studying the correlations of $w_0, w_1$ shown in Fig.~\ref{correlations}. 

\subsection{Comparison of Power Spectra and Summary Parameters}
\begin{figure*}[!p]
\begin{centering}
\includegraphics[width=0.45\textwidth]{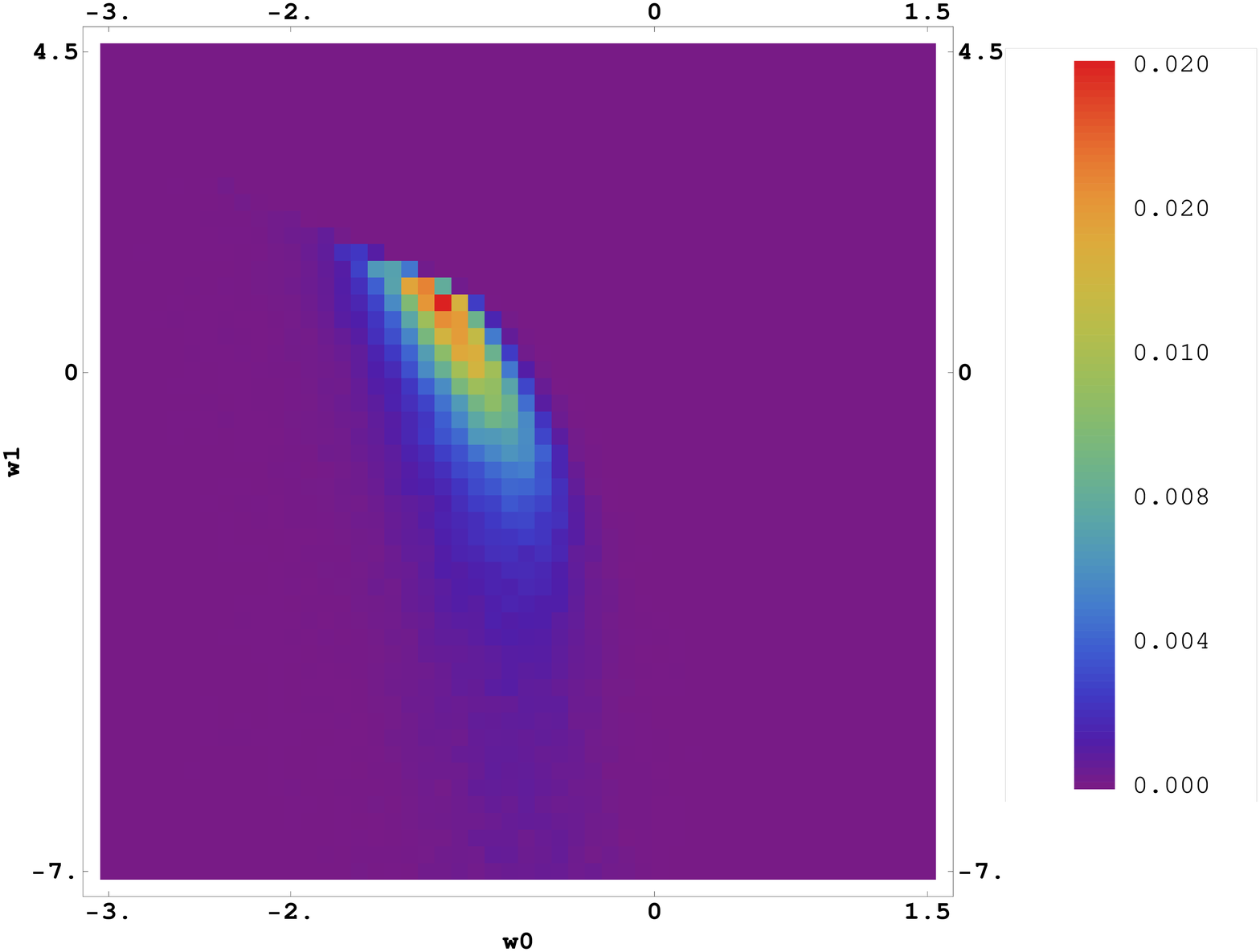}
\hfill
\includegraphics[width=0.45\textwidth]{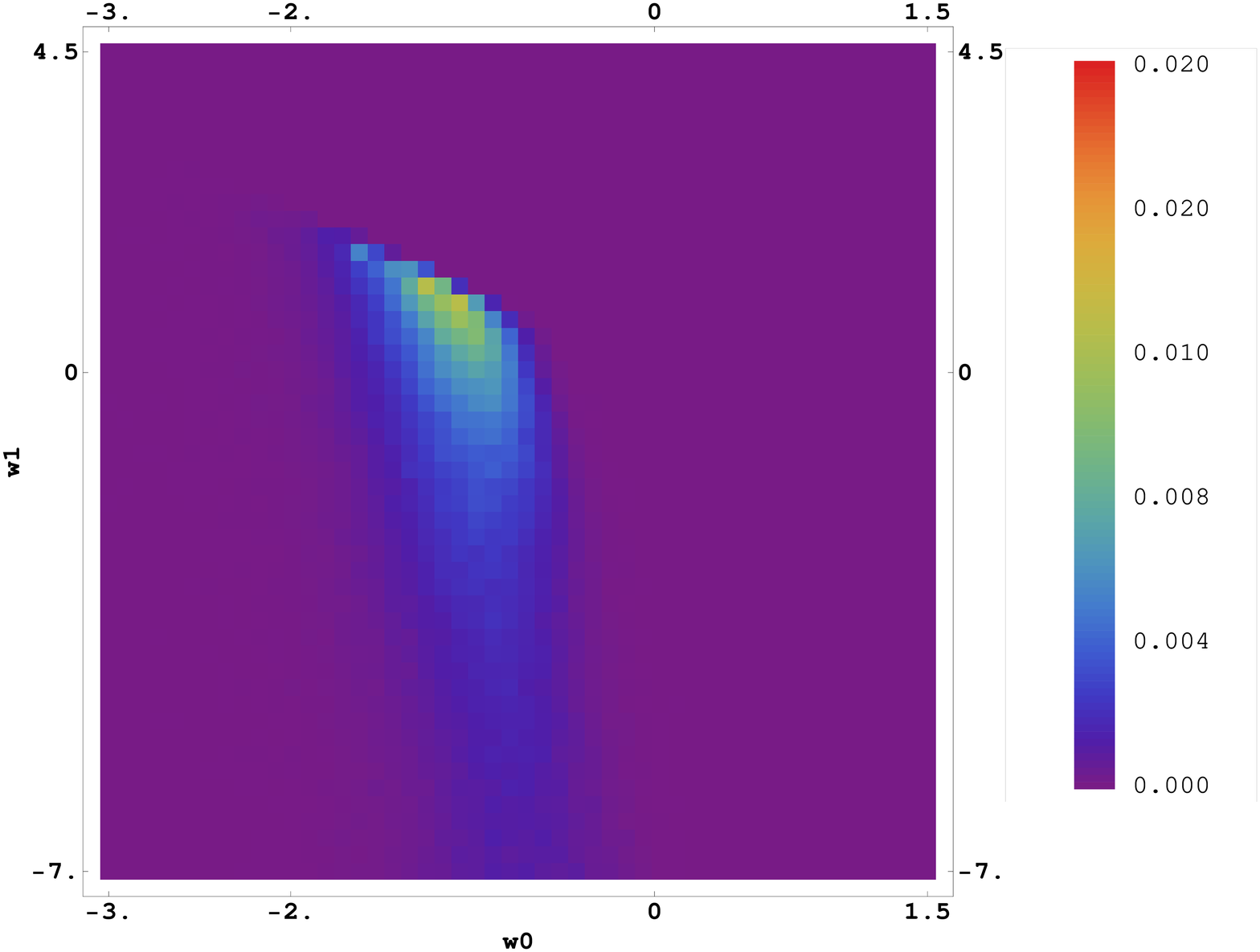}
\hfill
\includegraphics[width=0.45\textwidth]{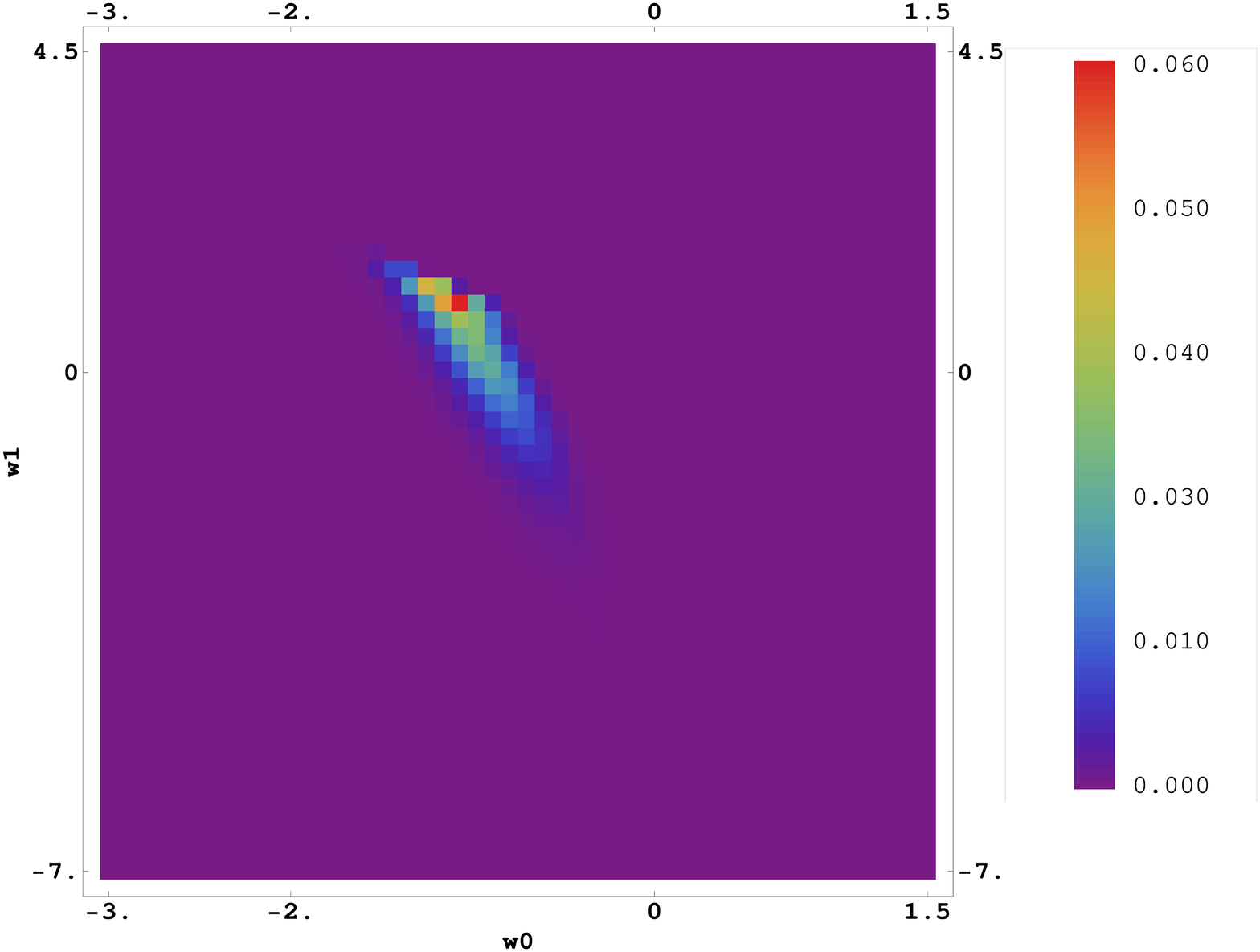}
\hfill
\includegraphics[width=0.45\textwidth]{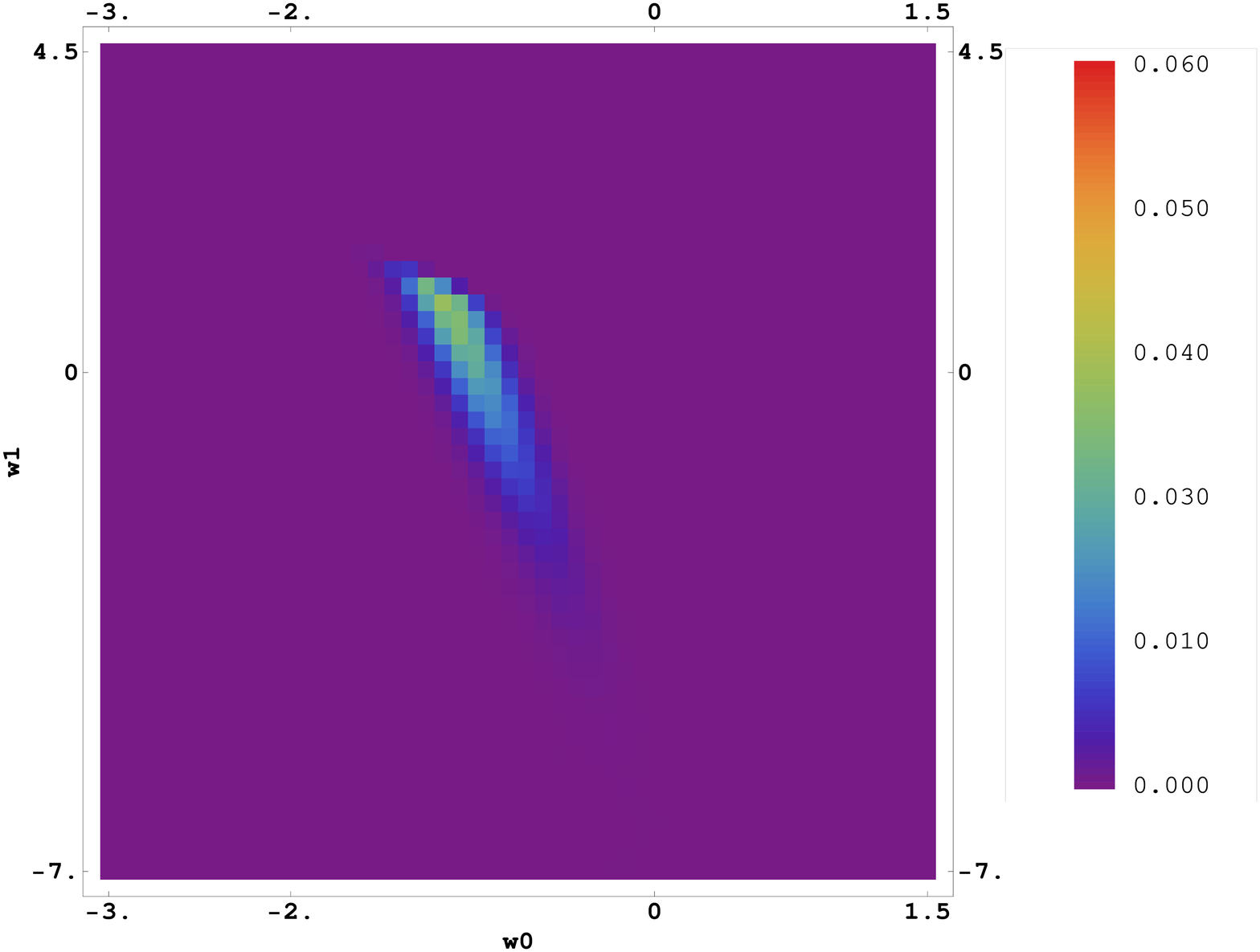}
\hfill
\caption{~Density Plots showing the Joint Posterior Distribution on 
$w_0,w_1$ given the data set I (upper panels), and data set II (lower panels) 
for likelihoods A (left) and B (right).
}
\label{DensityPlots}
\end{centering}
\end{figure*}

The differences between the posterior distributions due to the use of 
different likelihoods A and B, are most clearly 
studied in a binned (un-smoothed) density plot over all the chains, since
various differences can arise due to smoothing prescriptions inherent in 
making contours. We choose to study the joint posterior in the $w_0,w_1$ 
plane due to its importance in classifying dark energy experiments
~\citep{2006astro.ph..9591A}. We present the differences as a density plots for 
both data set combinations I (top panel) and II (bottom panel)
in Fig.~\ref{DensityPlots}. We can see that the differences 
are most appreciable for models, where the equation of state is close to 
$0$ (like matter) at asymptotically early times, and for the tail. 
where $w_1$ has low values. 
The
density plots on the left due to likelihood A are much sharper and narrower
than the corresponding plots using B on the right. 

We also study the marginalized one dimensional posteriors in 
Fig.~\ref{OneDPosts} on all other parameters. From Fig.~\ref{w1Posteriors} 
we see that with the maximal 
data set, the posterior due to Likelihood B (black) 
extends to about twice the 
tail of Likelihood A (blue) in $w_1.$ 
From Fig.~\ref{OneDPosts}, we show the 1 dimensional marginalized 
distributions on all other parameters from the comparison of spectra 
(ie. by using likelihood A)
(blue) and the summary parameters (red)
(ie. by using likelihood B). 
We see that the 
posterior distributions for $w_0,\theta$ are well approximated by the 
summary parameters. On the other hand, comparing the blue and red curves 
in Fig.~\ref{OneDPosts}, we see that the posteriors on 
all other quantities 
($\Omega_m,\Omega_k, \Omega_{DE}, \omega_b, H_0$)
are shifted significantly in comparison to the width of the distribution 
when the summary parameters used. As expected, the parameters related to the optical 
depth of re-ionization, scalar spectral index, and the amplitudes of the 
scalar fluctuations ($\tau, n_s, A_s$) are not constrained at all by the 
use of summary parameters resulting in a distribution about as broad as the
prior, while the spectra constrain them reasonably  well.
This is not surprising
as all of them affect the shape and normalization of the 
CMB power spectrum without affecting the distance to the 
surface of last scattering, or the angular position of the peaks.
\subsection{Impact of Relaxing the Dark Energy Parametrization}
\begin{figure*}
\begin{centering}
\includegraphics[scale=0.2,angle=-90]{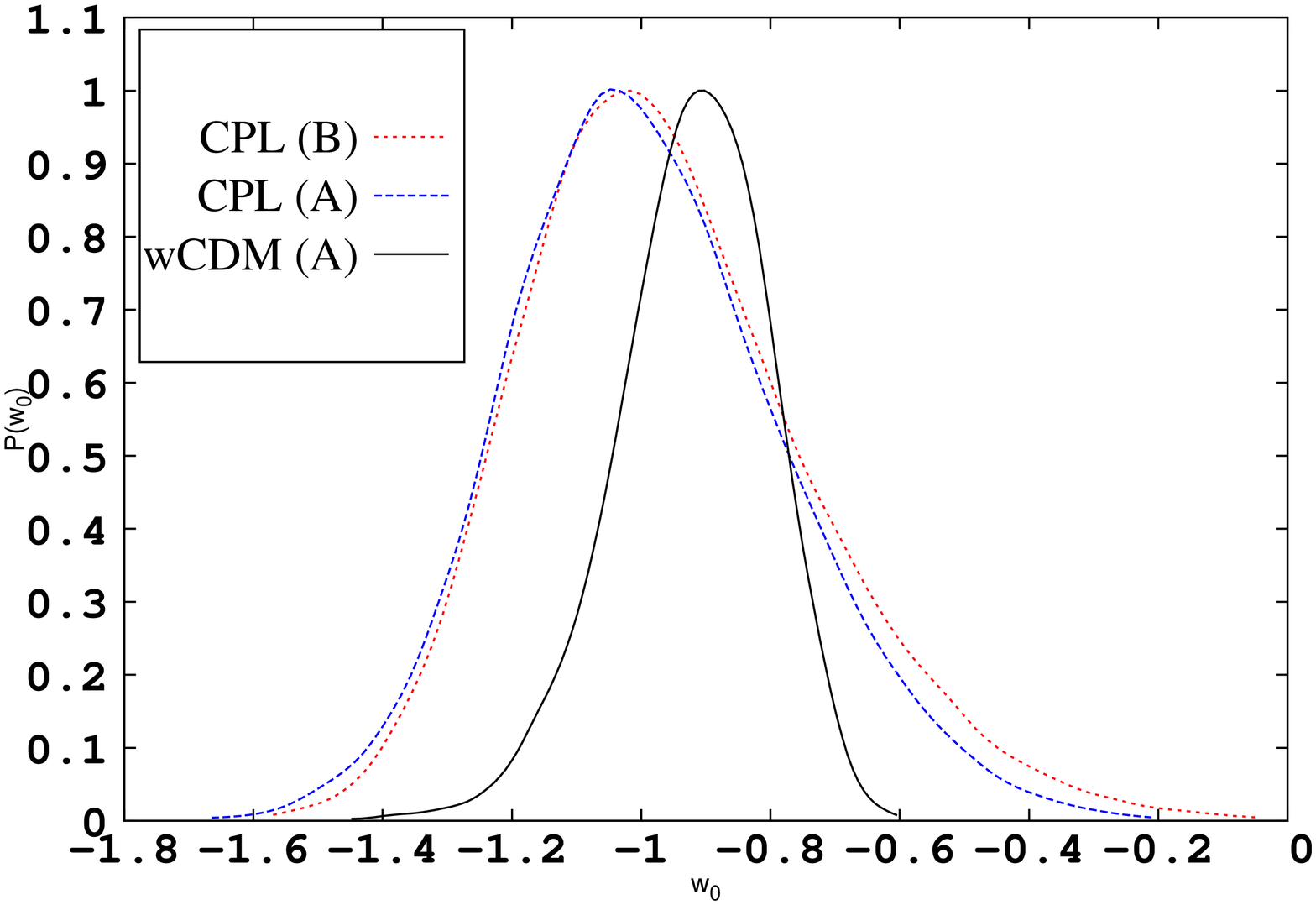}
\hfill
\includegraphics[scale=0.2,angle=-90]{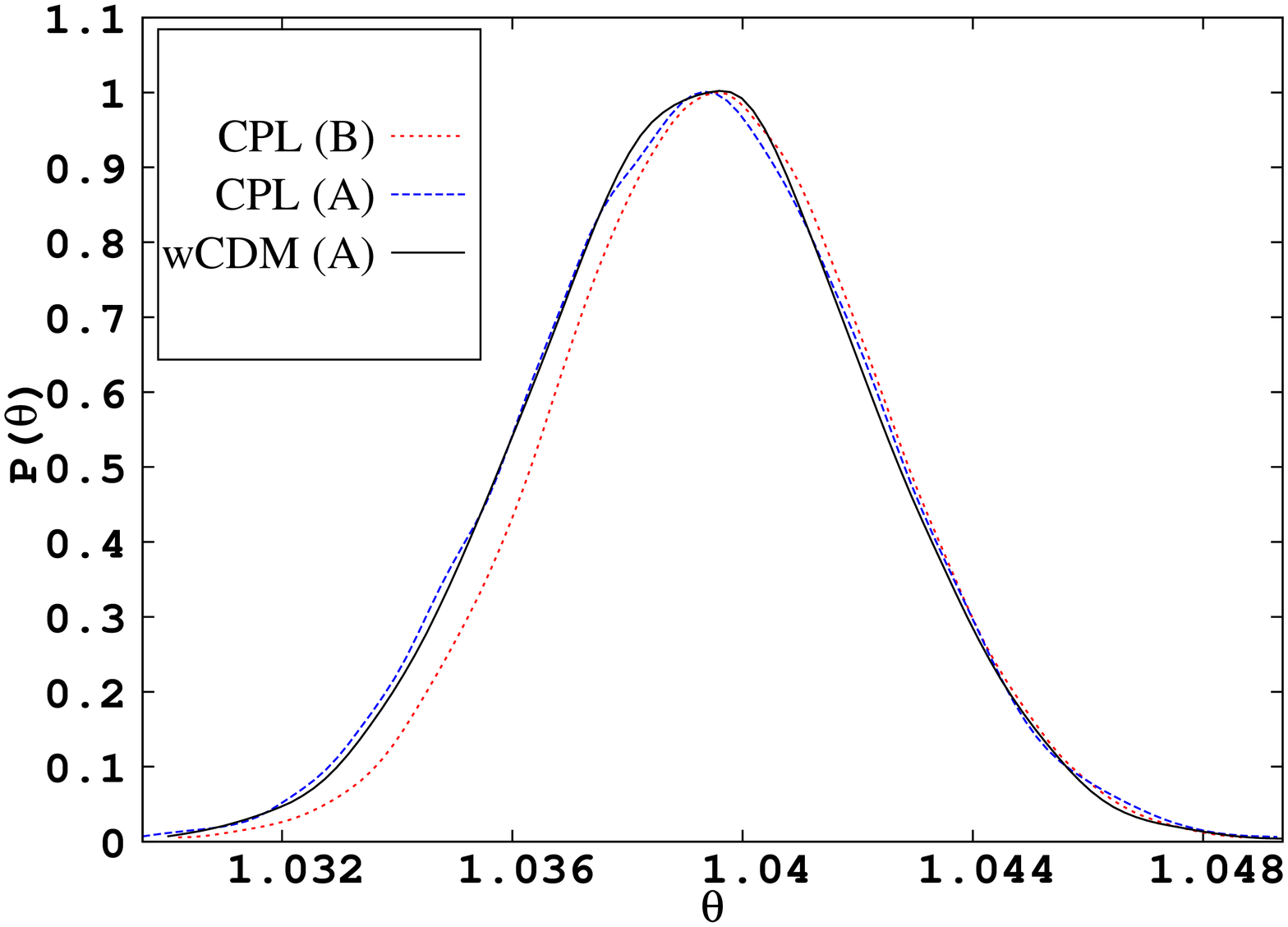}
\hfill
\includegraphics[scale=0.2,angle=-90]{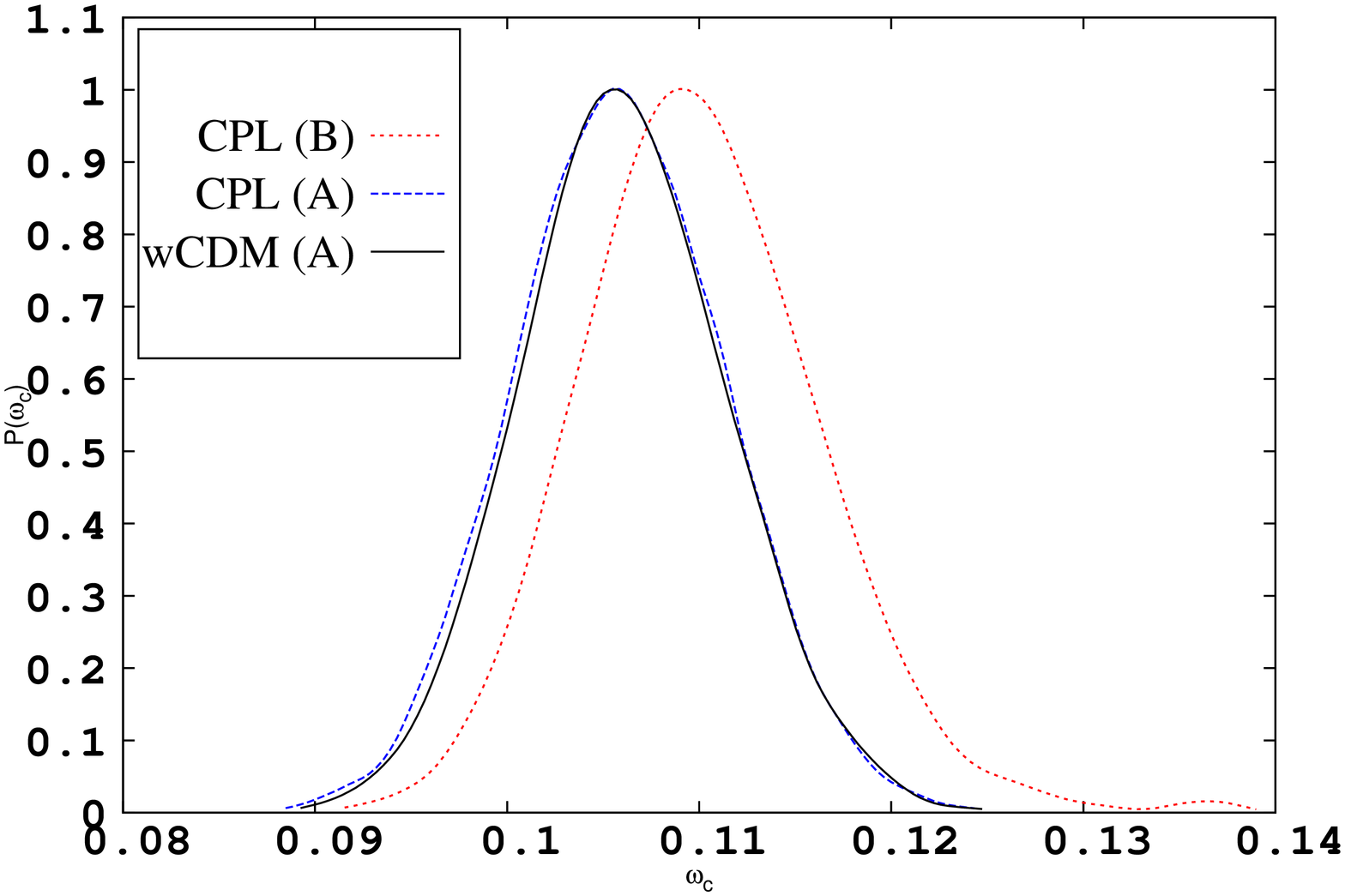}
\hfill
\includegraphics[scale=0.2,angle=-90]{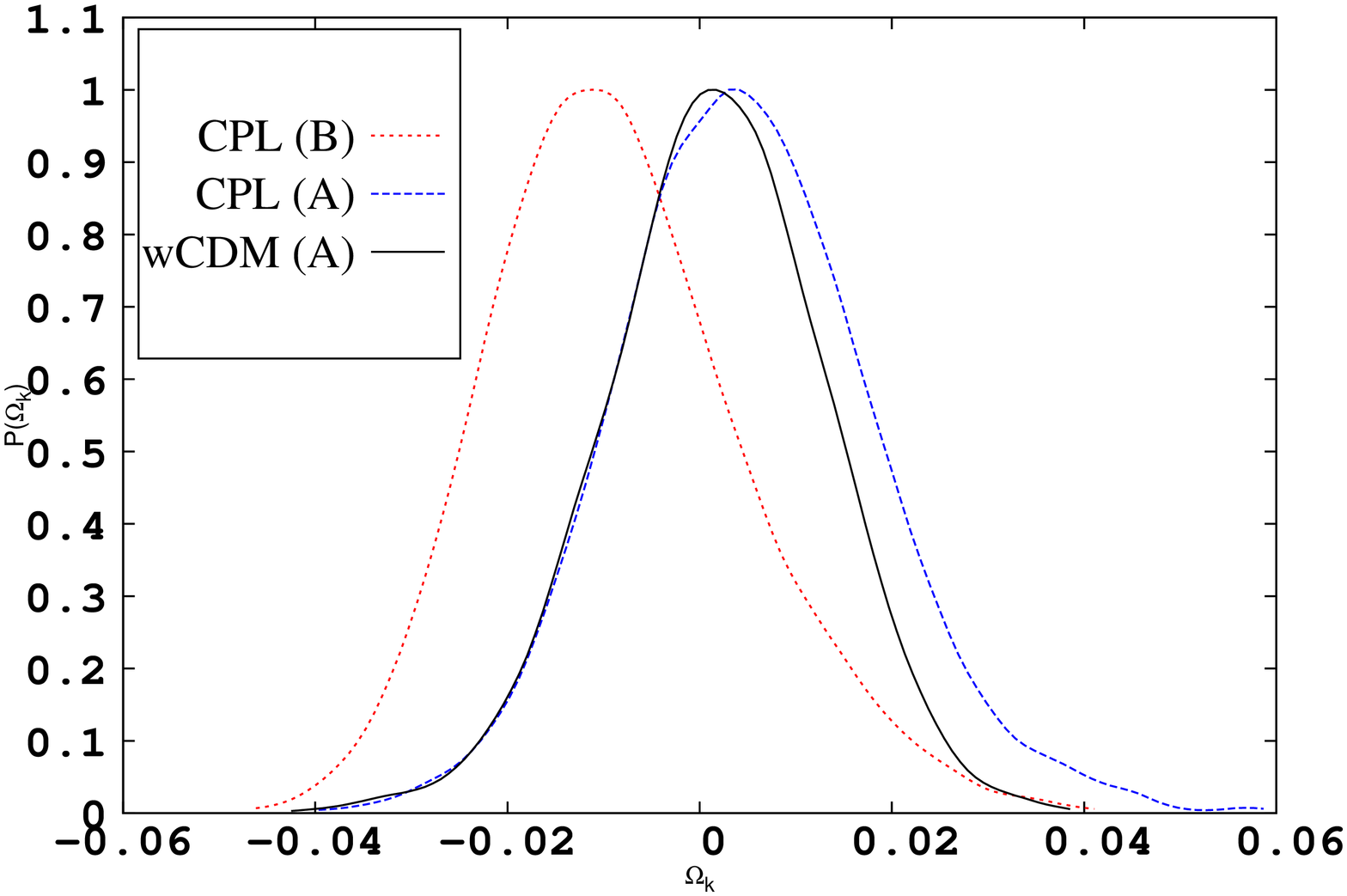}
\hfill
\includegraphics[scale=0.2,angle=-90]{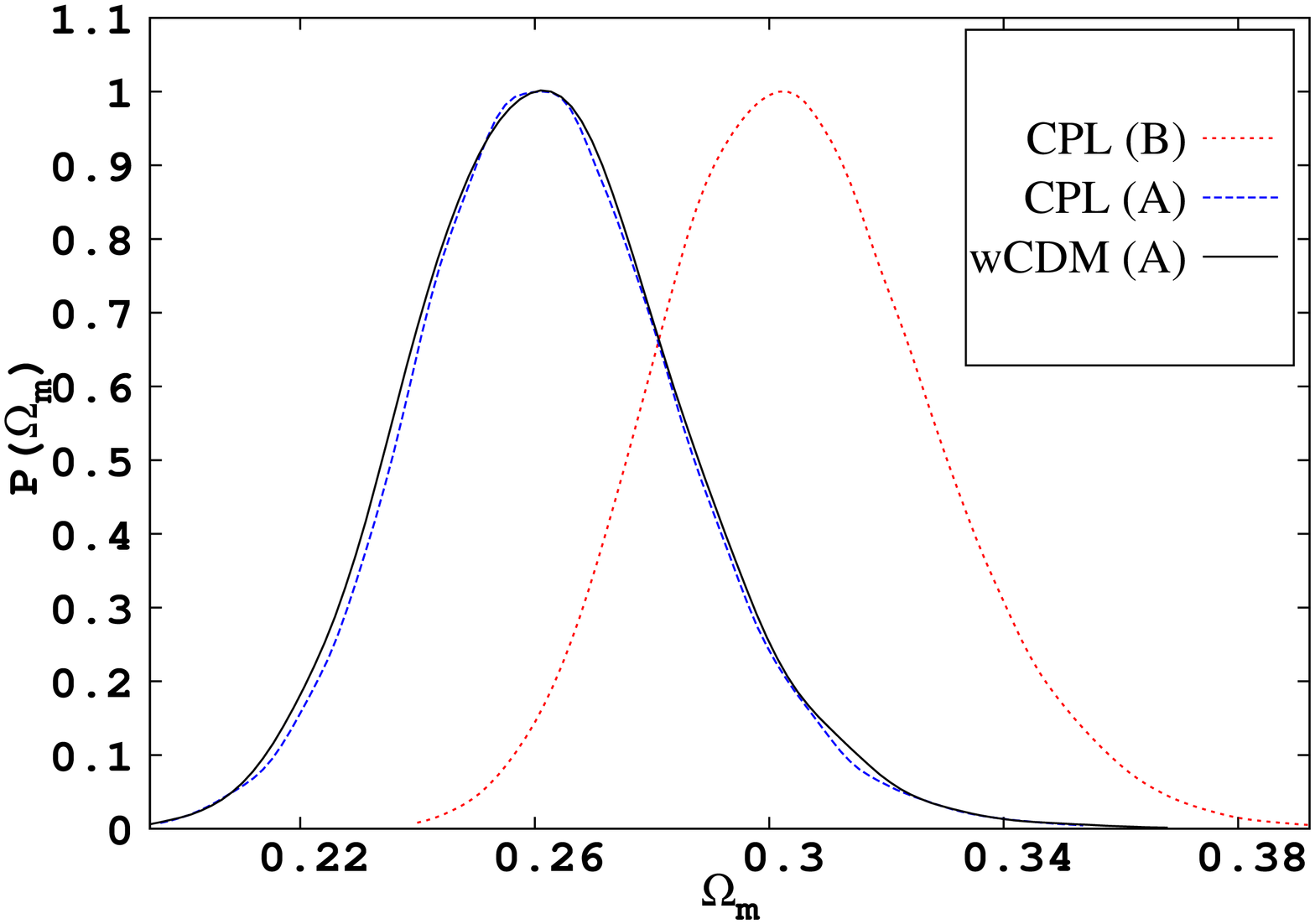}
\hfill
\includegraphics[scale=0.2,angle=-90]{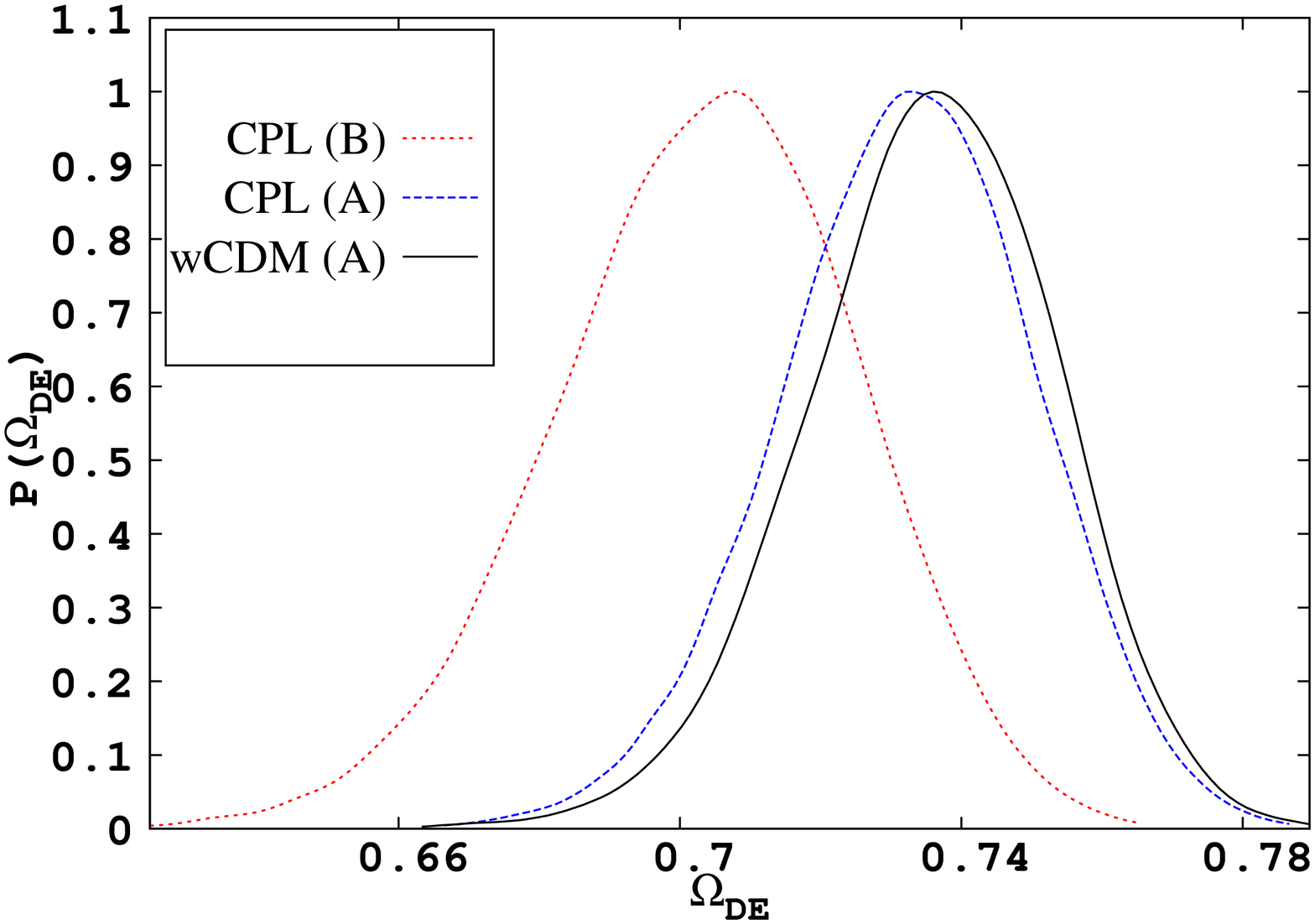}
\hfill
\includegraphics[scale=0.2,angle=-90]{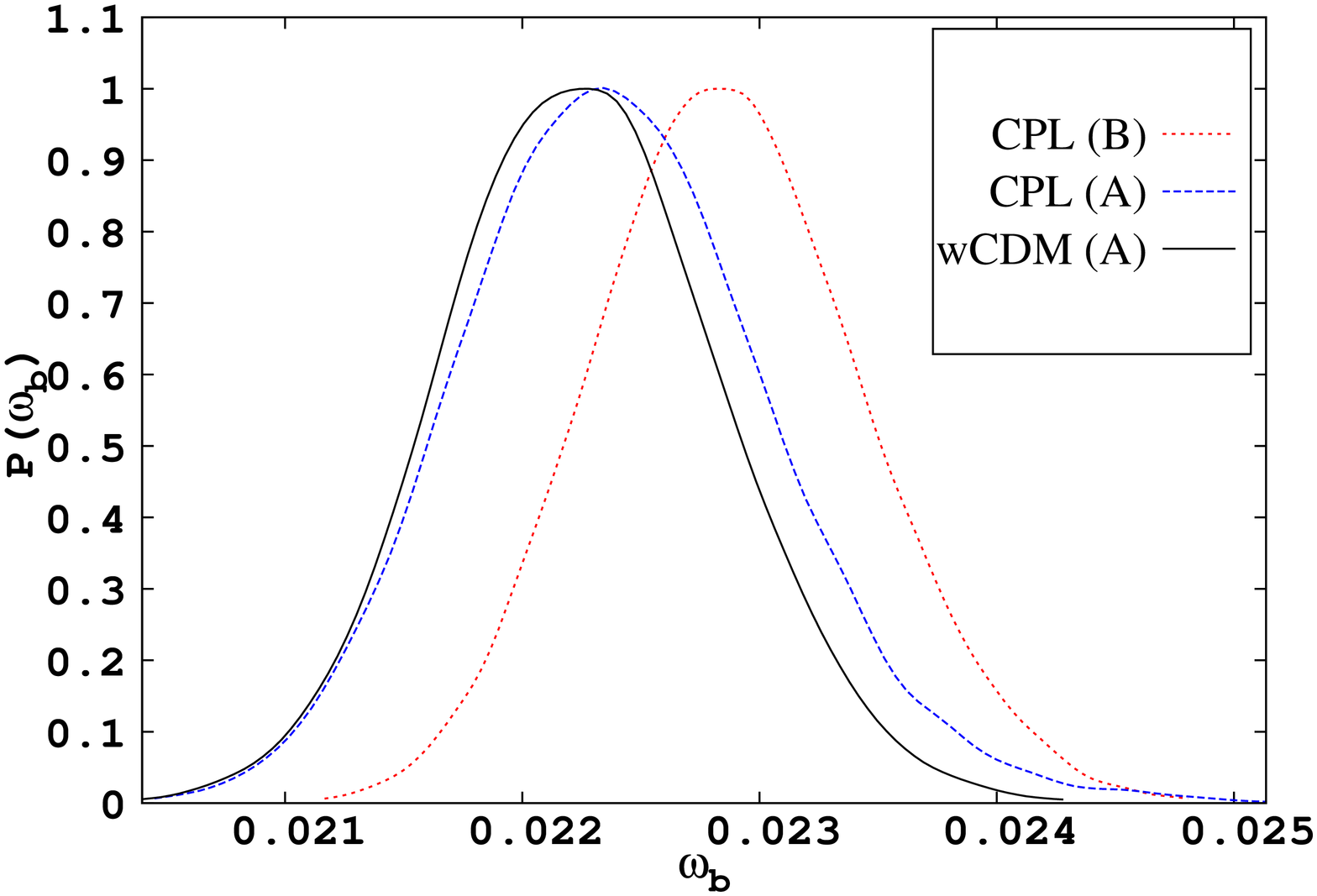}
\hfill
\includegraphics[scale=0.2,angle=-90]{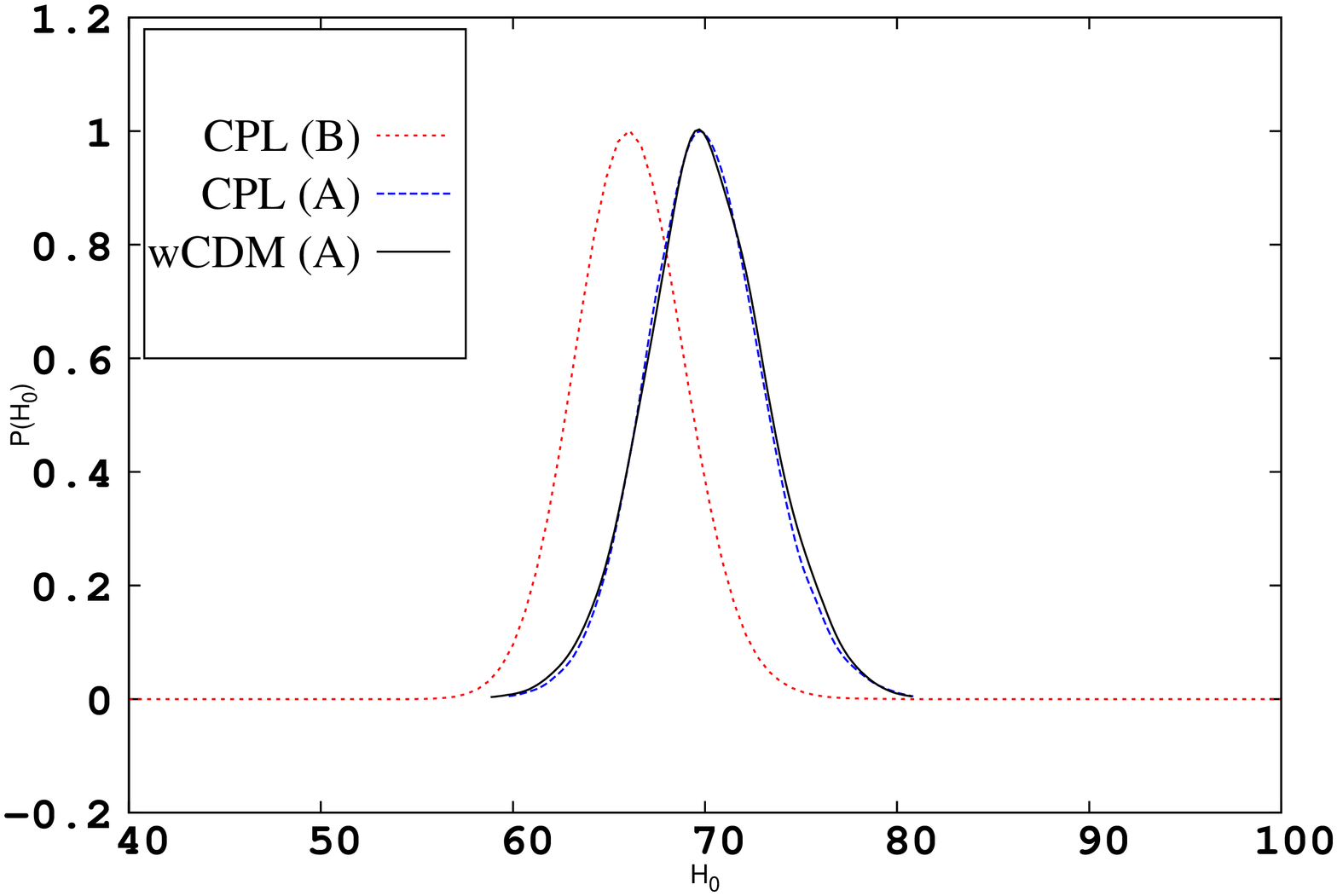}
\hfill
\includegraphics[scale=0.2,angle=-90]{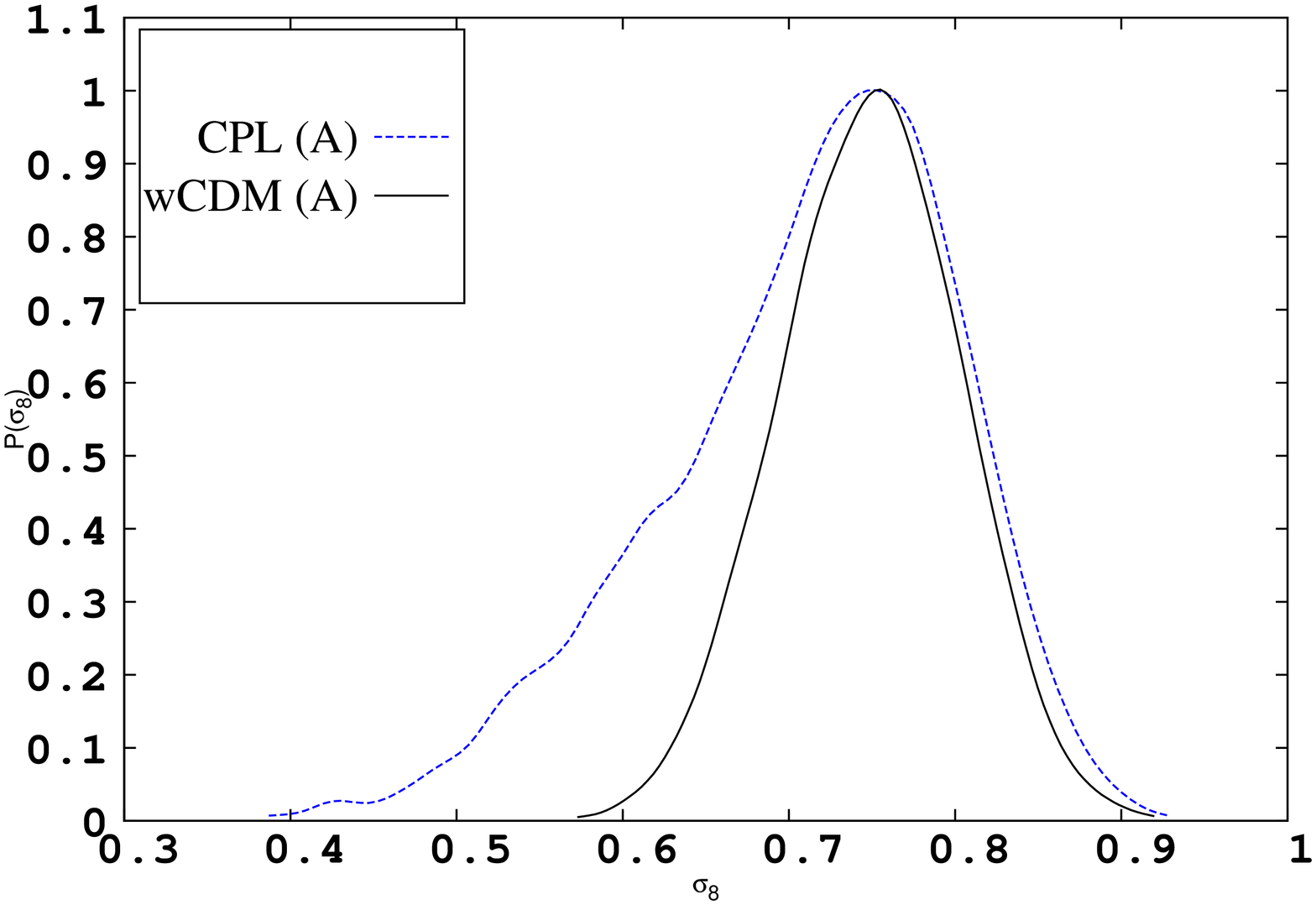}
\hfill
\includegraphics[scale=0.2,angle=-90]{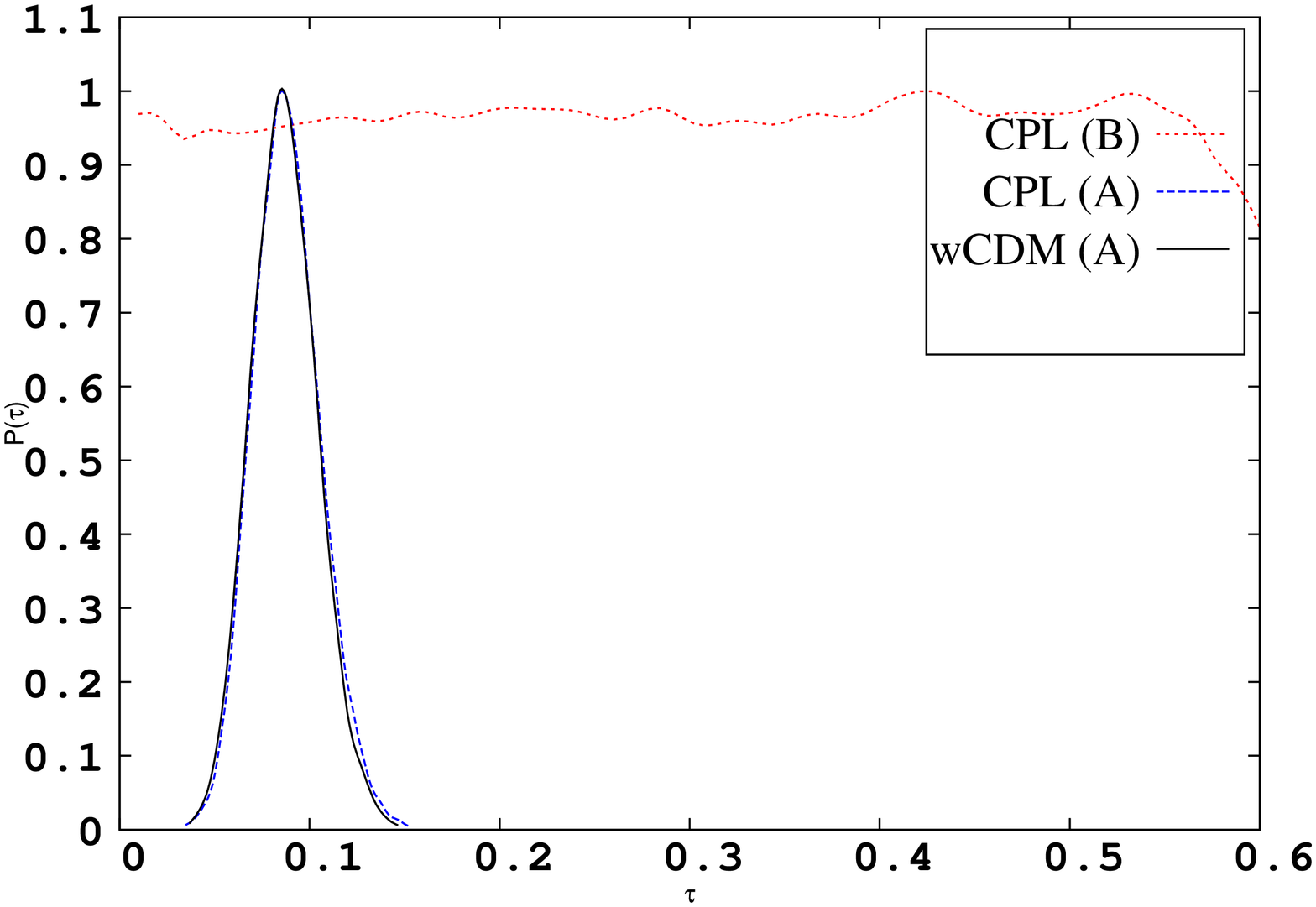}
\hfill
\includegraphics[scale=0.2,angle=-90]{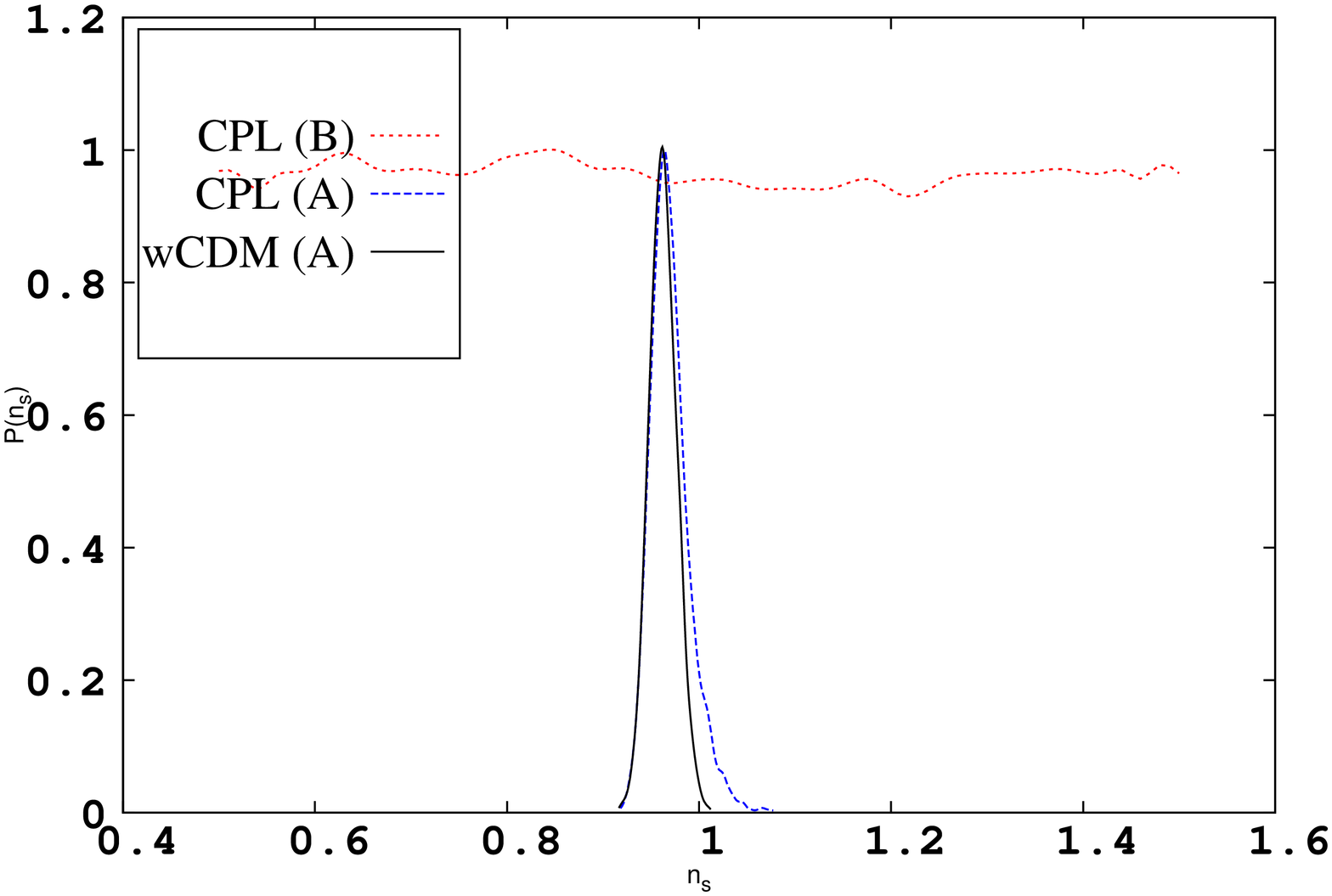}
\hfill
\includegraphics[scale=0.2,angle=-90]{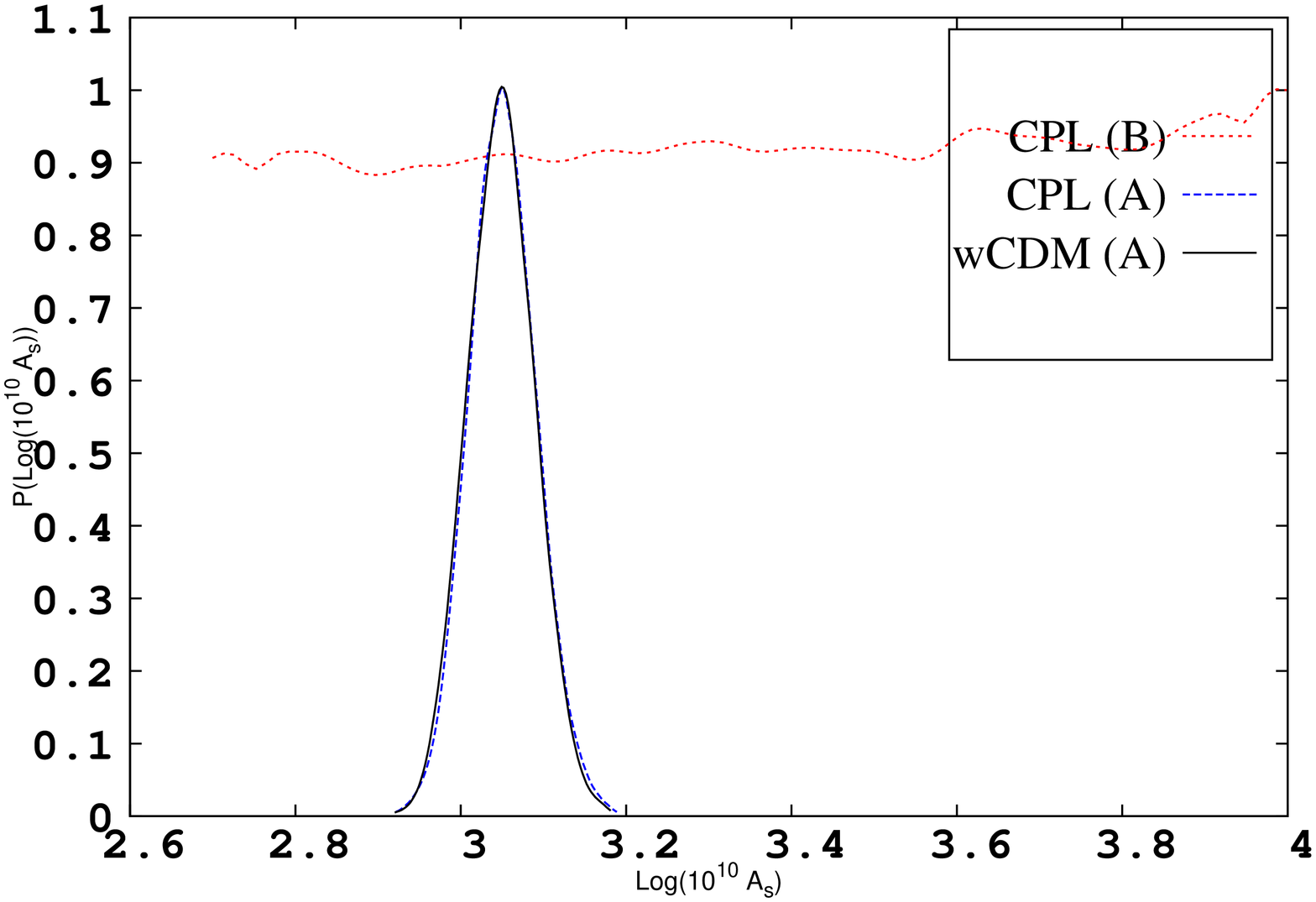}
\caption{~Marginalized one dimensional constraints on cosmological 
parameters given all current data sets (II) (WMAP5, SDSS, SNe, HST, BBN)  using Likelihoods (A) for a wCDM model (solid, black), (A) for a CPL model (dashed, blue), and (B) for a CPL model (dotted, red)}
\label{OneDPosts}
\end{centering}
\end{figure*}

\begin{figure*}
\begin{centering}
\includegraphics[width=5cm]{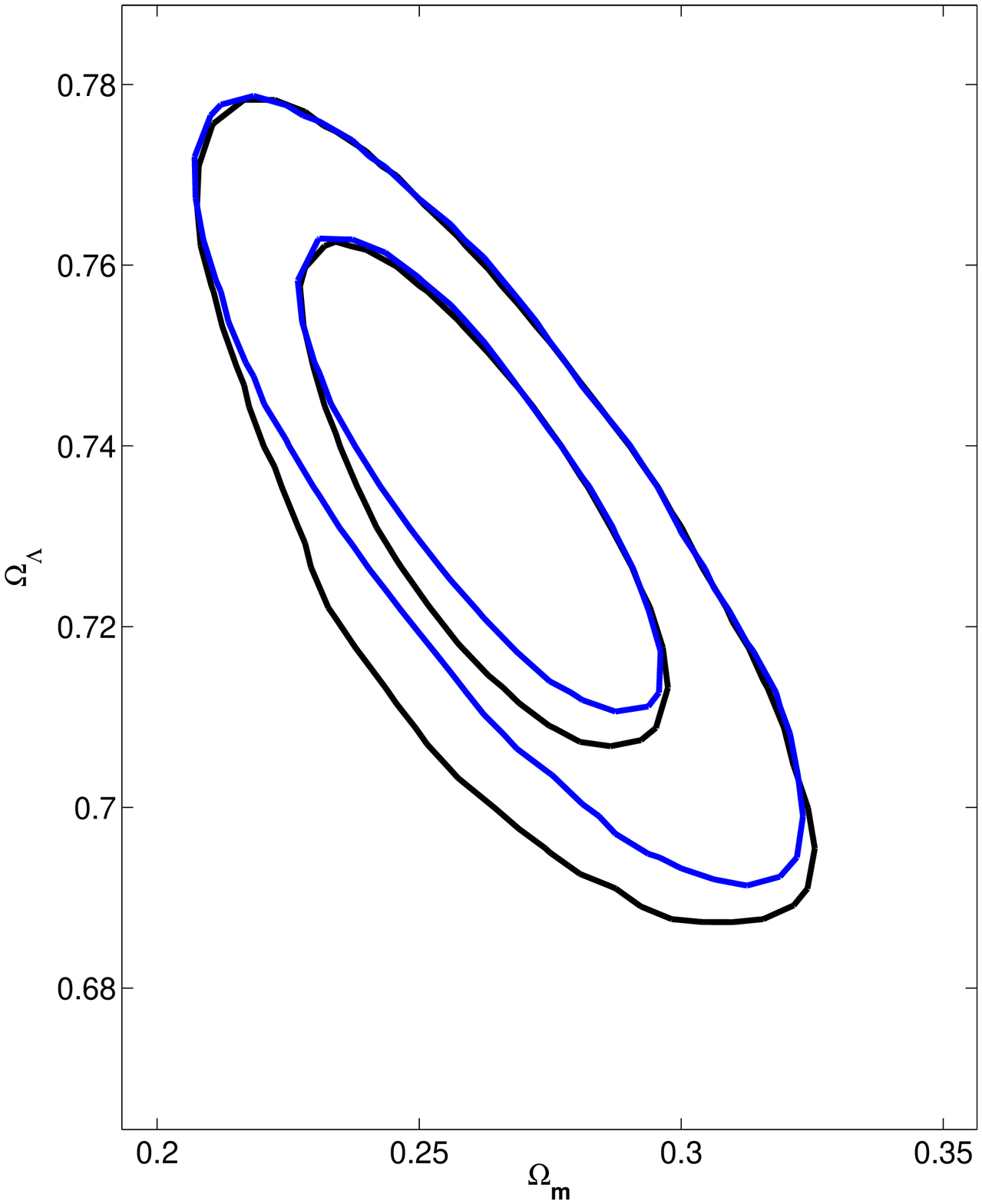}
\hfill
\includegraphics[width=5cm]{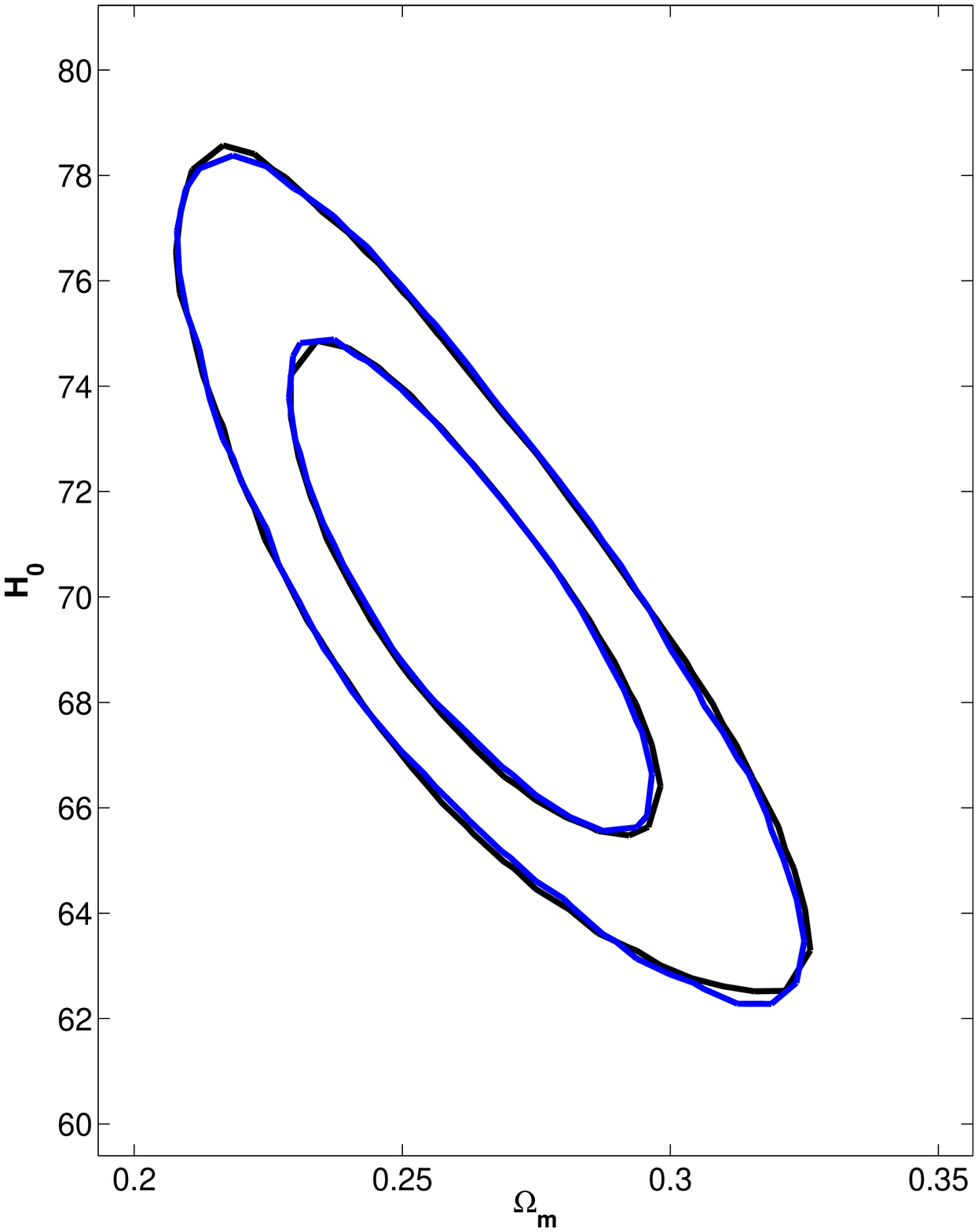}
\hfill
\includegraphics[width=5cm]{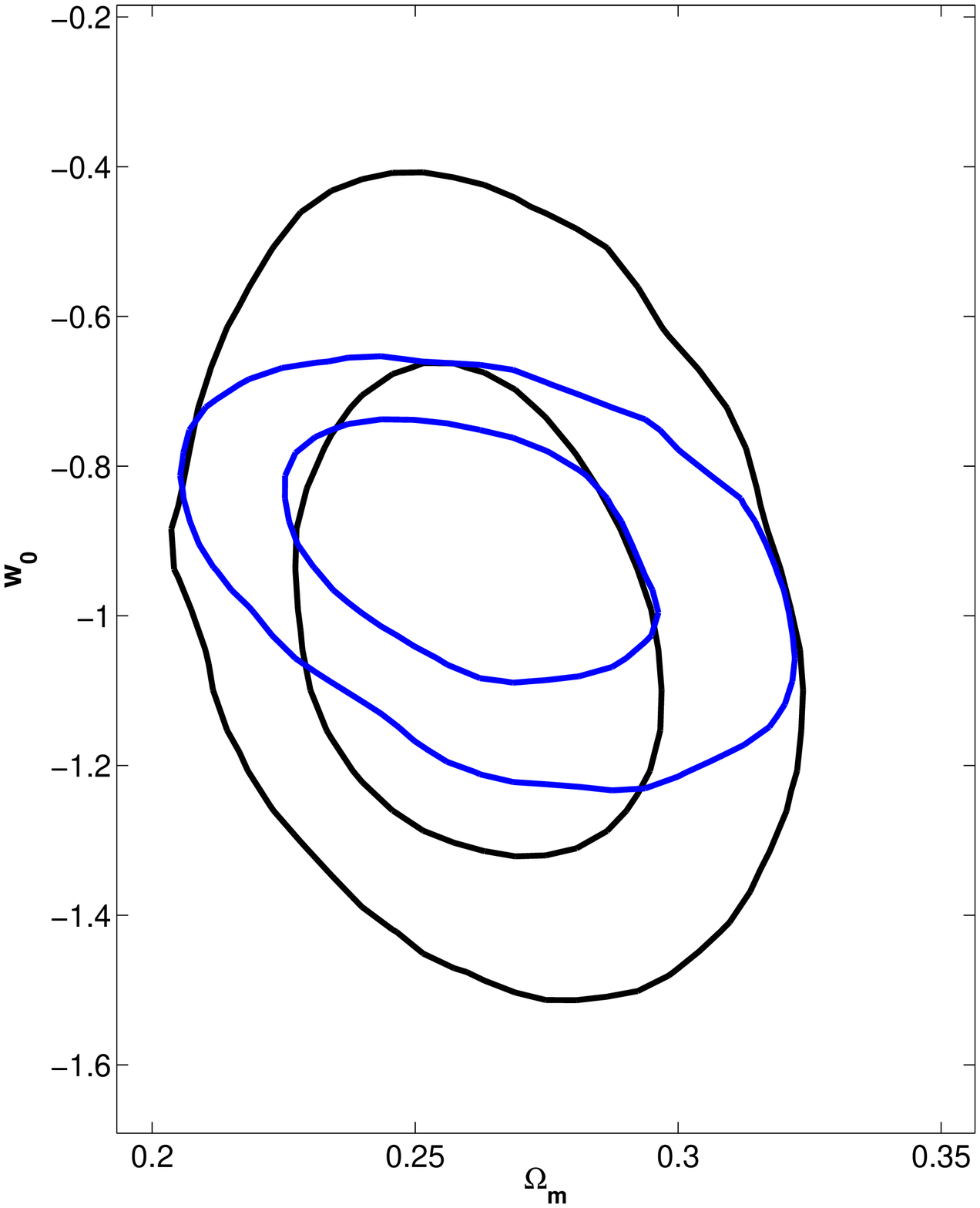}
\hfill
\includegraphics[width=5cm]{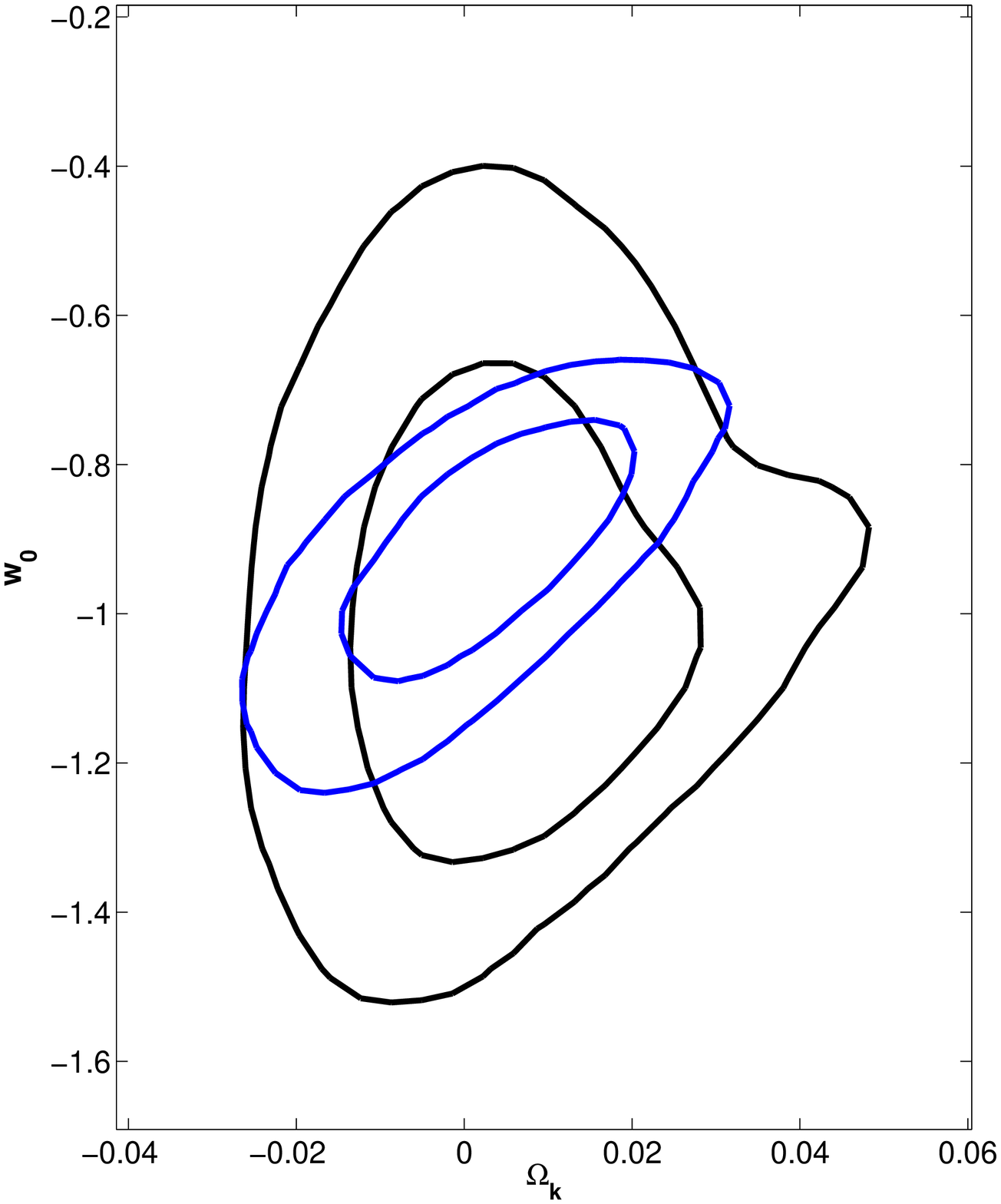}
\hfill
\includegraphics[width=5cm]{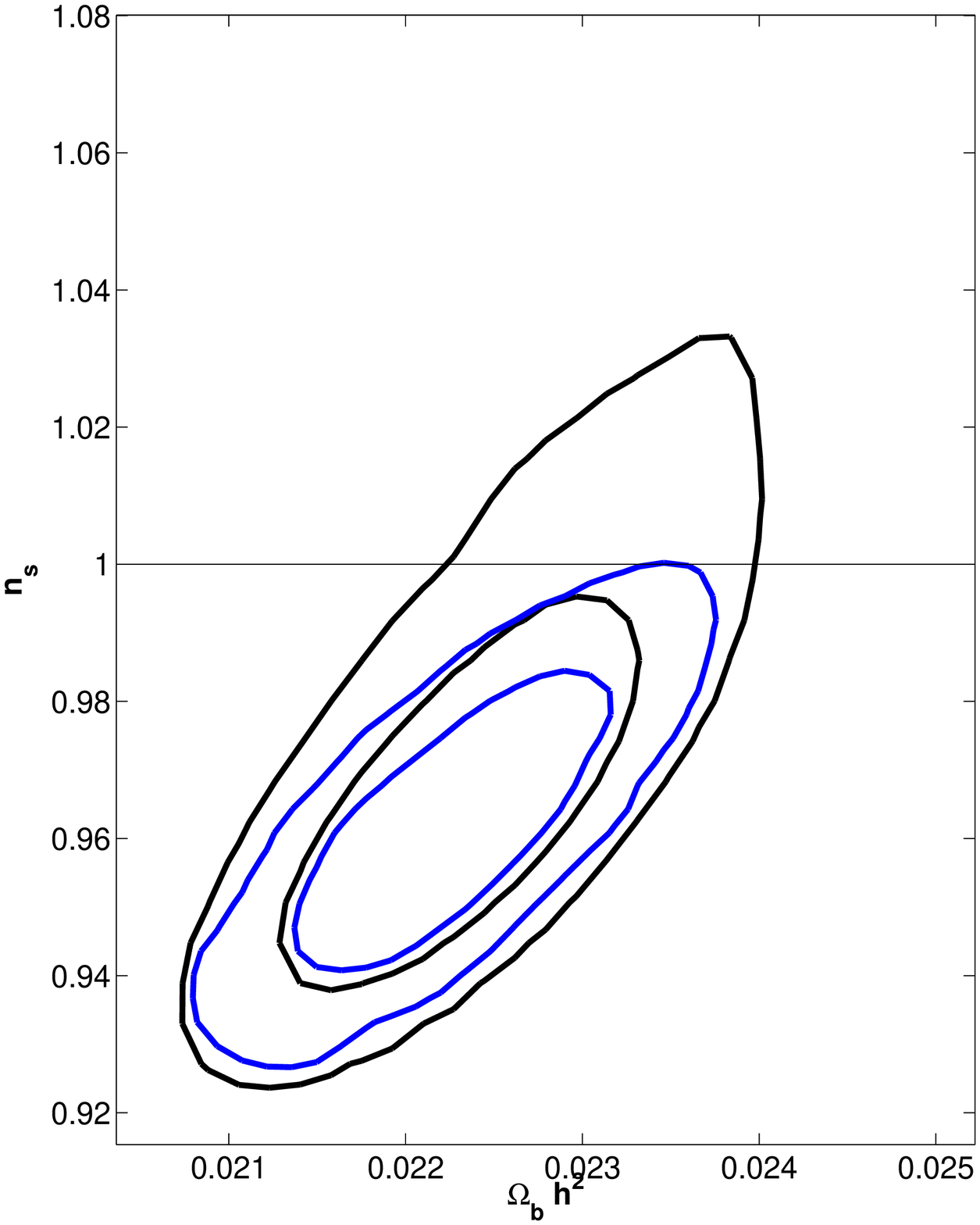}
\hfill
\includegraphics[width=5cm]{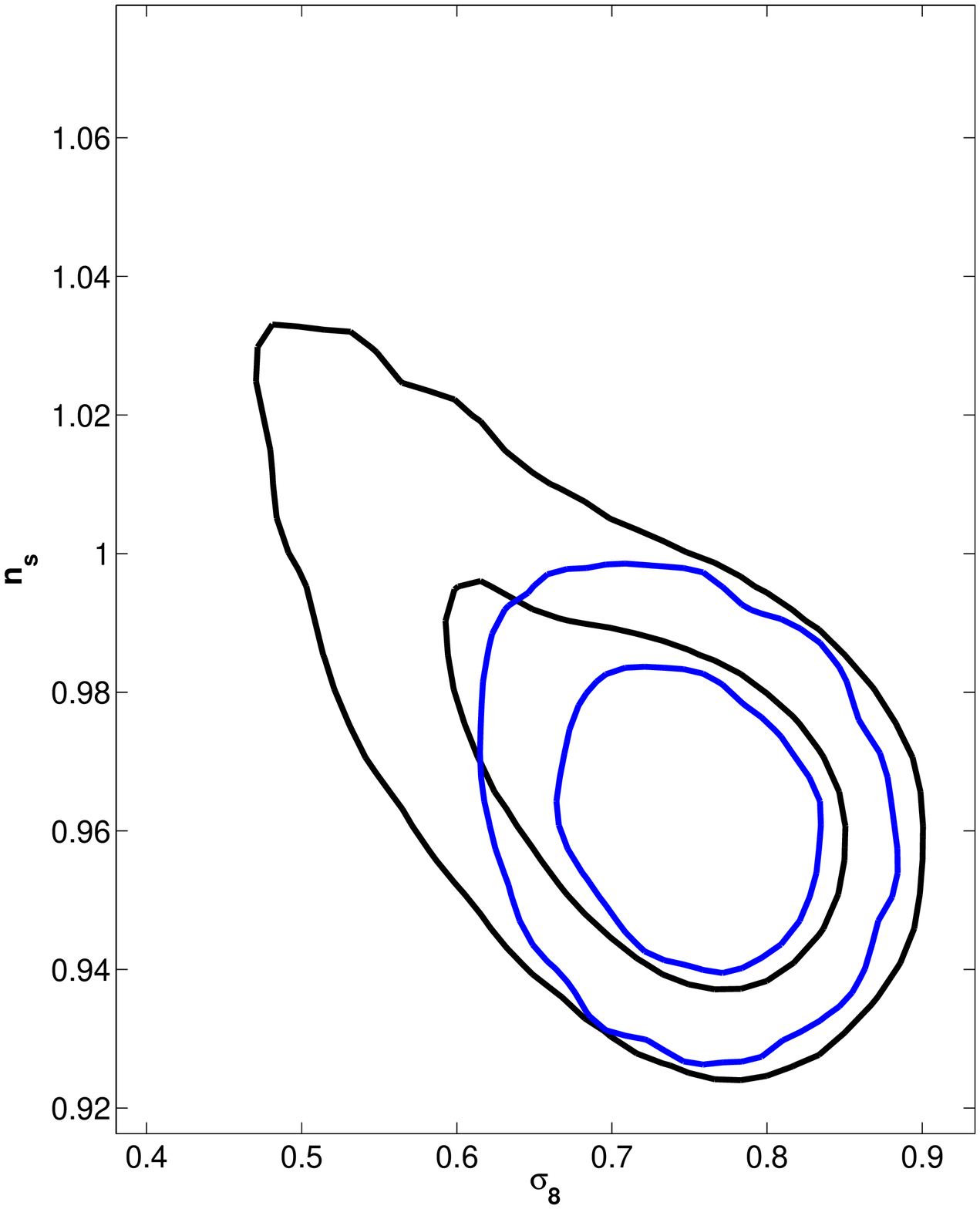}
\hfill
\includegraphics[width=5cm]{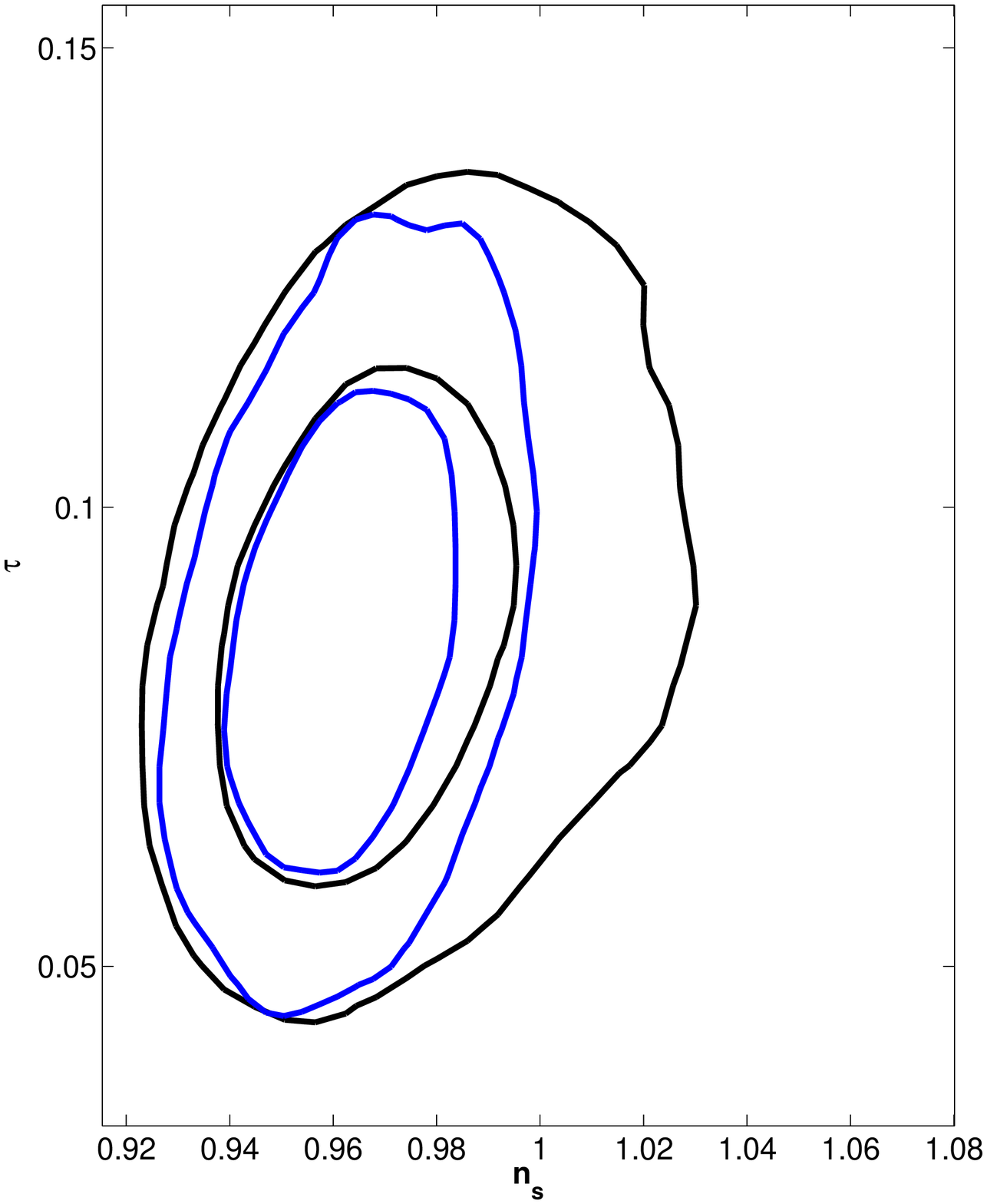}
\hfill
\includegraphics[width=5cm]{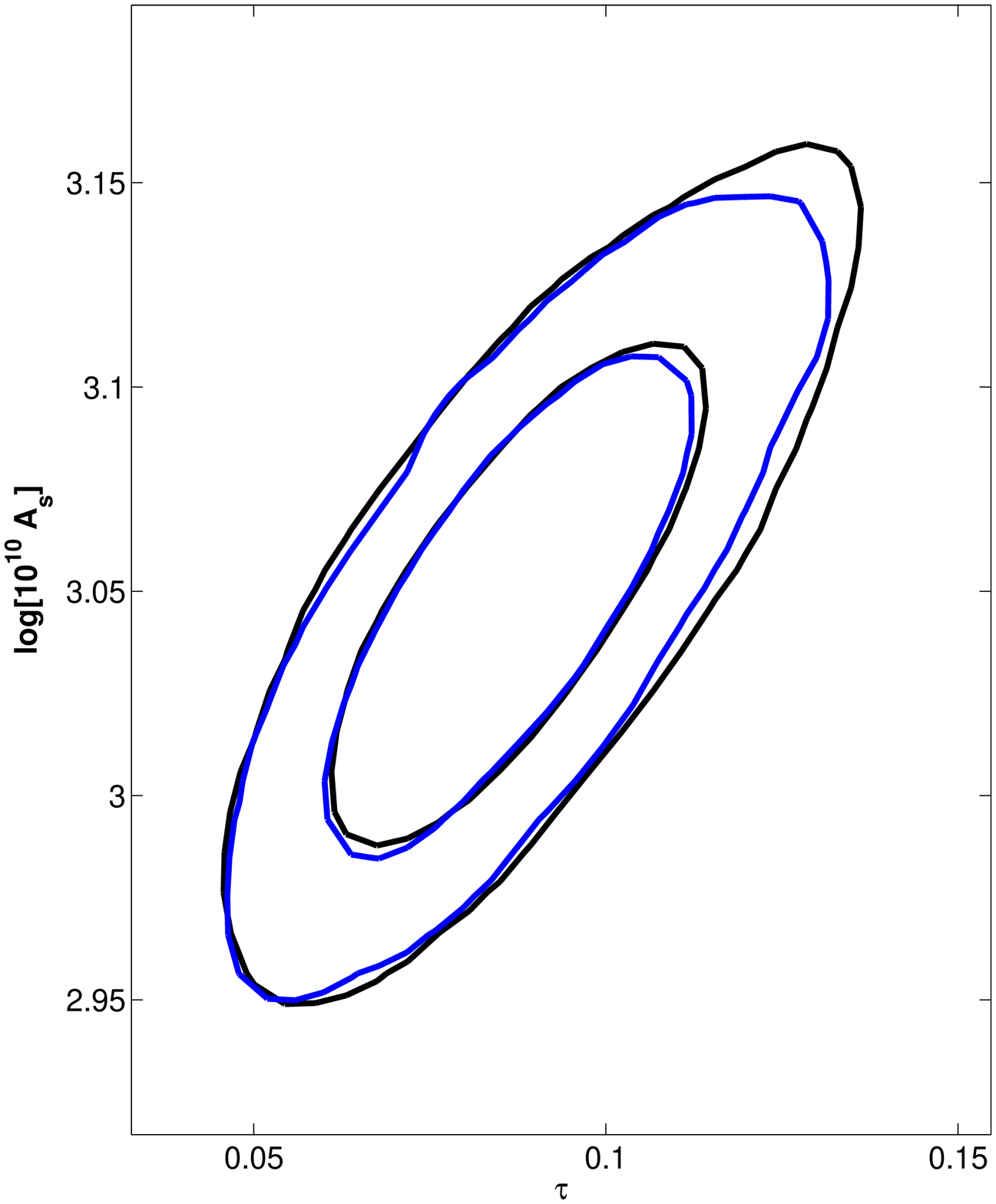}
\hfill
\includegraphics[width=5cm]{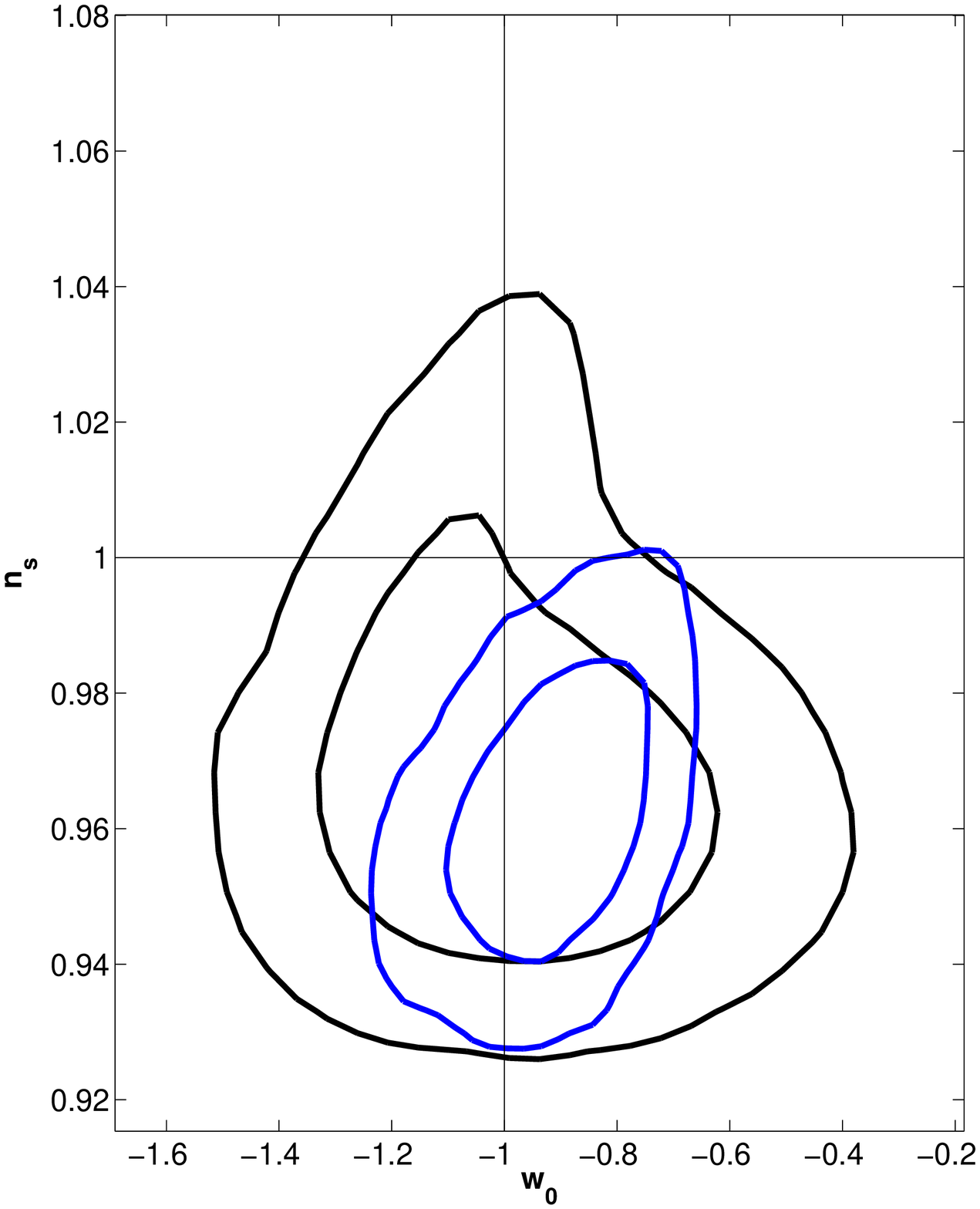}
\caption{~2D marginalized  joint posteriors of parameters 
not involving $w_1$ in a 
$wCDM$ model (blue) and a CPL model (black) 
from a full comparison of  power spectra (Likelihood A) using data set
II (WMAP5, SDSS, SNe, HST)}
\label{wCDMCorrelations}
\end{centering}
\end{figure*}

In general, when a model is relaxed to a more general model, one expects
that the constraints on the model parameters could become broader. This
happens, when the model parameters are correlated to the parameters that 
are allowed to float in the more general model, but were fixed in the 
special model. In our example, $\Lambda CDM$ and $wCDM$ are special cases 
of the CPL model. Hence, we look at how the constraints get broader when 
$w_1$ is allowed to vary. 
From Fig.~\ref{OneDPosts}, we note the differences in posteriors 
between a $wCDM$ model (black) and a CPL model (blue). We see that for
parameters like ($\theta, \omega_c, \tau, A_s$) and derived parameters 
like ($\Omega_m, H_0$),
the constraints from both
a $wCDM$ model and a CPL model are similar. 
However, for the parameters
($w_0,\Omega_K,\omega_b, n_s$) and derived parameters ($\Omega_K, \sigma_8$),
there are differences between the the posteriors of the $wCDM$ model, and 
the CPL model. Of these, only the constraints on $w_0$ are well 
approximated by the summary parameters. 
We note one cannot study the 
impact of dark energy on the standard cosmological parameters using 
summary parameters. 
These differences are reflected in 
the joint two dimensional posteriors of the $wCDM$ model (blue) and the 
CPL model (black) in Fig.~\ref{wCDMCorrelations}, which show that the 
joint constraints on the CPL model are significantly different. 

Of these parameters, it is interesting to note that the values of 
$\Omega_K$ and $n_s$ allowed by the CPL model. Both these parameters
are important theoretically as they are strongly related to parameters in 
inflation.The tail of $n_s$ using the data set (II) are usually considered
outside of the $2 \sigma$ limits. Similarly the values of $\Omega_K$ 
shown in Fig.~\ref{correlations} allowed by the data set (I) are usually 
considered ruled out by the CMB alone. 
The usual statements are on the basis of $\Lambda CDM$ or $wCDM$ models, 
and this shows that relaxing
constraints on the dark energy models can over-ride some of our intuition. 
The impact of the SDSS and BBN constraints again underline the 
importance of complementary data sets in this context. 

\section{Summary and Discussion}
\label{Disc}
Dark energy alters the background expansion of the universe leading
to geometric effects on the CMB and matter power spectrum. It also
changes the growth of gravitational potentials leading to dynamical effects
 that modify the magnitudes of the power spectrum. For a constant EoS
dark energy, the observed acceleration requires rapid decay of 
dark energy density with redshift, precluding dynamical effects. However
while dynamical effects are relatively unimportant for constant EoS dark
energy, they may be important for dark energy with time varying EoS. 
Hence, for such models, the comparison of power spectra may be able 
to distinguish between regions of parameter space that are degenerate to
dynamical effects-blind summary parameters used in previous studies. 

We study this by 
estimating parameter constraints in a CPL cosmology using two combinations
of data sets I and II (see Table~\ref{LikesTable})
WMAP five year data, SDSS LRG data, 
the Union compilation of supernovae data, 
the results of HST, and using the fits to BBN constraints. We do this
in two different methods (see Table~\ref{LikesTable}) (A) where we compare the observed CMB and matter 
power spectra to our theoretical computations of these quantities in a 
cosmology with a CPL parametrized dark energy using modified versions of
the publicly available WMAP and SDSS Likelihood codes, and 
(B) where we use a 
Gaussian likelihood in the three summary parameters for CMB, and the BAO 
summary parameters as reported in previous studies in the literature. 

\noindent
{\it{Differences between Constraints~:}}
Qualitative features in the $w_0, w_1$ joint posteriors are similar 
when either likelihood (A) or likelihood (B) are 
used, though there are quantitative differences. 
The differences in the $w_0, w_1$ joint posterior which is 
used to compute the dark energy Figure of Merit recommended by the Dark 
Energy Task force to rank the importance of experiments are studied in 
Fig.~\ref{DensityPlots}: This is an un-smoothed density plot to clearly 
show the differences between using Likelihoods A and B.
Clearly, likelihood B using summary parameters is good as
an approximate likelihood for the CPL model for the purpose of studying
dark energy parameters; however the likelihood (A) 
gives sharper and narrower posterior distributions. 
We study the differences in one dimensional marginalized constraints 
on the cosmological parameters when one uses the Likelihoods A and B in
Fig.~\ref{OneDPosts}. We find that the distributions due to B are mostly 
broader, but also include significant biases in some cases, 
further Likelihood B 
does not constrain the parameters related to $n_{\rm{s}}, \tau,A_{\rm{s}}$ at all. It is not surprising that the likelihood comparing power spectra
is more informative than one using summary parameters, as the latter only 
use a subset of the information available in the former.

\noindent
{\it{Features of the Posterior Distributions of the CPL model~:}}
Using the method (A), we show the features of the current constraints on 
the CPL dark energy parameters $w_0,$ and $ w_1$ in Fig.~\ref{w0w1Posteriors}. 
The 
main features are (i) a sharp drop in the posterior for models that allow 
significant early dark energy demonstrated in Fig.~\ref{EarlyEOS}, and (ii)
a long tail for the combination of data sets (I) in the direction of 
low $w_1$ values, for which the dark energy equation of state changes 
decreases in the past, resulting in a specific fractional density of dark 
energy causing more acceleration in the past; 
but also implying that dark energy itself decays away more rapidly, and 
(iii) the distribution is fairly non-Gaussian. 
The drop in the posterior for parameters, described in (i) 
that allow for significant dark energy happens for a combination of 
geometrical and dynamical effects.
Dynamical effects come from an Early ISW effect, and a different sourcing 
of the growth of matter perturbations due to different growth behavior of
the potential.
In Fig.~\ref{PhaseDiagram}, dark energy parameters of the kind 
described in (ii) were called ``burst dark energy" since they lead to 
short burst of acceleration as shown in Fig.~\ref{EffectonBackground}. 
Studying the correlations of the dark energy parameters with other 
background parameters in Fig.~\ref{correlations}, suggests that such models
are allowed if the flatness and Hubble constant values are also low, which 
are almost ruled out when we include the SDSS data. The non-Gaussian 
banana-shaped posterior on $w_0,w_1$ due to the data sets (I) are pinched 
off when data set (II) is used, however the distribution is still not very 
ellipsoidal. This is typical of posteriors of ill-constrained parameters, 
and can change with the addition of higher quality data. However, in
this case the distribution is unlikely to be ellipsoidal if it extends to 
early dark energy cutoff described in (i), which is not too far from the 
peak of the distribution.
These features are inherited in the marginalized one dimensional posterior
of $w_1,$ where we see an asymmetric distribution with a sharp drop in the
distribution at high values of $w_1$ and a long tail in the low values of 
$w_1$ in Fig.~\ref{w1Posteriors}.

\noindent
{\it{Impact on Standard Cosmological Parameters~:}}
We study the differences in one dimensional marginalized constraints on 
the parameters in a $wCDM$ model, when the model is relaxed to a CPL model
in Fig.~\ref{OneDPosts}. We find that the constraints on $
w_0, n_s, \sigma_8, \omega_b$ are different in a CPL model (blue) from 
their counterparts in a wCDM model (Black).
We also
compare the two dimensional joint posteriors of these 
cosmological parameters for a $wCDM$ model and a CPL model in 
Fig.~\ref{wCDMCorrelations}, where the posteriors also change when these 
parameters are involved. In particular, we note from Fig.~\ref{OneDPosts} 
that in a CPL model, there is a tail on the higher side of the 
posterior distributions of our maximal data set. Values above unity are 
allowed, in contrast to the situation for a $\Lambda CDM$ model. 
In Fig.~\ref{correlations}, we show that, in contrast to the situation in 
$\Lambda CDM$ model with spatial curvature, where data set I (WMAP5, HST, SNe) 
constrains the flatness parameters, this is not possible in $CPL$ models; 
Addition of the SDSS and BBN constraints are crucial to pinching off the 
banana in Fig.~\ref{correlations}.
As discussed, the posteriors on these parameters 
using the summary are not very good; so these questions cannot be addressed
by summary parameters.

While the analysis presented is for the CPL parametrization, one should 
remember that EoS of dark energy might be quite different. We note that 
the dynamical effects are likely to be more pronounced for equations of 
state that have a stiffer evolution with redshift. On the other hand,
if the EoS has a functional form significantly different from a CPL 
parametrization, the computed constraints may be biased~\citep{2008PhRvD..78b3526L}. Attempts to 
circumvent this problem have been made in terms of increasing the 
number of parameters describing EoS to enlarge the model space further, 
with an ultimate goal of making the
description `model independent'. This number of parameters that can be 
added is limited by the absence of tracers of cosmic evolution at 
high redshifts ~\citep{2005PhRvD..72d3509L,2008PhRvL.100x1302S}
even with the addition of futuristic SNe data. Typically, these analyzes 
are made possible by the 
degeneracies between other background parameters and the EoS by either 
invoking CMB constraints at very high redshift 
in terms of the summary parameters 
~\citep{2008PhRvL.100x1302S,2007PhRvD..76j3533W}
or by taking the cosmology to be flat
~\citep{2006PhRvD..74h3513F,2004MNRAS.354..275A,2006MNRAS.366.1081S}.
While our results suggest that such a 
use of summary parameters may be safe for CPL like parametrizations, 
it is unclear how good they are for other models intended to be included
in these enlarged sets. While the assumption of flatness can be 
justified on the basis of an inflation prior, we show that it cannot be 
justified using the data. The WMAP5 constraints on flatness for a 
universe with a cosmological constant significantly worsen in a CPL model
 and one intuitively expects them to be still broader for a dark energy with a 
parameter independent EoS. 

In comparing the matter power spectrum, we use analytic marginalization 
over a linear scale independent bias; thus parameter estimation from the 
matter power spectrum might be biased if there is really a scale dependent 
bias ~\citep{2008arXiv0810.0003R}. It is clear from their analysis as 
well as our results, that there is information in the power spectrum that 
is not encoded in the Baryon Acoustic Oscillations. Hence, it is yet 
another reason to study the bias of galaxies, so that one may be able to
extract the maximum information from galaxy surveys.

While we have showed that a likelihood comparing power spectra is 
more informative that likelihoods comparing summary parameters even for
CPL models, the computation of power spectra involves running 
computationally intensive Boltzmann code repeatedly to obtain a large 
number of posterior samples, while the computation of summary parameters is
extremely fast. 
To give a quantitative idea, it takes about six hours to get a 
thousand chain
steps on a single processor using Likelihood A, while it takes about six 
minutes to do the same using Likelihood B.  
Since such computations are quite doable for a particular model, one should
use a comparison of power spectra to do precision cosmology, while
summary parameters can still be used to explore new models and get rough 
estimates.
It might be possible to make the likelihood using summary parameters more 
informative by adding further summary parameters; 
an example of this approach has been followed in ~\citep{2007ApJ...664..633W}, where several additional summary parameters have been used. 
However, if it might be expected that a particular parametrization 
(such as the CPL model today) will have repeated use, one can also train
Boltzmann accelerators~\citep{2007arXiv0712.0194F, 2007ApJ...654....2F} to 
efficiently and accurately compute the CMB and matter power spectrum. If the parametrization is studied enough, the work in training the Boltzmann accelerators  would be 
offset by the gain in time for computation of accurate constraints.
\section{Acknowledgments}
This work was partially funded by NSF grants
AST 05-07676 and AST 07-08849, by NASA contract
JPL1236748.
This work was partially supported by the National Center for 
Supercomputing Applications under MRAC MCA04N015 and 
utilized NCSA IA-64 TeraGrid cluster (mercury). 
We would like to 
acknowledge the use of CAMB and CosmoMC in our work. 
BDW would like to
thank the Galileo Galilei Institute for hospitality while this work
was being completed. 
RB would like to thank W. Fendt for useful discussions.
\bibliography{CurrentConstraints_arXiv_v1}

\begin{thebibliography}{62}
\expandafter\ifx\csname natexlab\endcsname\relax\def\natexlab#1{#1}\fi

\bibitem[{{Alam} {et~al.}(2004){Alam}, {Sahni}, {Deep Saini}, \&
  {Starobinsky}}]{2004MNRAS.354..275A}
{Alam}, U., {Sahni}, V., {Deep Saini}, T., \& {Starobinsky}, A.~A. 2004,
  \mnras, 354, 275, arXiv:astro-ph/0311364

\bibitem[{{Albrecht} {et~al.}(2006){Albrecht}, {Bernstein}, {Cahn}, {Freedman},
  {Hewitt}, {Hu}, {Huth}, {Kamionkowski}, {Kolb}, {Knox}, {Mather}, {Staggs},
  \& {Suntzeff}}]{2006astro.ph..9591A}
{Albrecht}, A. {et~al.} 2006, ArXiv Astrophysics e-prints,
  arXiv:astro-ph/0609591

\bibitem[{Amsler {et~al.}(2008)}]{Amsler:2008zzb}
Amsler, C., {et~al.} 2008, Phys. Lett., B667, 1

\bibitem[{{Astier} {et~al.}(2006){Astier}, {Guy}, {Regnault}, {Pain},
  {Aubourg}, {Balam}, {Basa}, {Carlberg}, {Fabbro}, {Fouchez}, {Hook},
  {Howell}, {Lafoux}, {Neill}, {Palanque-Delabrouille}, {Perrett}, {Pritchet},
  {Rich}, {Sullivan}, {Taillet}, {Aldering}, {Antilogus}, {Arsenijevic},
  {Balland}, {Baumont}, {Bronder}, {Courtois}, {Ellis}, {Filiol}, {Gon{\c
  c}alves}, {Goobar}, {Guide}, {Hardin}, {Lusset}, {Lidman}, {McMahon},
  {Mouchet}, {Mourao}, {Perlmutter}, {Ripoche}, {Tao}, \&
  {Walton}}]{2006A&A...447...31A}
{Astier}, P. {et~al.} 2006, \aap, 447, 31, arXiv:astro-ph/0510447

\bibitem[{{Bond} {et~al.}(1997){Bond}, {Efstathiou}, \&
  {Tegmark}}]{1997MNRAS.291L..33B}
{Bond}, J.~R., {Efstathiou}, G., \& {Tegmark}, M. 1997, \mnras, 291, L33,
  arXiv:astro-ph/9702100

\bibitem[{{Caldwell}(2002)}]{2002PhLB..545...23C}
{Caldwell}, R.~R. 2002, Physics Letters B, 545, 23, arXiv:astro-ph/9908168

\bibitem[{{Caldwell} \& {Doran}(2005)}]{2005PhRvD..72d3527C}
{Caldwell}, R.~R., \& {Doran}, M. 2005, \prd, 72, 043527,
  arXiv:astro-ph/0501104

\bibitem[{{Carroll} {et~al.}(2003){Carroll}, {Hoffman}, \&
  {Trodden}}]{2003PhRvD..68b3509C}
{Carroll}, S.~M., {Hoffman}, M., \& {Trodden}, M. 2003, \prd, 68, 023509,
  arXiv:astro-ph/0301273

\bibitem[{{Chevallier} \& {Polarski}(2001)}]{2001IJMPD..10..213C}
{Chevallier}, M., \& {Polarski}, D. 2001, International Journal of Modern
  Physics D, 10, 213, arXiv:gr-qc/0009008

\bibitem[{{Cyburt} {et~al.}(2008){Cyburt}, {Fields}, \&
  {Olive}}]{2008JCAP...11..012C}
{Cyburt}, R.~H., {Fields}, B.~D., \& {Olive}, K.~A. 2008, Journal of Cosmology
  and Astro-Particle Physics, 11, 12

\bibitem[{{Cyburt} {et~al.}(2002){Cyburt}, {Fields}, {Pavlidou}, \&
  {Wandelt}}]{2002PhRvD..65l3503C}
{Cyburt}, R.~H., {Fields}, B.~D., {Pavlidou}, V., \& {Wandelt}, B. 2002, \prd,
  65, 123503, arXiv:astro-ph/0203240

\bibitem[{{Davis} {et~al.}(2007){Davis}, {M{\"o}rtsell}, {Sollerman}, {Becker},
  {Blondin}, {Challis}, {Clocchiatti}, {Filippenko}, {Foley}, {Garnavich},
  {Jha}, {Krisciunas}, {Kirshner}, {Leibundgut}, {Li}, {Matheson}, {Miknaitis},
  {Pignata}, {Rest}, {Riess}, {Schmidt}, {Smith}, {Spyromilio}, {Stubbs},
  {Suntzeff}, {Tonry}, {Wood-Vasey}, \& {Zenteno}}]{2007ApJ...666..716D}
{Davis}, T.~M. {et~al.} 2007, \apj, 666, 716, arXiv:astro-ph/0701510

\bibitem[{{Doran} \& {Lilley}(2002)}]{2002MNRAS.330..965D}
{Doran}, M., \& {Lilley}, M. 2002, \mnras, 330, 965, arXiv:astro-ph/0104486

\bibitem[{{Doran} {et~al.}(2001{\natexlab{a}}){Doran}, {Lilley}, {Schwindt}, \&
  {Wetterich}}]{2001ApJ...559..501D}
{Doran}, M., {Lilley}, M., {Schwindt}, J., \& {Wetterich}, C.
  2001{\natexlab{a}}, \apj, 559, 501, arXiv:astro-ph/0012139

\bibitem[{{Doran} {et~al.}(2001{\natexlab{b}}){Doran}, {Schwindt}, \&
  {Wetterich}}]{2001PhRvD..64l3520D}
{Doran}, M., {Schwindt}, J.-M., \& {Wetterich}, C. 2001{\natexlab{b}}, \prd,
  64, 123520, arXiv:astro-ph/0107525

\bibitem[{Dunkley {et~al.}(2008)}]{Dunkley:2008ie}
Dunkley, J., {et~al.} 2008, 0803.0586

\bibitem[{{Elgar{\o}y} \& {Multam{\"a}ki}(2007)}]{2007A&A...471...65E}
{Elgar{\o}y}, O., \& {Multam{\"a}ki}, T. 2007, \aap, 471, 65,
  arXiv:astro-ph/0702343

\bibitem[{{Fang} {et~al.}(2008){Fang}, {Hu}, \& {Lewis}}]{2008PhRvD..78h7303F}
{Fang}, W., {Hu}, W., \& {Lewis}, A. 2008, \prd, 78, 087303, 0808.3125

\bibitem[{{Fay} \& {Tavakol}(2006)}]{2006PhRvD..74h3513F}
{Fay}, S., \& {Tavakol}, R. 2006, \prd, 74, 083513, arXiv:astro-ph/0606431

\bibitem[{{Fendt} \& {Wandelt}(2007{\natexlab{a}})}]{2007arXiv0712.0194F}
{Fendt}, W.~A., \& {Wandelt}, B.~D. 2007{\natexlab{a}}, ArXiv e-prints,
  0712.0194

\bibitem[{{Fendt} \& {Wandelt}(2007{\natexlab{b}})}]{2007ApJ...654....2F}
------. 2007{\natexlab{b}}, \apj, 654, 2, arXiv:astro-ph/0606709

\bibitem[{{Freedman} {et~al.}(2001){Freedman}, {Madore}, {Gibson}, {Ferrarese},
  {Kelson}, {Sakai}, {Mould}, {Kennicutt}, {Ford}, {Graham}, {Huchra},
  {Hughes}, {Illingworth}, {Macri}, \& {Stetson}}]{2001ApJ...553...47F}
{Freedman}, W.~L. {et~al.} 2001, \apj, 553, 47, arXiv:astro-ph/0012376

\bibitem[{{Garnavich} {et~al.}(1998){Garnavich}, {Jha}, {Challis},
  {Clocchiatti}, {Diercks}, {Filippenko}, {Gilliland}, {Hogan}, {Kirshner},
  {Leibundgut}, {Phillips}, {Reiss}, {Riess}, {Schmidt}, {Schommer}, {Smith},
  {Spyromilio}, {Stubbs}, {Suntzeff}, {Tonry}, \&
  {Carroll}}]{1998ApJ...509...74G}
{Garnavich}, P.~M. {et~al.} 1998, \apj, 509, 74, arXiv:astro-ph/9806396

\bibitem[{{Hicken} {et~al.}(2009){Hicken}, {Wood-Vasey}, {Blondin}, {Challis},
  {Jha}, {Kelly}, {Rest}, \& {Kirshner}}]{2009arXiv0901.4804H}
{Hicken}, M., {Wood-Vasey}, W.~M., {Blondin}, S., {Challis}, P., {Jha}, S.,
  {Kelly}, P.~L., {Rest}, A., \& {Kirshner}, R.~P. 2009, ArXiv e-prints,
  0901.4804

\bibitem[{{Hinshaw} {et~al.}(2008){Hinshaw}, {Weiland}, {Hill}, {Odegard},
  {Larson}, {Bennett}, {Dunkley}, {Gold}, {Greason}, {Jarosik}, {Komatsu},
  {Nolta}, {Page}, {Spergel}, {Wollack}, {Halpern}, {Kogut}, {Limon}, {Meyer},
  {Tucker}, \& {Wright}}]{2008arXiv0803.0732H}
{Hinshaw}, G. {et~al.} 2008, ArXiv e-prints, 0803.0732

\bibitem[{{Hu}(2005)}]{2005PhRvD..71d7301H}
{Hu}, W. 2005, \prd, 71, 047301, arXiv:astro-ph/0410680

\bibitem[{{Hu} \& {Dodelson}(2002)}]{2002ARA&A..40..171H}
{Hu}, W., \& {Dodelson}, S. 2002, \araa, 40, 171, arXiv:astro-ph/0110414

\bibitem[{{Hu} \& {Eisenstein}(1998)}]{1998ApJ...498..497H}
{Hu}, W., \& {Eisenstein}, D.~J. 1998, \apj, 498, 497, arXiv:astro-ph/9710216

\bibitem[{{Hu} {et~al.}(2001){Hu}, {Fukugita}, {Zaldarriaga}, \&
  {Tegmark}}]{2001ApJ...549..669H}
{Hu}, W., {Fukugita}, M., {Zaldarriaga}, M., \& {Tegmark}, M. 2001, \apj, 549,
  669, arXiv:astro-ph/0006436

\bibitem[{{Hu} \& {Sugiyama}(1996)}]{1996ApJ...471..542H}
{Hu}, W., \& {Sugiyama}, N. 1996, \apj, 471, 542, arXiv:astro-ph/9510117

\bibitem[{{Huey}(2004)}]{2004astro.ph.11102H}
{Huey}, G. 2004, ArXiv Astrophysics e-prints, astro-ph/0411102

\bibitem[{{Huey} \& {Wandelt}(2006)}]{2006PhRvD..74b3519H}
{Huey}, G., \& {Wandelt}, B.~D. 2006, \prd, 74, 023519, arXiv:astro-ph/0407196

\bibitem[{{Knop} {et~al.}(2003){Knop}, {Aldering}, {Amanullah}, {Astier},
  {Blanc}, {Burns}, {Conley}, {Deustua}, {Doi}, {Ellis}, {Fabbro}, {Folatelli},
  {Fruchter}, {Garavini}, {Garmond}, {Garton}, {Gibbons}, {Goldhaber},
  {Goobar}, {Groom}, {Hardin}, {Hook}, {Howell}, {Kim}, {Lee}, {Lidman},
  {Mendez}, {Nobili}, {Nugent}, {Pain}, {Panagia}, {Pennypacker}, {Perlmutter},
  {Quimby}, {Raux}, {Regnault}, {Ruiz-Lapuente}, {Sainton}, {Schaefer},
  {Schahmaneche}, {Smith}, {Spadafora}, {Stanishev}, {Sullivan}, {Walton},
  {Wang}, {Wood-Vasey}, \& {Yasuda}}]{2003ApJ...598..102K}
{Knop}, R.~A. {et~al.} 2003, \apj, 598, 102, arXiv:astro-ph/0309368

\bibitem[{{Komatsu} {et~al.}(2008){Komatsu}, {Dunkley}, {Nolta}, {Bennett},
  {Gold}, {Hinshaw}, {Jarosik}, {Larson}, {Limon}, {Page}, {Spergel},
  {Halpern}, {Hill}, {Kogut}, {Meyer}, {Tucker}, {Weiland}, {Wollack}, \&
  {Wright}}]{2008arXiv0803.0547K}
{Komatsu}, E. {et~al.} 2008, ArXiv e-prints, 0803.0547

\bibitem[{{Kowalski} {et~al.}(2008){Kowalski}, {Rubin}, {Aldering},
  {Agostinho}, {Amadon}, {Amanullah}, {Balland}, {Barbary}, {Blanc}, {Challis},
  {Conley}, {Connolly}, {Covarrubias}, {Dawson}, {Deustua}, {Ellis}, {Fabbro},
  {Fadeyev}, {Fan}, {Farris}, {Folatelli}, {Frye}, {Garavini}, {Gates},
  {Germany}, {Goldhaber}, {Goldman}, {Goobar}, {Groom}, {Haissinski}, {Hardin},
  {Hook}, {Kent}, {Kim}, {Knop}, {Lidman}, {Linder}, {Mendez}, {Meyers},
  {Miller}, {Moniez}, {Mour{\~a}o}, {Newberg}, {Nobili}, {Nugent}, {Pain},
  {Perdereau}, {Perlmutter}, {Phillips}, {Prasad}, {Quimby}, {Regnault},
  {Rich}, {Rubenstein}, {Ruiz-Lapuente}, {Santos}, {Schaefer}, {Schommer},
  {Smith}, {Soderberg}, {Spadafora}, {Strolger}, {Strovink}, {Suntzeff},
  {Suzuki}, {Thomas}, {Walton}, {Wang}, {Wood-Vasey}, \&
  {Yun}}]{2008ApJ...686..749K}
{Kowalski}, M. {et~al.} 2008, \apj, 686, 749, 0804.4142

\bibitem[{{Lazkoz} {et~al.}(2008){Lazkoz}, {Nesseris}, \&
  {Perivolaropoulos}}]{2008JCAP...07..012L}
{Lazkoz}, R., {Nesseris}, S., \& {Perivolaropoulos}, L. 2008, Journal of
  Cosmology and Astro-Particle Physics, 7, 12, 0712.1232

\bibitem[{{Lewis} \& {Bridle}(2002)}]{2002PhRvD..66j3511L}
{Lewis}, A., \& {Bridle}, S. 2002, \prd, 66, 103511, arXiv:astro-ph/0205436

\bibitem[{{Lewis} {et~al.}(2000){Lewis}, {Challinor}, \&
  {Lasenby}}]{2000ApJ...538..473L}
{Lewis}, A., {Challinor}, A., \& {Lasenby}, A. 2000, \apj, 538, 473,
  arXiv:astro-ph/9911177

\bibitem[{Li {et~al.}(2005)Li, Feng, \& Zhang}]{Li:2005fm}
Li, M.-z., Feng, B., \& Zhang, X.-m. 2005, JCAP, 0512, 002, hep-ph/0503268

\bibitem[{{Linden} \& {Virey}(2008)}]{2008PhRvD..78b3526L}
{Linden}, S., \& {Virey}, J.-M. 2008, \prd, 78, 023526, 0804.0389

\bibitem[{{Linder}(2003)}]{2003PhRvL..90i1301L}
{Linder}, E.~V. 2003, Physical Review Letters, 90, 091301,
  arXiv:astro-ph/0208512

\bibitem[{{Linder} \& {Huterer}(2005)}]{2005PhRvD..72d3509L}
{Linder}, E.~V., \& {Huterer}, D. 2005, \prd, 72, 043509,
  arXiv:astro-ph/0505330

\bibitem[{{Nolta} {et~al.}(2008){Nolta}, {Dunkley}, {Hill}, {Hinshaw},
  {Komatsu}, {Larson}, {Page}, {Spergel}, {Bennett}, {Gold}, {Jarosik},
  {Odegard}, {Weiland}, {Wollack}, {Halpern}, {Kogut}, {Limon}, {Meyer},
  {Tucker}, \& {Wright}}]{2008arXiv0803.0593N}
{Nolta}, M.~R. {et~al.} 2008, ArXiv e-prints, 0803.0593

\bibitem[{{Perlmutter} {et~al.}(1999){Perlmutter}, {Aldering}, {Goldhaber},
  {Knop}, {Nugent}, {Castro}, {Deustua}, {Fabbro}, {Goobar}, {Groom}, {Hook},
  {Kim}, {Kim}, {Lee}, {Nunes}, {Pain}, {Pennypacker}, {Quimby}, {Lidman},
  {Ellis}, {Irwin}, {McMahon}, {Ruiz-Lapuente}, {Walton}, {Schaefer}, {Boyle},
  {Filippenko}, {Matheson}, {Fruchter}, {Panagia}, {Newberg}, {Couch}, \& {The
  Supernova Cosmology Project}}]{1999ApJ...517..565P}
{Perlmutter}, S. {et~al.} 1999, \apj, 517, 565, arXiv:astro-ph/9812133

\bibitem[{{Rassat} {et~al.}(2008){Rassat}, {Amara}, {Amendola}, {Castander},
  {Kitching}, {Kunz}, {Refregier}, {Wang}, \& {Weller}}]{2008arXiv0810.0003R}
{Rassat}, A. {et~al.} 2008, ArXiv e-prints, 0810.0003

\bibitem[{{Riess} {et~al.}(1998){Riess}, {Filippenko}, {Challis},
  {Clocchiatti}, {Diercks}, {Garnavich}, {Gilliland}, {Hogan}, {Jha},
  {Kirshner}, {Leibundgut}, {Phillips}, {Reiss}, {Schmidt}, {Schommer},
  {Smith}, {Spyromilio}, {Stubbs}, {Suntzeff}, \&
  {Tonry}}]{1998AJ....116.1009R}
{Riess}, A.~G. {et~al.} 1998, \aj, 116, 1009, arXiv:astro-ph/9805201

\bibitem[{{Riess} {et~al.}(2004){Riess}, {Strolger}, {Tonry}, {Casertano},
  {Ferguson}, {Mobasher}, {Challis}, {Filippenko}, {Jha}, {Li}, {Chornock},
  {Kirshner}, {Leibundgut}, {Dickinson}, {Livio}, {Giavalisco}, {Steidel},
  {Ben{\'{\i}}tez}, \& {Tsvetanov}}]{2004ApJ...607..665R}
------. 2004, \apj, 607, 665, arXiv:astro-ph/0402512

\bibitem[{{Sarkar} {et~al.}(2008){Sarkar}, {Sullivan}, {Joudaki}, {Amblard},
  {Holz}, \& {Cooray}}]{2008PhRvL.100x1302S}
{Sarkar}, D., {Sullivan}, S., {Joudaki}, S., {Amblard}, A., {Holz}, D.~E., \&
  {Cooray}, A. 2008, Physical Review Letters, 100, 241302, 0709.1150

\bibitem[{Seager {et~al.}(1999)Seager, Sasselov, \& Scott}]{Seager:1999bc}
Seager, S., Sasselov, D.~D., \& Scott, D. 1999, astro-ph/9909275

\bibitem[{Seager {et~al.}(2000)Seager, Sasselov, \& Scott}]{Seager:1999km}
------. 2000, Astrophys. J. Suppl., 128, 407, astro-ph/9912182

\bibitem[{{Shafieloo} {et~al.}(2006){Shafieloo}, {Alam}, {Sahni}, \&
  {Starobinsky}}]{2006MNRAS.366.1081S}
{Shafieloo}, A., {Alam}, U., {Sahni}, V., \& {Starobinsky}, A.~A. 2006, \mnras,
  366, 1081, arXiv:astro-ph/0505329

\bibitem[{{Simha} \& {Steigman}(2008)}]{2008JCAP...06..016S}
{Simha}, V., \& {Steigman}, G. 2008, Journal of Cosmology and Astro-Particle
  Physics, 6, 16, 0803.3465

\bibitem[{{Tegmark} {et~al.}(2006){Tegmark}, {Eisenstein}, {Strauss},
  {Weinberg}, {Blanton}, {Frieman}, {Fukugita}, {Gunn}, {Hamilton}, {Knapp},
  {Nichol}, {Ostriker}, {Padmanabhan}, {Percival}, {Schlegel}, {Schneider},
  {Scoccimarro}, {Seljak}, {Seo}, {Swanson}, {Szalay}, {Vogeley}, {Yoo},
  {Zehavi}, {Abazajian}, {Anderson}, {Annis}, {Bahcall}, {Bassett}, {Berlind},
  {Brinkmann}, {Budavari}, {Castander}, {Connolly}, {Csabai}, {Doi},
  {Finkbeiner}, {Gillespie}, {Glazebrook}, {Hennessy}, {Hogg}, {Ivezi{\'c}},
  {Jain}, {Johnston}, {Kent}, {Lamb}, {Lee}, {Lin}, {Loveday}, {Lupton},
  {Munn}, {Pan}, {Park}, {Peoples}, {Pier}, {Pope}, {Richmond}, {Rockosi},
  {Scranton}, {Sheth}, {Stebbins}, {Stoughton}, {Szapudi}, {Tucker}, {Berk},
  {Yanny}, \& {York}}]{2006PhRvD..74l3507T}
{Tegmark}, M. {et~al.} 2006, \prd, 74, 123507, arXiv:astro-ph/0608632

\bibitem[{{Tegmark} {et~al.}(2004){Tegmark}, {Strauss}, {Blanton}, {Abazajian},
  {Dodelson}, {Sandvik}, {Wang}, {Weinberg}, {Zehavi}, {Bahcall}, {Hoyle},
  {Schlegel}, {Scoccimarro}, {Vogeley}, {Berlind}, {Budavari}, {Connolly},
  {Eisenstein}, {Finkbeiner}, {Frieman}, {Gunn}, {Hui}, {Jain}, {Johnston},
  {Kent}, {Lin}, {Nakajima}, {Nichol}, {Ostriker}, {Pope}, {Scranton},
  {Seljak}, {Sheth}, {Stebbins}, {Szalay}, {Szapudi}, {Xu}, {Annis},
  {Brinkmann}, {Burles}, {Castander}, {Csabai}, {Loveday}, {Doi}, {Fukugita},
  {Gillespie}, {Hennessy}, {Hogg}, {Ivezi{\'c}}, {Knapp}, {Lamb}, {Lee},
  {Lupton}, {McKay}, {Kunszt}, {Munn}, {O'Connell}, {Peoples}, {Pier},
  {Richmond}, {Rockosi}, {Schneider}, {Stoughton}, {Tucker}, {vanden Berk},
  {Yanny}, \& {York}}]{2004PhRvD..69j3501T}
------. 2004, \prd, 69, 103501, arXiv:astro-ph/0310723

\bibitem[{{Tonry} {et~al.}(2003){Tonry}, {Schmidt}, {Barris}, {Candia},
  {Challis}, {Clocchiatti}, {Coil}, {Filippenko}, {Garnavich}, {Hogan},
  {Holland}, {Jha}, {Kirshner}, {Krisciunas}, {Leibundgut}, {Li}, {Matheson},
  {Phillips}, {Riess}, {Schommer}, {Smith}, {Sollerman}, {Spyromilio},
  {Stubbs}, \& {Suntzeff}}]{2003ApJ...594....1T}
{Tonry}, J.~L. {et~al.} 2003, \apj, 594, 1, arXiv:astro-ph/0305008

\bibitem[{{Wang}(2008)}]{2008PhRvD..77l3525W}
{Wang}, Y. 2008, \prd, 77, 123525, 0803.4295

\bibitem[{{Wang} \& {Mukherjee}(2007)}]{2007PhRvD..76j3533W}
{Wang}, Y., \& {Mukherjee}, P. 2007, \prd, 76, 103533, arXiv:astro-ph/0703780

\bibitem[{{Weller} \& {Lewis}(2003)}]{2003MNRAS.346..987W}
{Weller}, J., \& {Lewis}, A.~M. 2003, \mnras, 346, 987, arXiv:astro-ph/0307104

\bibitem[{Wong {et~al.}(2007)Wong, Moss, \& Scott}]{Wong:2007ym}
Wong, W.~Y., Moss, A., \& Scott, D. 2007, 0711.1357

\bibitem[{{Wood-Vasey} {et~al.}(2007){Wood-Vasey}, {Miknaitis}, {Stubbs},
  {Jha}, {Riess}, {Garnavich}, {Kirshner}, {Aguilera}, {Becker}, {Blackman},
  {Blondin}, {Challis}, {Clocchiatti}, {Conley}, {Covarrubias}, {Davis},
  {Filippenko}, {Foley}, {Garg}, {Hicken}, {Krisciunas}, {Leibundgut}, {Li},
  {Matheson}, {Miceli}, {Narayan}, {Pignata}, {Prieto}, {Rest}, {Salvo},
  {Schmidt}, {Smith}, {Sollerman}, {Spyromilio}, {Tonry}, {Suntzeff}, \&
  {Zenteno}}]{2007ApJ...666..694W}
{Wood-Vasey}, W.~M. {et~al.} 2007, \apj, 666, 694, arXiv:astro-ph/0701041

\bibitem[{{Wright}(2007)}]{2007ApJ...664..633W}
{Wright}, E.~L. 2007, \apj, 664, 633, arXiv:astro-ph/0701584

\bibitem[{{Zhang} {et~al.}(2006){Zhang}, {Li}, {Piao}, \&
  {Zhang}}]{2006MPLA...21..231Z}
{Zhang}, X.-F., {Li}, H., {Piao}, Y.-S., \& {Zhang}, X. 2006, Modern Physics
  Letters A, 21, 231, arXiv:astro-ph/0501652

\end{thebibliography}
\end{document}